\documentclass[
reprint,
longbibliography,
superscriptaddress,
preprintnumbers,
nobibnotes,
amsmath,amssymb,
aps,
pra,
floatfix
]{revtex4-1}

\usepackage{gensymb}
\usepackage{mathrsfs}
\usepackage{graphicx}
\usepackage{dcolumn}
\usepackage{bm}
\usepackage{floatrow}
\usepackage{titlesec}
\usepackage{cancel}
\usepackage{physics}
\usepackage{bbm}

\usepackage{multirow}

\titleformat*{\section}{\normalsize\bfseries\filcenter}
\titlespacing*{\section}
{0pt}{3.0ex}{2ex}

\usepackage[%
  colorlinks=true,
  urlcolor=blue,
  linkcolor=blue,
  citecolor=blue,
]{hyperref}
\usepackage{xcolor}
\usepackage{textcomp}
\usepackage{amsmath}
\usepackage{siunitx}
\usepackage[artemisia]{textgreek}
\usepackage[normalem]{ulem}
\usepackage{upgreek}
\usepackage{amssymb}
\usepackage{titlesec}
\titleformat*{\section}{\normalsize\bfseries\filcenter}

\titlespacing*{\section}
{0pt}{3.0ex}{1.5ex}

\titlespacing*{\subsection}
{0pt}{2.5ex}{0.5ex}
\titlespacing*{\subsubsection}
{0pt}{2.0ex}{0.5ex}

\usepackage{etoolbox}

\newcommand{\unm}{Center for High Technology Materials and Department of Physics and Astronomy, University of New Mexico, Albuquerque, NM, USA}

\begin{document}

\title{Nuclear quadrupole resonance spectroscopy with a femtotesla diamond magnetometer}

\author{Yaser Silani}
\email{yaser.silani@gmail.com}
\affiliation{\unm}

\author{Janis Smits}
\affiliation{\unm}

\author{Ilja Fescenko}
\affiliation{\unm}
\affiliation{Laser Center, University of Latvia, Riga, Latvia}

\author{Michael W. Malone}
\affiliation{Los Alamos National Laboratory, Los Alamos, NM, USA}

\author{Andrew F. McDowell}
\affiliation{NuevoMR, Albuquerque, NM, USA}

\author{Andrey Jarmola}
\affiliation{ODMR Technologies Inc., El Cerrito, CA, USA}
\affiliation{Department of Physics, University of California, Berkeley, CA, USA}

\author{Pauli Kehayias}
\affiliation{Sandia National Laboratories, Albuquerque, NM, USA}

\author{Bryan Richards}
\affiliation{\unm}

\author{Nazanin Mosavian}
\affiliation{\unm}

\author{Nathaniel Ristoff}
\affiliation{\unm}

\author{Victor M. Acosta}
\email{vmacosta@unm.edu}
\affiliation{\unm}


\begin{abstract} 
Sensitive Radio-Frequency (RF) magnetometers that can detect oscillating magnetic fields at the femtotesla level are needed for demanding applications such as Nuclear Quadrupole Resonance (NQR) spectroscopy. RF magnetometers based on Nitrogen-Vacancy (NV) centers in diamond have been predicted to offer femtotesla sensitivity, but published experiments have largely been limited to the picotesla level. Here, we demonstrate a femtotesla RF magnetometer based on an NV-doped diamond membrane inserted between two ferrite flux concentrators. The device operates in bias magnetic fields of $2\mbox{-}10~{\rm \upmu T}$ and provides a ${\sim}300$-fold amplitude enhancement within the diamond for RF magnetic fields in the $0.07\mbox{-}3.6~{\rm MHz}$ range. The magnetometer's sensitivity is ${\sim} 70~{\rm fT\,s^{1/2}}$ at $0.35~{\rm MHz}$, and the noise floor decreases to below $2~{\rm fT}$ after $1~{\rm hour}$ of acquisition. We used this sensor to detect the $3.6~{\rm MHz}$ NQR signal of $^{14}{\rm N}$ in sodium nitrite powder at room temperature. NQR signals are amplified by a resonant RF coil wrapped around the sample, allowing for higher signal-to-noise ratio detection. The diamond RF magnetometer's recovery time after a strong RF pulse is ${\sim} 35~{\rm \upmu s}$, limited by the coil ring-down time. The sodium-nitrite NQR frequency shifts linearly with temperature as $\mbox{-}1.00\,{\pm}\,0.02~{\rm kHz/K}$, the magnetization dephasing time is $T_2^{\ast}=887\pm51~{\rm \upmu s}$, and a spin-lock spin-echo pulse sequence extends the signal lifetime to $332\,{\pm}\,23~{\rm ms}$, all consistent with coil-based NQR studies. Our results expand the sensitivity frontier of diamond magnetometers to the femtotesla range, with potential applications in security, medical imaging, and materials science.

\end{abstract}

\maketitle

\section{\label{sec:Introduction}Introduction}
Magnetometers based on negatively-charged Nitrogen-Vacancy (NV) centers in diamond are promising room-temperature sensors for detecting magnetic phenomena across a wide range of frequencies~\cite{Bar2020}. Over the last decade, various sensing protocols and fabrication methods have been developed to improve the sensitivity of diamond magnetometers in the sub-$10\mbox{-}{\rm kHz}$~\cite{Bar2016,Fes2020,Eis2021,Zha2021}, Radio-Frequency (RF)~\cite{Mas2018,Gle2018,Smi2019}, and microwave (MW)~\cite{Wan2022,Als2022} frequency ranges. However, the best reported sensitivities, ${\sim}1~{\rm pT\,s^{1/2}}$~\cite{Fes2020}, still trail the achievable levels in magnetometers based on alkali-metal vapor~\cite{Sav2005,Led2007,Gri2010,Was2010,Dho2022,Cha2012,Ked2014} and superconducting quantum interference devices~\cite{Cla1980,Sch2016,Sto2017}. 

Recently, magnetic flux concentrators have been used to improve the performance of diamond magnetometers~\cite{Fes2020,Zha2021,Li2021,Xie2022}, and a sub-picotesla sensitivity has been realized for low frequencies (${\lesssim 1\,{\rm kHz}}$) by inserting an NV-doped diamond membrane between two ferrite cones~\cite{Fes2020}. We hypothesized that the same approach can be used for improving the sensitivity in the RF range (kHz-MHz) if the flux concentrator's magnetic properties do not degrade at such frequencies. In fact, better sensitivity may be expected in the RF range, since the diamond RF magnetometer can operate in a pulsed regime where the NV spin coherence time is significantly longer~\cite{Tay2008}.

A diamond RF magnetometer with femtotesla sensitivity may find immediate application as a non-inductive detector for Nuclear Quadrupole Resonance (NQR) spectroscopy~\cite{Con1990,Aug1998,Lee2006}. NQR spectroscopy is a solid-state analysis technique that provides a unique chemical fingerprint based on the coupling of nuclear quadrupole moments to their local electric field gradients~\cite{Das1958,Zax1985}. NQR spectroscopy is used to identify powder substances in ambient conditions for security~\cite{Gar2001,Kim2014,Mal2020} and pharmaceutical~\cite{Bal2005,Luz2013,Kyr2015} applications and to study the temperature-dependent properties of single-crystal materials~\cite{Lyf1976,Sha1986,Hor1990,Hue1999,Lan2010}. These applications typically require the ability to detect kHz-MHz frequencies, at low bias magnetic fields $\lesssim1~{\rm mT}$, with femtotesla sensitivity~\cite{Mal2016}. Previously, NV centers were used to detect NQR signals arising from nanoscale statistical polarization in single-crystal boron-nitride layers in direct contact with the diamond~\cite{Lov2017,Hen2022}. However, using NV centers to remotely detect powders remains an open challenge due to the need for high sensitivity~\cite{Shi2014}.

In this paper, we demonstrate a frequency-tunable diamond RF magnetometer with a sensitivity of ${\sim}70~{\rm fT\,s^{1/2}}$ at $0.35~{\rm MHz}$, using ferrite flux concentrators~\cite{Fes2020} and a multi-pulse synchronized readout scheme~\cite{Gle2018,Smi2019}. The sensitivity remains within a factor of three of this value for the $0.07{-}3.6~{\rm MHz}$ frequency range. We used the magnetometer to detect the $3.6~{\rm MHz}$ NQR signal of $^{14}{\rm N}$ in sodium nitrite powder samples~\cite{Oja1967,Pet1976}. Our work expands the sensitivity frontier of diamond magnetometry to the femtotesla range and introduces a new method for remote detection of solid-state magnetic resonance.

\section{\label{sec:exp}Experimental design}
\begin{figure}[t]
    \centering
\includegraphics[width=\columnwidth]{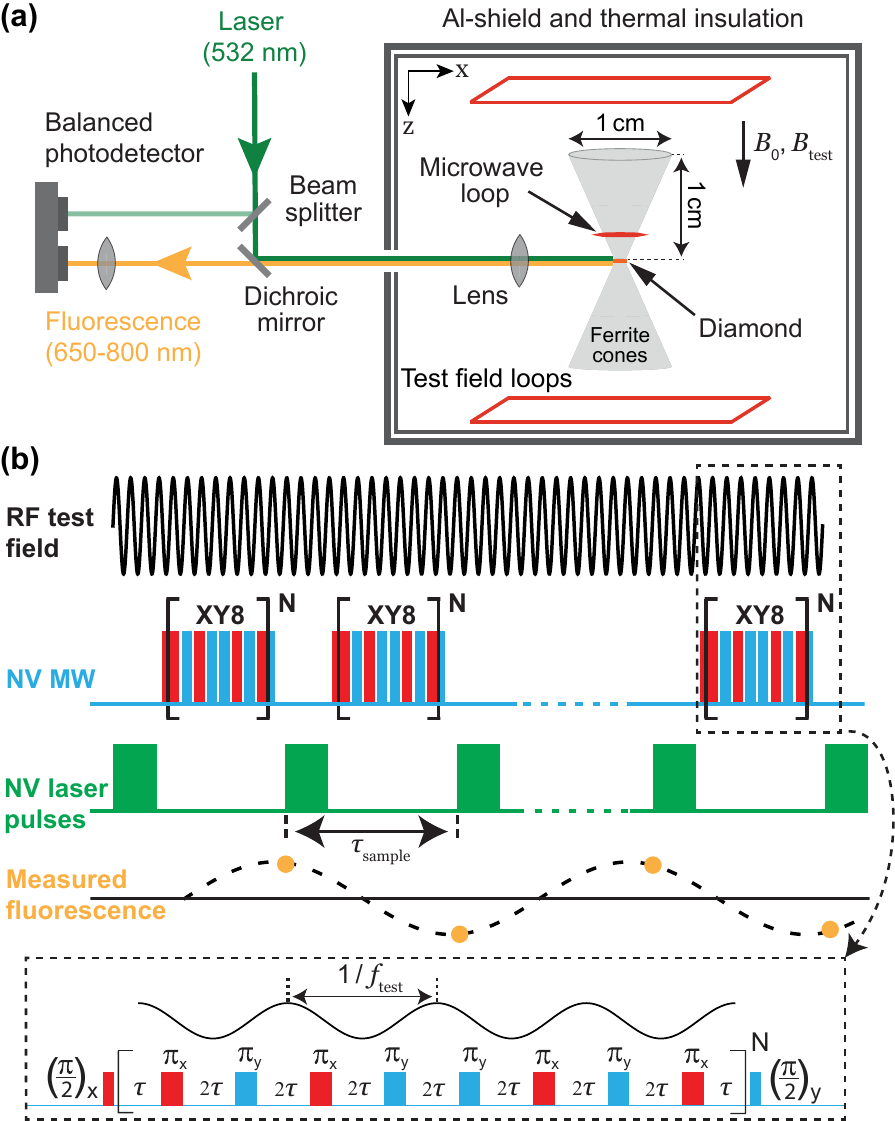}\hfill
\caption{\textbf{Experimental setup and RF detection scheme.} (a) Schematic of the experimental setup for detecting RF test fields. A permanent magnet (not shown) outside of the Al-shield was used to apply weak bias magnetic fields along the $z$-axis, $B_{\rm 0}\,{=}\, 2\mbox{-}10~{\rm \upmu T}$. Additional details of the setup can be found in \ref{sec:AppxSetup} and Ref.~\cite{Fes2020}. (b) Diamond RF magnetometry is performed with a continuous series of repeated XY8-$N$ pulse sequences on the NV electron spins~\cite{Smi2019}. Each XY8-$N$ sequence begins and ends with a resonant microwave (MW) $\pi/2$ pulse. Between the $\pi/2$ pulses, $8N$ resonant $\pi$-pulses, spaced by $2\tau=1/(2f_{\rm test}$), are applied with alternating phase. Following each XY8-$N$ sequence, a $12\mbox{-}{\rm \upmu s}$ laser pulse ($0.2~{\rm W}$, $532~{\rm nm}$) is applied for optical readout and repolarization of the NV centers. The resulting time trace of NV fluorescence readouts is proportional to an aliased version of the RF test field.}
\label{fig:AC setup}
\end{figure}

A schematic of the diamond RF magnetometer is shown in Fig.~\ref{fig:AC setup}(a). The apparatus is similar to one previously used for low-frequency magnetometry~\cite{Fes2020}, with modifications (\ref{sec:AppxSetup}) made for RF magnetometry and NQR spectroscopy. For example, the mu-metal shield is replaced by an aluminum shield to suppress RF interference. A thermal-insulation housing with thermoelectric temperature control is added around the shield to improve thermal stability. Calibrated RF test fields with a magnitude $B_{\rm test}$ and frequency $f_{\rm test}$ are applied along the magnetometer sensing axis ($z$-axis) using a pair of low-inductance rectangular wire loops.

A permanent magnet is used to compensate the ambient magnetic field and apply a weak bias field, $B_0=2\mbox{-}10~{\rm \upmu T}$, approximately along the $z$-axis. The flux concentrators enhance the $z$-component of the magnetic field in the gap between the cone tips. An NV-doped diamond membrane is positioned in the gap with its (100) crystal faces normal to the $z$-axis. This geometry results in two NV spin transition frequencies $f_{\pm}\approx D\,\pm\,\epsilon\gamma_{\rm nv}B_0/\sqrt{3}$, where $D=2.87~{\rm GHz}$ is the NV zero field splitting, $\epsilon$ is the flux-concentrator enhancement factor ($\epsilon\approx300$ for DC fields)~\cite{Fes2020}, $\gamma_{\rm nv}=28~{\rm GHz/T}$ is the NV gyromagnetic ratio, and the factor of $\sqrt{3}$ comes from the $55\degree$ angle of each NV axis with respect to the $z$-axis.

Figure~\ref{fig:AC setup}(b) shows the XY8-$N$ synchronized readout pulse sequence used for diamond RF magnetometry~\cite{Smi2019}. This pulse sequence contains a series of short MW pulses ($10\mbox{-}70~{\rm ns}$), of alternating phase, that are tuned to one of the $f_{\pm}$ resonances in the $2.7\mbox{-}3.0~{\rm GHz}$ range. Each XY8-$N$ sequence is followed by a $12\mbox{-}{\rm \upmu s}$ laser pulse ($0.2~{\rm W}$, $532~{\rm nm}$), generated with an acousto-optic modulator, for optical readout and spin polarization of the NV centers. The sequence is repeated continuously, and the resulting time trace of NV fluorescence readouts is approximately proportional to an aliased version of the applied RF field sampled at the time of the first $\pi/2$ pulse of each XY8-$N$ sequence (\ref{sec:AppxPSN1}). For a magnetic field oscillating with frequency $f$, the NV fluorescence signal oscillates with an alias frequency $f_{\rm alias}=f-f_{\rm ref}$. The reference frequency, $f_{\rm ref}$,  depends on the sampling rate $1/\tau_{\rm sample}$, which is adjusted by varying a sub-${\rm \upmu s}$ dead time after each XY8-$N$ sequence~\cite{Smi2019}.

A theoretical bound on the diamond RF magnetometer sensitivity is set by photoelectron shot noise as:
\begin{equation}
\label{eq:psn}\eta_{\rm psn}~{\approx}~\frac{\sqrt{3}\,\xi}{\epsilon\,\sqrt{\delta}}\frac{1}{4\,\gamma_{\rm nv}\,C\sqrt{n_{\rm nv}\,V_{\rm sen}\,\phi\,\tau_{\rm tot}}}.
\end{equation}
In the first term of Eq.~\ref{eq:psn}, the $\sqrt{3}$ factor comes from the $55\degree$ angle of each NV axis with respect to the $z$-axis, $\xi\,{=}\,1.6$ accounts for additional photoelectron noise arising from the balanced detection and normalization procedure (\ref{sec:AppxPSN1}), and $\delta\,{=}\,\tau_{\rm tot}/\tau_{\rm sample}\,{\approx}\,0.75$ is a duty cycle, where $\tau_{\rm tot}$ is the NV phase accumulation time and $\tau_{\rm sample}$ is the sequence repetition time. In the second term, $C$ is the XY8-$N$ fluorescence contrast (\ref{sec:AppxPSN1}), $n_{\rm nv}$ is the NV concentration, $V_{\rm sen}$ is the illuminated NV sensor volume, and $\phi$ is the probability of detecting a photoelectron per NV center for a single readout. 

\begin{figure*}[ht]
\includegraphics[width=1.0\textwidth]{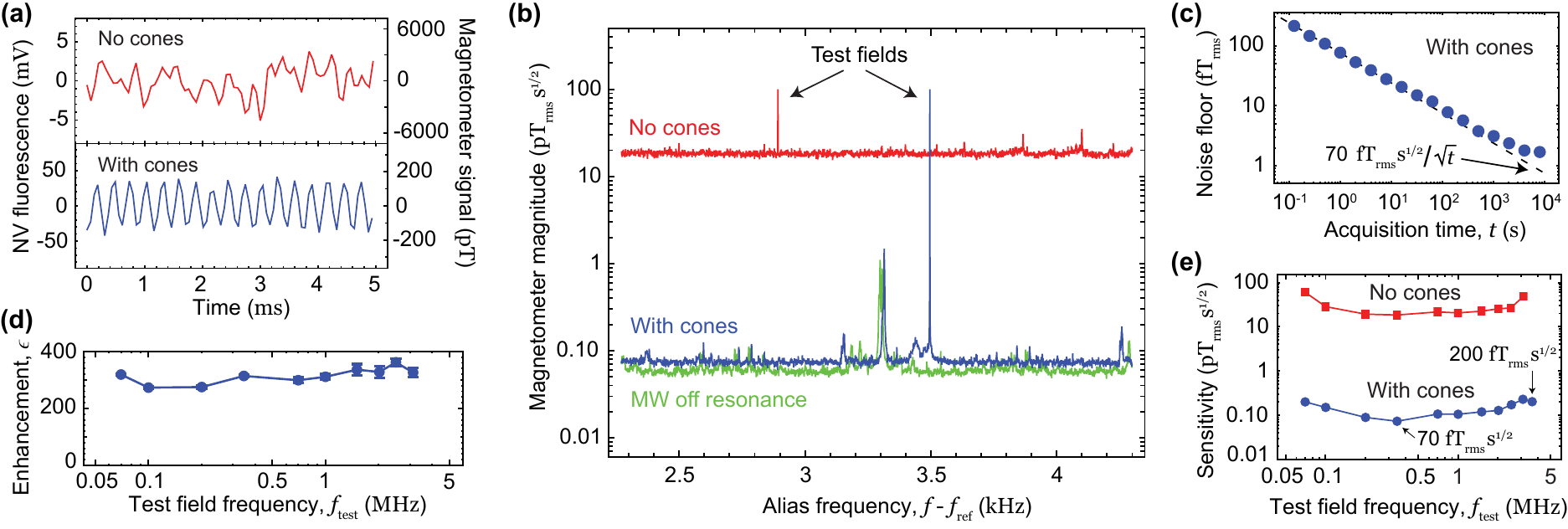}\hfill
\caption{\textbf{Femtotesla diamond RF magnetometry.} (a) Real-time NV fluorescence signal with (blue) and without (red) ferrite cones in the magnetometer assembly. In each case, an XY8-4 synchronized readout sequence was used, and a $0.35~{\rm MHz}$ test field with $100~{\rm pT_{rms}}$ amplitude was applied. The NV fluorescence photodetector voltage is converted to magnetic field units using the procedure described in \ref{sec:AppxCalib1}. For the same applied field amplitude, the photodetector voltage signal with cones is ${\sim}220$ times larger than that without cones, due primarily to the flux concentrator enhancement (and, to a lesser extent, differences in the photon collection efficiency). A digital $4.5~{\rm kHz}$ low-pass filter was applied to the data without cones for better visualization. (b) Fourier transform spectra of the NV signals with and without the cones. No digital filtering was applied. The noise floor reaches ${\sim}70~{\rm fT\,s^{1/2}}$ with the cones and ${\sim}18~{\rm pT\,s^{1/2}}$ without them. A reference spectrum (green), obtained by detuning the MW frequency $200~{\rm MHz}$ off the NV resonance, shows an effective noise floor of ${\sim}60~{\rm fT\,s^{1/2}}$. (c) Noise floor with the cones present as a function of acquisition time, $t$. 
(d) Ferrite cones enhancement factor versus the test field frequency (\ref{sec:Appxepsilon}). (e) Diamond RF magnetometer sensitivity as a function of test field frequency with (blue circles) and without (red squares) the ferrite cones (\ref{sec:AppxContrst}). For all data in this figure, except for the ``MW off resonance'' spectrum in (b), the MW frequency was tuned to one of the NV $f_{\pm}$ resonances.
}
\label{fig:fT sensor}
\end{figure*}

The natural-isotopic-abundance single-crystal diamond membrane used in this work contained an initial nitrogen concentration of ${\sim}20~{\rm ppm}$. The diamond was doped with NV centers by electron irradiation and annealing (\ref{sec:AppxSetup}), resulting in an NV concentration of $n_{\rm nv}\approx3~{\rm ppm}$ and a transverse NV spin coherence time of ${\sim}50~{\rm \upmu s}$ for an XY8-4 sequence. The diamond was subsequently cut and polished into a (100)-oriented membrane with dimensions ${\sim}300\times300\times35~{\rm \upmu m^3}$. The NV sensor volume is defined by the area of the illuminating laser beam and the length of its path in the diamond as $V_{\rm sen}\approx2\times10^5~{\rm \upmu m^3}$. The peak detected photoelectron current was typically ${\sim}0.17~{\rm mA}$ over a $2~{\rm \upmu s}$ readout window, which indicates the detection efficiency of our setup is $\phi\approx0.02$. We found the best sensitivity of our setup to occur for XY8-4 sequences at $f_{\rm test}=0.35~{\rm MHz}$, where $\tau_{\rm tot}=44~{\rm \upmu s}$ and $C\approx0.01$. With these values, and assuming the flux-concentrator enhancement factor $\epsilon\approx300$ is the same as for DC fields, Eq.~\eqref{eq:psn} predicts a photoelectron-shot-noise-limited sensitivity $\eta_{\rm psn}\approx30~{\rm fT\,s^{1/2}}$.

\section{\label{sec:Magnetometer} Magnetometer characterization}
We measured the magnetometer's sensitivity as a function of test field frequency and acquisition time. First, we applied a test field with frequency $f_{\rm test}=0.35~{\rm MHz}$ and magnitude $B_{\rm test}=100~{\rm pT_{rms}}$ (\ref{sec:AppxCalib1}) and recorded the magnetometer signal under an XY8-4 synchronized readout sequence for 100~{\rm s} with and without the ferrite cones. Figure~\ref{fig:fT sensor}(a) shows the real-time NV fluorescence signal for a segment of each time trace. While the magnetometer signal without ferrite cones is dominated by noise, a clear oscillation at frequency $f_{\rm test}-f_{\rm ref}=3.5~{\rm kHz}$ is observed with the cones present. 

To determine the magnetometer sensitivity, each $100~{\rm s}$ NV time trace is divided into one hundred $1~{\rm s}$ segments, a spectrum is obtained for each segment by taking the absolute value of the Fourier transform, and the 100 spectra are averaged together. Figure~\ref{fig:fT sensor}(b) shows the resulting magnetic spectra for the recordings with and without the ferrite cones. The magnetic field sensitivity, defined here as the average noise floor for $1\mbox{-}{\rm s}$ acquisition time in a few-hundred-Hz band near the signal frequency (\ref{sec:AppxContrst}), is ${\sim}70~{\rm fT_{rms}\,s^{1/2}}$ with the cones and ${\sim}18~{\rm pT_{rms}\,s^{1/2}}$ without them. A reference spectrum (with ferrite cones) was obtained by detuning the MW frequency $200~{\rm MHz}$ off the NV resonance, revealing an effective noise floor of ${\sim}60~{\rm fT_{rms}\,s^{1/2}}$. While the ${\sim}70~{\rm fT_{rms}\,s^{1/2}}$ measured noise floor with ferrite cones is ${\sim}2$ times greater than the photoelectron-shot-noise estimate (\ref{sec:AppxPSN2}), the experimental sensitivity represents a ${\gtrsim}10$-fold improvement over previous diamond magnetometry studies~\cite{Fes2020} and a ${\gtrsim}100$-fold improvement over previous diamond studies in the RF range~\cite{Mas2018,Gle2018,Smi2019}.

To characterize temporal stability, we continuously recorded the ferrite-cones diamond RF magnetometer signal for several hours. Figure~\ref{fig:fT sensor}(c) shows the magnetic noise floor as a function of averaging time, $t$. We find that the noise floor scales with the expected ${\sim}70~{\rm fT\,s^{1/2}}/{\sqrt{t}}$ behavior out to $t\gtrsim10^3~{\rm s}$. The noise floor decreases to below $2~{\rm fT}$ after 1 hour of acquisition, before leveling off.

We used ferrite cones made of a manganese-zinc material (MN60) that is usually considered more suitable for low-frequency (${\lesssim}\,1~{\rm MHz}$) applications~\cite{MN60}, and it was initially unclear whether large enhancement factors would be possible at higher frequency~\cite{Bol2016}. To probe the frequency dependence, we recorded the diamond RF magnetometer signal for different values of $f_{\rm test}$. The enhancement factor provided by the ferrite cones is defined as $\epsilon=B_{\rm gap}/B_{\rm test}$, where $B_{\rm gap}$ is the magnetic field amplitude within the diamond when the cones are present. For each value of $f_{\rm test}$, we calibrated $B_{\rm test}$ by recording the NV signal amplitude (without cones) as a function of the current amplitude applied to the test field loops. We repeated the process with the cones present to calibrate $B_{\rm gap}$. At each frequency, $\epsilon$ was estimated from the ratio of the response curves (\ref{sec:Appxepsilon}). The results are shown in Fig.~\ref{fig:fT sensor}(d). Surprisingly, we find the enhancement factor is nearly constant ($\epsilon\approx300$) in the $0.07{-}3.12~{\rm MHz}$ frequency range. 

We also determined the magnetic sensitivity as a function of frequency, using the process described for Fig.~\ref{fig:fT sensor}(a,b). Figure~\ref{fig:fT sensor}(e) shows the sensitivity as a function of $f_{\rm test}$. For each frequency, the duration and spacing of the MW pulses and the length of the XY8-$N$ sequence were modified to maximize sensitivity (see Table~\ref{tab:sens} in \ref{sec:AppxContrst}). With the cones, the best sensitivity is ${\sim}70~{\rm fT\,s^{1/2}}$ at $f_{\rm test}=0.35~{\rm MHz}$, and the sensitivity remains within a factor of 3 of this value throughout the range $0.07\mbox{-}3.62~{\rm MHz}$. A diamond with a lower nitrogen concentration, and thus longer coherence time, can be used to extend the frequency range down to ${\sim}1~{\rm kHz}$~\cite{Bar2020}. A stronger MW field, combined with a higher-NV-concentration diamond (to limit the number of MW pulses without sacrificing sensitivity), could extend the frequency up to ${\gtrsim}10~{\rm MHz}$, as long as the ferrite's permeability and relative loss factor do not degrade~\cite{Fes2020}.

\begin{figure}[t]
    \centering
\includegraphics[width=\columnwidth]{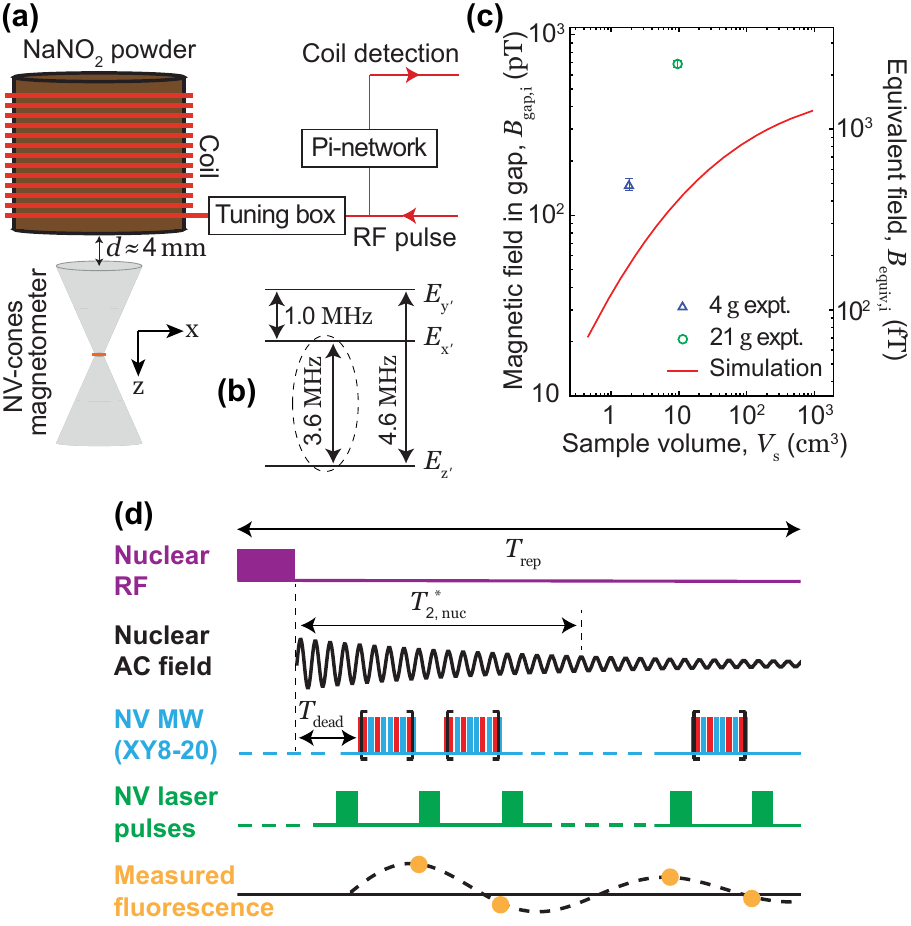}\hfill
\caption{\textbf{NQR setup.} (a) Schematic of the setup used for NQR spectroscopy. A bias field $B_0\approx10~{\rm \upmu T}$ is applied along the $z$-axis. An NaNO$_2$ sample is housed in a plastic cylinder container and placed $d\approx4~{\rm mm}$ above the ferrite-cones diamond RF magnetometer. A resonant RF coil is wrapped around the sample container. The $3.6~{\rm MHz}$ NQR transition of $^{14}{\rm N}$ nuclei in ${\rm NaNO}_{\rm 2}$ is excited by applying RF pulses along the $z$-axis. The resulting oscillating nuclear magnetic field is also along the $z$-axis and is simultaneously detected by the ferrite-cones diamond RF magnetometer and the resonant RF coil (see~\ref{sec:AppxNQRcoil}). (b) Energy levels and nuclear spin transitions of $^{14}{\rm N}$ in ${\rm NaNO}_{\rm 2}$ at room temperature and low ($\lesssim1~{\rm mT}$) magnetic field. (c) Initial magnetic field amplitude within the diamond, $B_{\rm gap,i}$ (left axis), and equivalent magnetic field, $B_{\rm equiv,i}=B_{\rm gap,i}/300$ (right axis), as a function of sample volume. expt. -- experiment. (d) Pulse sequence used for NV NQR detection (\ref{appxnqrtime}). After an RF excitation pulse, an XY8-20 synchronized readout pulse sequence is used to detect an aliased version of the nuclear AC magnetic field. The entire sequence is repeated every $T_{\rm rep}=0.5~{\rm s}$. }
\label{fig:NQR setup}
\end{figure}

\section{\label{sec:NQR} NQR spectroscopy of NaNO$_2$ powder} 
\subsection{\label{sec:nqrtheory} Experimental design and theoretical estimates}
Having demonstrated femtotesla sensitivity in the RF range, we next used our ferrite-cones diamond RF magnetometer as a detector in NQR spectroscopy. The sample we selected to study is sodium nitrite (NaNO$_2$) powder (\ref{sec:appxNQRNaNO2}), a well-studied standard for $^{14}$N NQR spectroscopy~\cite{Fis1999,Hib2008}. Figure~\ref{fig:NQR setup}(a) shows a schematic of the NQR detection setup. A resonant RF coil is wrapped around a NaNO$_2$ powder sample, and the sample is placed ${\sim}4~{\rm mm}$ above the ferrite-cones diamond RF magnetometer. Two coil assemblies are used: one for a $4\mbox{-}{\rm gram}$ sample and the other for a $21\mbox{-}{\rm gram}$ sample. A capacitor tuning circuit and pi-network are used for conventional inductive detection (\ref{sec:AppxNQRcoil}).

\begin{figure*}[hbt]
\includegraphics[width=1.0\textwidth]{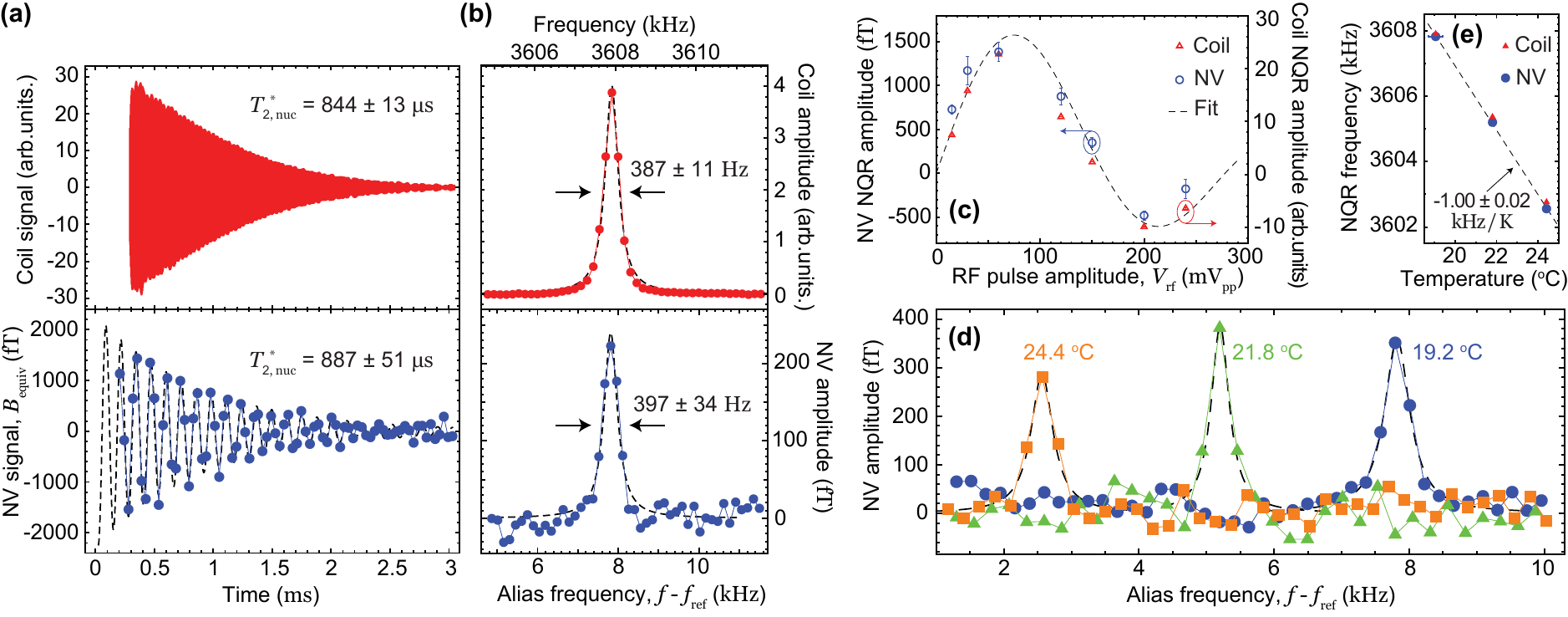}\hfill
\caption{\textbf{NQR spectroscopy of $\boldsymbol{^{14}{\rm N}}$ in NaNO$_2$.} (a) Room-temperature time-domain NQR signal of $^{14}{\rm N}$ in a 21-gram NaNO$_2$ powder sample acquired by the resonant RF coil (top) and diamond RF magnetometer (bottom). A $50\mbox{-}{\rm \upmu s}$ RF pulse at $3605~{\rm kHz}$ was used to excite the sample via the resonant RF coil with loaded quality factor $Q\,{\approx}\,23$ (see \ref{sec:AppxNQRcoil}). The sequence was repeated every $500~{\rm ms}$, and the signal was averaged over 86000 repetitions. A digital bandpass filter is applied for better visualization: $3.58\mbox{-}3.63~{\rm MHz}$ for the coil signal and $4.6\mbox{-}11.7~{\rm kHz}$ for the NV signal. A decaying sinusoidal function fit to the NV signal reveals an initial amplitude $B_{\rm equiv,i}=2300\pm 115~{\rm fT}$ and nuclear dephasing time $T_{\rm 2,\,nuc}^{\ast}=887\pm 51~{\rm \upmu s}$. (b) Imaginary part of the Fourier transform of the time-domain NQR signals shown in (a), along with Lorentzian fits. (c) NQR signal amplitude as a function of RF pulse amplitude, $V_{\rm rf}$ (measured in volts, prior to amplification), applied at $3607.5~{\rm kHz}$ for $300~{\rm \upmu s}$. For each RF pulse amplitude, the imaginary part of the Fourier transform of the first $810~{\rm \upmu s}$ of the signals is calculated, such that the NQR resonance is contained in a single frequency point. The value of that point is taken as the NQR amplitude, and the error bars are the standard deviation of points within a $5~{\rm kHz}$ band near resonance. The dashed black line is a fit to a function $J_{\rm 3/2}({\alpha})/{\sqrt{\alpha}}$ with the first peak occurring at the nutation angle $\alpha= 119^{\degree}$~\cite{Veg1974}. (d) NV NQR spectra (imaginary part of Fourier transform) obtained for three different ambient temperatures, along with Lorentzian fits. RF pulses were applied at $3605~{\rm kHz}$ for $50\mbox{-}{\rm \upmu s}$. (e) NQR resonance frequency as a function of ambient temperature, along with linear fit.}
\label{fig:21gr NQR}
\end{figure*}

The nuclear quadrupole Hamiltonian is given by~\cite{Sui2006}:
\begin{equation}
\setlength\abovedisplayskip{5pt}
\label{eq:Hnqr}H_{\rm Q}\,{=}\,f_{\rm Q}\,[I_{\rm z^{'}}^{2}+\frac{\eta}{3}(I_{\rm x^{'}}^{2}-I_{\rm y^{'}}^{2})],
\setlength\belowdisplayskip{5pt}
\end{equation}
where $f_{\rm Q}$ is the quadrupole coupling frequency, $\eta$ is the asymmetry parameter, and $\{I_{\rm x^{'}},I_{\rm y^{'}},I_{\rm z^{'}}\}$ are the spin components along the principle axes of a given crystallite. As depicted in Fig.~\ref{fig:NQR setup}(b), at low magnetic field ($B_0\lesssim1~{\rm mT}$), the $^{14}{\rm N}$ nucleus in NaNO$_2$ ($I{=}1$, $f_{\rm Q}{=}4.1~{\rm MHz}$, $\eta{=}0.38$) has three non-degenerate energy levels, $\{E_{\rm z^{'}},E_{\rm x^{'}},E_{\rm y^{'}}\}$, and magnetic-dipole transitions are allowed between each level~\cite{Oja1967}. We used our sensor to detect the $E_{\rm z^{'}}\leftrightarrow E_{\rm x^{'}}$ transition at $f_{\rm nqr}=f_{\rm Q}(1-\eta/3)=3.6~{\rm MHz}$. For powder samples, where many crystallites are randomly oriented, application of a resonant RF pulse along the $z$-axis produces a net oscillating magnetization (frequency $f_{\rm nqr}$) along the $z$-axis~\cite{Blo1955}, see \ref{sec:AppxNQRmagn}. Thus, to maximize the NQR signal, the RF coil axis was aligned with the magnetometer detection axis, Fig.~\ref{fig:NQR setup}(a).

We carried out simulations to estimate the oscillating magnetic field amplitude produced by cylindrical NaNO$_2$ samples following an optimal RF excitation pulse on the $3.6~{\rm MHz}$ transition. The initial amplitude of the oscillating sample magnetization was estimated to be $M_0=3.3~{\rm \upmu A/m}$ along the $z$-axis (see \ref{sec:AppxNQRmagn}). A finite-element model was used to make a preliminary estimate of the resulting initial magnetic field amplitude in the diamond when the ferrite cones were present, $B_{\rm gap,i}$. For this and all subsequent NQR measurements, $B_{\rm gap}$ is converted to an equivalent magnetic field $B_{\rm equiv}=B_{\rm gap}/\epsilon$, using $\epsilon=300$ (see~\ref{sec:Appxepsilon}), to compare with the case of uniform magnetic fields. 

For a given sample volume $V_s$, we swept the cylinder aspect ratio to estimate the maximum possible nuclear field amplitude~(\ref{sec:AppxSimuMagnCyl}). Figure~\ref{fig:NQR setup}(c) shows the maximum simulated $B_{\rm equiv,i}$ as a function of sample volume. The estimated $B_{\rm equiv,i}$ values are at the few-hundred femtotesla level for the $V_s=1\mbox{-}10^3~{\rm cm^3}$ range. Figure~\ref{fig:NQR setup}(c) also shows experimentally-measured NQR signals from two sample masses (these measurements are described below). The experimental values are $3\mbox{-}6$ times larger than the simulated estimates, despite several optimistic assumptions such as perfect powder packing, optimal RF excitation pulse, and ideal sample aspect ratio. As discussed below, this difference is due to signal amplification from the resonant RF coil used in the experiment.

Figure~\ref{fig:NQR setup}(d) shows the pulse sequence used for NQR spectroscopy. RF excitation pulses (typically $50\mbox{-}200~{\rm \upmu s}$) are applied to the resonant RF coil, and the oscillating nuclear magnetic field is detected by the diamond RF magnetometer using a series of repeated XY8-20 pulse sequences with $f_{\rm ref}=3600.07~{\rm kHz}$. After a duration $T_{\rm rep}\approx0.5~{\rm s}$, chosen to be comparable to the $^{14}$N thermal relaxation time, $T_{\rm 1,\,nuc}\approx0.3~{\rm s}$~\cite{Pet1976}, the entire sequence is repeated. To compare to conventional NQR detection, the same coil is also used to detect the signal inductively after it is passed through a pi-network and amplified by a low-noise pre-amplifier, (see Fig.~\ref{fig:NQR setup}(a), \ref{sec:AppxNQRcoil}).

\subsection{\label{sec:nqr1pulse} Experimental results with a single RF pulse}
Figure~\ref{fig:21gr NQR}(a) shows the NQR signals of the 21-gram NaNO$_2$ powder sample detected by the resonant RF coil and ferrite-cones diamond RF magnetometer. The signals are fit with exponentially-decaying sinusoidal functions, and the fitted $1/e$ decay times are $T_{\rm 2,\,nuc}^{\ast}\,{=}\,844\pm13~{\rm \upmu s}$ for the coil signal and $T_{\rm 2,\,nuc}^{\ast}\,{=}\,887\pm51~{\rm \upmu s}$ for the NV signal. These values are in good agreement with each other and consistent with literature values~\cite{Pet1976}.

The NV signal in Fig.~\ref{fig:21gr NQR}(a) has an initial amplitude $B_{\rm equiv,i}\,{=}\,2300\pm 115~{\rm fT}$, which is a factor of $6$ higher than the simulation in Fig.~\ref{fig:NQR setup}(c). The discrepancy comes from induction in the resonant RF coil wrapped around the sample in the experiment~\cite{Qiu2007,Gre2021}. The oscillating sample magnetization induces an oscillating current in the coil. The current is resonantly amplified and produces a larger oscillating field with a phase shift. This AC magnetic field can be described as arising from an effective magnetization throughout the resonant RF coil:
\begin{equation}
\label{eq:EffMag}
\setlength\abovedisplayskip{5pt}
M_{\rm eff}\approx Q\frac{m_s}{V_c\, \rho_s} M_0,
\setlength\belowdisplayskip{5pt}
\end{equation}
where $Q\gg1$ is the coil's quality factor, $V_c$ is the coil volume, $m_s$ is the sample mass, $\rho_s$ is the sample's crystal density, and $M_0$ is the initial amplitude of the oscillating sample magnetization. Taking the parameters used for the experiments in Fig.~\ref{fig:21gr NQR} ($Q=23$, $V_c=20~{\rm cm^3}$, $m_s=21~{\rm g}$, $\rho_s=2.17~{\rm g/cm^3})$ and assuming the maximum initial magnetization, $M_0=3.3~{\rm \upmu A/m}$ (\ref{sec:AppxNQRmagn}), we find $M_{\rm eff}\approx37~{\rm \upmu A/m}$. We used the finite-element model to evaluate the magnetic field within the diamond, assuming $M_{\rm eff}$ is uniform throughout the excitation coil volume (\ref{sec:AppxSimuMagnCyl}). After converting to the effective magnetic field, the estimated initial amplitude is $B_{\rm equiv,i}\,{\approx}\,3100~{\rm fT}$. This is only a factor of ${\sim}1.35$ larger than the experimental value, and the remaining difference may be due to imperfect RF excitation.

Figure~\ref{fig:21gr NQR}(b) shows the NQR frequency spectra detected by both coil and NV sensors along with Lorentzian fits. For the NV NQR spectrum, the fitted resonance frequency is $f_{\rm nqr}=f_{\rm alias}+f_{\rm ref}=3607.883\pm0.013~{\rm kHz}$, which is in reasonable agreement with the fitted coil-detected NQR frequency of $3607.908\pm0.004~{\rm kHz}$.

Figure~\ref{fig:21gr NQR}(c) shows the NQR signal amplitude as a function of the RF pulse amplitude $V_{\rm rf}$, for a pulse length $t_{\rm rf}\,{=}\,300~{\rm \upmu s}$. The signal amplitude is well described by the function $S_{\rm nqr}(V_{\rm rf})\,{=}\,S_{\rm max}\,J_{\rm 3/2}({\alpha})/(0.436\,{\sqrt{2\alpha/\pi}})$ (see Ref.~\cite{Veg1974} and \ref{sec:AppxNQRmagn}), where $J_{\rm 3/2}$ is the Bessel function of order 3/2, $\alpha\,{=}\,2\pi\gamma_{\rm n} K V_{\rm rf}t_{\rm rf}$ is the nutation angle, $\gamma_{\rm n}$ is the nuclear spin gyromagnetic ratio, and $K V_{\rm rf}=B_{\rm rf}$ is the applied RF magnetic field amplitude with fitted conversion factor $K$ (\ref{sec:AppxRFamplify}). The maximum signal amplitude, $S_{\rm max}$, occurs following an RF excitation pulse with $\alpha\,{\approx}\,2.08\,{\rm rad}\,{=}\,119^{\degree}$~\cite{Veg1974}. 

\subsection{\label{sec:nqrtemp} Temperature dependence of NQR frequency}
The temperature dependence of NQR frequencies provides insight into crystal structure~\cite{Lyf1976,Sha1986,Hor1990,Hue1999,Lan2010}, and it can be used to validate the interpretation of spectra.  We studied the temperature dependence of the $3.6~{\rm MHz}$ NaNO$_2$ NQR transition near room temperature by controlling the apparatus temperature with a Peltier element and measuring the ambient temperature with a thermistor located near the sample (\ref{sec:appxNQRtemp}). Figure~\ref{fig:21gr NQR}(d) shows the NV NQR spectrum for three different temperatures. Each ${\sim}2.6~{\rm K}$ step in temperature leads to a shift of the resonance of several linewidths. The NQR central frequencies are plotted as a function of temperature in Fig.~\ref{fig:21gr NQR}(e). A linear fit yields the coefficient $\mbox{-}1.00\,{\pm}\,0.02~{\rm kHz/K}$, which is consistent with previous measurements near room temperature ~\cite{Oja1967,Pet1976}.

\begin{figure}[t]
    \centering
\includegraphics[width=\columnwidth]{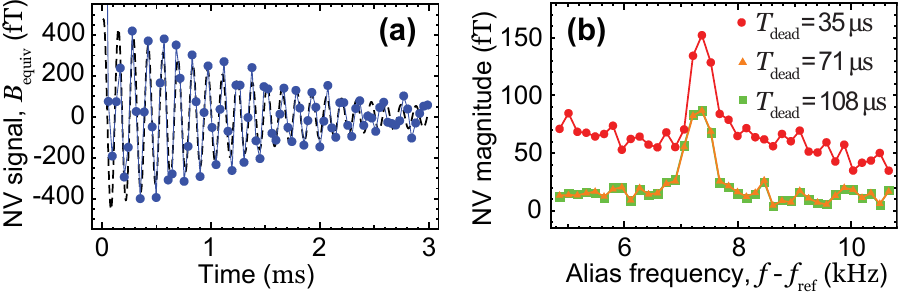}\hfill
\caption{\textbf{NV NQR recovery time.} (a) Time-domain NV NQR signal of 4 grams of ${\rm NaNO}_{\rm 2}$ powder enclosed in a $Q\,{\approx}\,8$ resonant RF coil. The RF pulse was applied at $3608~{\rm kHz}$ for $200~{\rm \upmu s}$. A digital bandpass filter ($4.7\mbox{-}10.8~{\rm kHz}$) was applied for better visualization. (b) NV NQR spectrum (absolute value of Fourier transform) for three different deadtimes. $T_{\rm dead}$ is defined in Fig.~\ref{fig:NQR setup}(d).}
\label{fig:NV_NQR deadtime}
\end{figure}

\subsection{\label{sec:nqrdead} Sensor recovery time following an RF pulse}
To detect NQR signals from samples with short dephasing times~\cite{Ver1962,Aug1998}, a sensor with a short recovery time following an RF excitation pulse is desired. For inductive detection, the recovery time is determined by the resonant RF coil's ring-down, and it can be ${\ll}100~{\rm \upmu s}$ for low-$Q$ coils. However, some non-inductive detectors like alkali-metal vapor magnetometers have significantly longer recovery times, ${\gtrsim}1~{\rm ms}$~\cite{Lee2006,Dho2022}. We hypothesized that the NV sensor's recovery time should be limited by either the coil ring-down or the NV polarization time ($12~{\rm \upmu s}$ in our experiment), whichever is longer. To measure the recovery time, we used a 4-gram ${\rm NaNO}_{\rm 2}$ powder sample and a resonant RF coil with a loaded quality factor $Q\,{\approx}\,8$ (\ref{sec:AppxNQRcoil}). The smaller sample and lower $Q$ were chosen to test the limits of mass sensitivity and recovery time of our apparatus. Figure~\ref{fig:NV_NQR deadtime}(a) shows the time-domain NV NQR signal following a $200\mbox{-}{\rm \upmu s}$ RF pulse (${\alpha}\,{\approx}\,119^{\degree}$). Fig.~\ref{fig:NV_NQR deadtime}(b) shows the Fourier transform spectrum for three different ``deadtimes'', computed by dropping the corresponding initial data points from the time-domain data in Fig.~\ref{fig:NV_NQR deadtime}(a). For a deadtime of ${\sim}35~{\rm \upmu s}$, the NQR peak is still prominent above the background. The coil-detected signal exhibits a similar recovery time, suggesting that this timescale is limited by the coil ring-down time and not by properties of the NV centers.

The initial amplitude in Fig.~\ref{fig:NV_NQR deadtime}(a), $B_{\rm equiv,i}\,{=}\,495\,{\pm}\,38~{\rm fT}$, is consistent with a modest amplification due to the resonant RF coil. Using Eq.~\ref{eq:EffMag}, and correcting for the coil standoff from the sensor, the upper bound on the initial signal amplitude from the $4\mbox{-}\rm g$ sample is $B_{\rm equiv,i}\approx760~{\rm fT}$ (\ref{sec:AppxSimuMagnCyl}), which is only a factor of ${\sim}1.5$ larger than the experimental value. 

\subsection{\label{sec:nqrslse} Spin-lock spin-echo spectroscopy}
\begin{figure}[t]
    \centering
\includegraphics[width=\columnwidth]{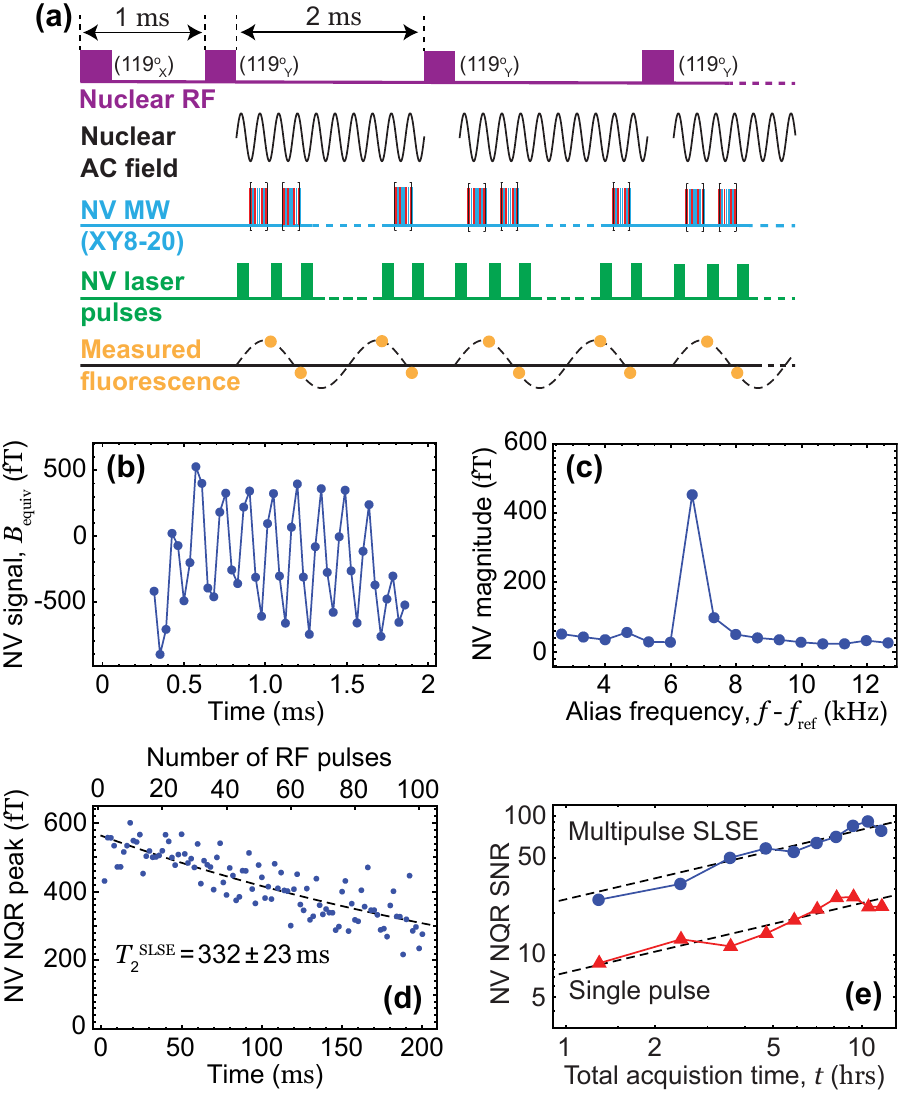}\hfill
\caption{\textbf{SLSE NV NQR signal.} (a) Spin-Lock Spin-Echo (SLSE) pulse sequence. Following an initial RF pulse, one hundred echo pulses are applied every $2~{\rm ms}$. All RF pulses are applied at $3608~{\rm kHz}$ for $50~{\rm \upmu s}$ ($\alpha\approx119^{\degree}$), and the initial pulse has a $90^{\degree}$ phase shift with respect to all the echo pulses. Meanwhile, a synchronized XY8-20 MW pulse sequence is applied to the NV centers, with $f_{\rm ref}=3600.07~{\rm kHz}$. The entire sequence is repeated every second. (b) Time-domain NV SLSE signal (coherent average of the first 20 echos) from the 21-g sample. (c) SLSE NQR spectrum obtained from the absolute value of the Fourier transform of data in (b). (d) SLSE signal magnitude as a function of the time passed since the first RF pulse. The fitted exponential decay constant is $T_{\rm 2}^{\rm SLSE}=332\pm23~{\rm ms}$. (e) Signal-to-Noise Ratio (SNR) of NV NQR signals as a function of total acquisition time, $t$, for both SLSE and single-RF-pulse protocol. The dashed black lines are fits to a $\sqrt{t}$ dependence.}
\label{fig:NQR SLSE}
\end{figure}

Nuclear spins in most room-temperature solids have the property $T_{\rm 2,\,nuc}^{\ast}\,{\ll}\,T_{\rm 1,\,nuc}$. This implies a low duty cycle for NQR readout, since $T_{\rm 2,\,nuc}^{\ast}$ limits the spin-precession acquisition time and $T_{\rm 1,\,nuc}$ bounds the re-thermalization time needed to repeat the sequence. One technique to increase the NQR readout duty cycle is to apply a Spin-Lock Spin-Echo (SLSE) pulse sequence~\cite{Mar1977}. In SLSE, Fig.~\ref{fig:NQR SLSE}(a), a series of phase-synchronized RF pulses are applied to extend the lifetime of the NQR signal out to a time $T_2^{\rm SLSE}\,{\gg}\, T_{\rm 2,\,nuc}^{\ast}$. Each echo pulse resets the phase of the nuclear spin precession, such that the NQR signals following each pulse can be coherently averaged together~\cite{Mar1977,Can1980,Mar1986}. Figures~\ref{fig:NQR SLSE}(b,c) show the NV-detected time-domain and frequency-domain SLSE signals from the 21-g sample averaged over the first 20 echos. A clear resonance at the expected NQR frequency ($f_{\rm alisas}+f_{\rm ref}=3606.7~{\rm kHz}$) is observed. Figure~\ref{fig:NQR SLSE}(d) shows the SLSE signal magnitude as a function of time since the first RF pulse. A fit to a single exponential decay reveals $T_{\rm 2}^{\rm SLSE}=332\pm23~{\rm ms}$, a timescale that is consistent with previous studies~\cite{Mar1977,Mal2011}. Figure~\ref{fig:NQR SLSE}(e) shows the SLSE Signal-to-Noise Ratio (SNR) as a function of the total experimental acquisition time, $t$. The SNR of the single-RF-pulse measurement in Fig.~\ref{fig:21gr NQR} is also shown for comparison. In both cases, the SNR scales as $\sqrt{t}$, but the SLSE SNR is $\sim3$ times greater. This is due to a combination of factors: the SLSE data has a ${\sim}3$-fold smaller NQR signal amplitude, but it has more than an order-of-magnitude higher readout duty cycle, and there was some additional RF noise in the single-pulse data that was not present in the SLSE measurement (\ref{sec:AppxSLSE}).

\section{\label{sec:Discussions}Discussion and conclusion} 
While this work realizes several benchmarks in the development of diamond quantum sensors, the present implementation of RF magnetometry and NQR detection operates far from fundamental limits. The diamond RF magnetometer sensitivity could be improved by illuminating a greater fraction of the diamond~\cite{Cle2015}, increasing the fluorescence collection efficiency~\cite{Xu2019,Les2012}, optimizing the MW pulse sequence~\cite{Zho2020,Aru2022}, and increasing the flux concentrator enhancement factor~\cite{Fes2020}. Each of these improvements could plausibly provide a $\gtrsim3$-fold improvement in sensitivity and taken together might allow a sensitivity $\lesssim1~{\rm fT\,s^{1/2}}$, provided the flux concentrator's magnetic noise remains sufficiently low~\cite{Lee2008}. Magnetic fields from localized samples may also be further enhanced by providing an additional flux return path using a closed cylinder or C-shaped ferrite clamp~\cite{Kim2016}.

To improve NQR detection, it is tempting to leverage the resonant-induction amplification method introduced here and increase the coil $Q$, see Eq.~\eqref{eq:EffMag}. However, the Johnson noise in the coil is also resonantly amplified and must be considered. The rms Johnson magnetic noise inside an impedance-matched solenoidal coil on resonance~\cite{Qiu2007} is:
\begin{equation}
\label{eq:JohNois} \eta_{\rm J}\,{\approx}\,\sqrt{\frac{2\,k_{\rm B}\,T\,\mu_{\rm c}\,Q}{\pi\,f_{0}\,V_{\rm c}}},
\end{equation}
where $k_{\rm B}$ is the Boltzmann constant, $T$ is the temperature, $\mu_{\rm c}$ is the permeability inside the coil, and $f_{0}$ is the coil's resonance frequency. When using a high-Q RF coil wrapped around the sample, the SNR of NQR detection is fundamentally limited by Johnson noise, regardless of the mode of detection. Assuming $f_0=f_{\rm nqr}$ and $\mu_{\rm c}=\mu_0$, where $\mu_0$ is the vacuum permeability, and neglecting nuclear-spin dephasing and experimental dead times, the Johnson-noise-limited SNR is given by Eqs.~\eqref{eq:EffMag},\eqref{eq:JohNois} as~\cite{Hou1976}:
\begin{equation}
\label{eq:JohLimSNR} {\rm SNR_{J}}\approx\sqrt{\frac{\pi\mu_{0\,} Q f_{\rm nqr}}{4 V_{\rm c\,} k_{\rm B\,} T}}\frac{m_{\rm s}}{\rho_{\rm s}}M_0.
\end{equation}
Using parameters from the experiment (\ref{sec:AppxNQRcoil}), we find ${\rm SNR_{J}}\approx210~{\rm Hz^{1/2}}$ for the 4-g coil and ${\rm SNR_{J}}\approx1020~{\rm Hz^{1/2}}$ for the 21-g coil. These values are ${\sim}2$ orders of magnitude higher than the SNR in our experiments (\ref{sec:AppxCoilEnhanc1}), but they represent an upper bound for future optimization. In either case, increasing $Q$ can improve the SNR in NQR experiments. For Johnson-noise-limited detection, ${\rm SNR_{J}}\propto\sqrt{Q}$, Eq.~\eqref{eq:JohLimSNR}. If the noise floor is not yet limited by Johnson noise, as in our experiments (\ref{sec:AppxCoilEnhanc2}), ${\rm SNR}\propto Q$, since $M_{\rm eff}$ still scales linearly with $Q$, Eq.~\eqref{eq:EffMag}. However, in both cases, higher $Q$ is likely to result in a longer recovery time, which is problematic for some applications~\cite{Ver1962,Aug1998,Mal2020}. Ultimately, the benefits of the resonant-induction amplification method are limited, as the SNR upper bound is the same for both diamond RF magnetometer and inductive-coil detection, and the method may not be compatible with remote detection. 

For remote NQR detection, an optimized diamond RF magnetometer is more likely to offer a clear advantage over inductive-coil detection. Consider the case where a low-$Q$ RF excitation loop is located sufficiently far from the sample that resonant-induction amplification can be neglected (\ref{sec:AppxCoilEnhanc3}). Suppose the ferrite-cones diamond RF magnetometer in the present experiment was replaced by a Johnson-noise-limited coil of equivalent volume ($V_c=1.5~{\rm cm^3}$) with $Q=10$. The equivalent magnetic sensitivity of such a sensor is $\eta_{\rm J}/Q=8~{\rm fT_{rms}\,Hz^{-1/2}}$, see Eq.~\eqref{eq:JohNois}. An order-of-magnitude improvement in diamond RF magnetometer sensitivity would already provide a superior SNR. Such a device could find application as a non-contact detector of pharmaceutical compounds, such as synthetic opioids like fentanyl, which require high sensitivity and a short recovery time~\cite{Mal2020}. Beyond NQR spectroscopy, our device may also be used in applications such as magnetic induction tomography~\cite{Cha2019,Rus2023}, underwater communication~\cite{Ger2017,Dea2018}, or the search for exotic spin interactions~\cite{Jac2016,Chu2022,Lia2022}.

In summary, we demonstrated a broadband ($0.07{-}3.62~{\rm MHz}$) ferrite-cones diamond RF magnetometer with a sensitivity of ${\sim}70~{\rm fT\,s^{1/2}}$ at $0.35~{\rm MHz}$. The magnetometer was used to detect the $3.6~{\rm MHz}$ NQR signal of $^{14}{\rm N}$ from room temperature ${\rm NaNO}_{\rm 2}$ powder samples. The short recovery time in our device after RF excitation pulse, ${\sim}35~{\rm \upmu s}$, may offer advantages over other sensitive magnetometers for NQR spectroscopy.

\begin{acknowledgments}
The authors acknowledge advice and support from J. Damron, M. Aiello, A. Berzins, M. Saleh Ziabari, A. M. Mounce, D. Budker, I. Savukov, and M. Conradi. This work was funded by NSF award CHE-1945148 and NIH awards DP2GM140921, R41GM145129, and R21EB027405. Research presented in this presentation was also supported, in part, by the Laboratory Directed Research and Development program of Los Alamos National Laboratory under project number 20220086DR. Los Alamos National Laboratory is operated by Triad National Security, LLC, for the National Nuclear Security Administration of U.S. Department of Energy (Contract No. 89233218CNA000001). I. F. acknowledges support from Latvian Council of Science project lzp-2021/1-0379.

\textbf{Competing interests}
A. Jarmola is a co-founder of ODMR Technologies and has financial interests in the company. A.F.M. is the founder of NuevoMR LLC and has financial interests in the company. The remaining authors declare no competing financial interests.

\textbf{Author contributions}
V.~M.~Acosta, Y.~Silani, and I.~Fescenko conceived the idea for the study in consultation with A.~Jarmola, P.~Kehayias, and M.~Malone. Y.~Silani carried out simulations, performed experiments, and analyzed the data with guidance from J.~Smits and V.~M.~Acosta. I.~Fescenko, M.~Malone, A.~F.~McDowell, A.~Jarmola, B.~Richards, N.~Mosavian, and N.~Ristoff contributed to experimental design and data analysis. All authors discussed results and helped to write the paper.

\end{acknowledgments}

\clearpage

\appendix

\setcounter{equation}{0}
\setcounter{section}{0}

\setcounter{table}{0}

\makeatletter
\renewcommand{\thetable}{A\arabic{table}}

\renewcommand{\theequation}{A\Roman{section}-\arabic{equation}}
\renewcommand{\thefigure}{A\arabic{figure}}
\renewcommand{\thesection}{Appendix~\Alph{section}}
\renewcommand{\thesubsection}{\hspace{-0.3mm}\arabic{subsection}}

\makeatother

\begin{center}
\section{\label{sec:AppxSetup}} 
\setlength{\parskip}{-0.8em}{
\textbf {Experimental setup}}
\end{center}

The apparatus used here, shown in Fig.~\ref{fig:AC setup}(a), was adapted from the one presented in Ref.~\cite{Fes2020}. Here we provide additional information, with a focus on the changes that were implemented for RF magnetometry. An acousto-optic modulator (Brimrose TEM-85-10-532), driven at $81~{\rm MHz}$ by an RF signal generator (RF-Consultant TPI-1001-B), is used to gate a continuous-wave 532 nm green laser beam (Lighthouse Photonics Sprout D-5W) and produce $12\mbox{-}{\rm \upmu s}$ laser pulses. Following the AOM, a half-wave plate (Thorlabs WPH10ME-633) is used to adjust the laser beam's polarization. The laser beam is then focused onto the edge facet of a diamond membrane using a 1-inch diameter aspheric condenser lens (${\rm NA}\,{=}\,0.79$, Thorlabs ACL25416U-B). The same condenser lens is used to collect the NV fluorescence. The fluorescence is spectrally filtered by a dichroic mirror (Thorlabs DMLP567R) and a $650~{\rm nm}$ long-pass filter (Thorlabs FELH0650). Finally, it is focused onto the ``fluorescence channel'' of the photodetector using a 2-inch diameter lens (Thorlabs ACL50832U-B). 

A two-channel balanced photodetector (Thorlabs PDB210A) with a fixed gain $G=175~{\rm kV/A}\approx1.1\times10^{24}~{\rm V/(photoelectron/s)}$ and a 3-dB bandwidth of ${\rm DC}{-}1~{\rm MHz}$ is used to record the NV fluorescence signal. A small portion of the laser beam is picked off prior to the condenser lens and is directed to the ``laser channel'' of the photodetector for balanced detection.

The laser beam's peak power is measured before the condenser lens (but after the pick-off) to be ${\sim}250~{\rm mW}$. We estimate the peak power entering the diamond to be ${\sim}200~{\rm mW}$, after taking into account the ${\sim}80\%$ transmission of the aspheric condenser at 532 nm wavelength. Using a camera imaging system, the full-width-at-half-maximum (FWHM) spot diameter of the laser beam on the diamond face was estimated to be ${\sim}35~{\rm \upmu m}$. The effective sensing volume, $V_{\rm sen}$, is taken as the product of the excitation beam area, ${\sim}\pi\times(35~{\rm \upmu m}/2)^{2}$ and the optical path length in the diamond, ${\sim}300~{\rm \upmu m}$. 

The diamond membrane used here was created from a natural-isotopic-abundance diamond substrate. grown by chemical-vapor deposition, with an initial nitrogen concentration $[{\rm N}]\,{\approx}\,20~{\rm ppm}$. The diamond was irradiated with 2-MeV electrons at a dose of ${\sim}10^{18}~{\rm cm^{-2}}$, and then it was annealed in a vacuum furnace at $800\mbox{-}1100\,^{\circ}{\rm C}$ to form NV centers. The diamond properties, irradiation, and annealing procedures are similar to that described in Ref.~\cite{Smi2019}. The diamond was subsequently cut and polished into a (100)-oriented membrane with dimensions ${\sim}300\times300\times35~{\rm \upmu m^3}$. 

A ${\sim}1$-cm thick rectangular aluminum shield with $18.4{\times}18.4{\times}30.8~{\rm cm^{3}}$ dimensions is placed around the magnetometer apparatus to reduce RF interference. A permanent magnet placed outside of the Al-shield, ${\sim}65~{\rm cm}$ above the ferrite cones, is used to compensate the laboratory's ambient magnetic field and to apply weak bias magnetic fields, $B_0=2\mbox{-}10~{\rm \upmu T}$, approximately along the $z$-axis. A vector magnetometer (Twinleaf VMR018) is used to map the bias magnetic field at the location of the cones. The field components along the $x$ and $y$ axes are minimized by moving the magnet around.

An I/Q-modulated microwave (MW) signal generator (Rohde$\,{\&}\,$Schwarz SMU200A) is used to drive the NV electron spin transitions in the $2.7\mbox{-}3.0~{\rm GHz}$ frequency range. DC voltages, gated by a pair of TTL-controlled switches, are used to modulate the MW carrier's phase, via the generator's analog I/Q modulation port. A third switch is used to gate the MW amplitude on a ${\sim}10~{\rm ns}$ timescale. Then, the MW pulses are passed through an amplifier (Mini-Circuits ZHL-16W-43-S+) and circulator, and the output is connected to a one-and-a-half-turn copper loop (AWG38). This loop is wound around one of the ferrite cones and positioned ${\sim}100~{\rm \upmu m}$ above the gap between the cones. For the measurements without the cones, a copper wire placed on top of the diamond, parallel to the excitation beam, was used to drive the NV spin transition.

A waveform signal generator (Teledyne LeCroy Wavestation 2012) is used as a source for low-frequency RF signals. One of its channels is used to create sinusoidal RF test signals. The output of this channel is connected to a pair of rectangular wire loops with ${\sim}1.5{\times}3.0{\times}1.6~{\rm cm^{3}}$ dimensions placed around the diamond-cones assembly. The rectangular loop pair has a low enough impedance that it produces RF magnetic fields with a relatively constant amplitude (at constant applied voltage amplitude) from $0\mbox{-}4~{\rm MHz}$ (see~\ref{sec:AppxCalib1}). The dimensions of the rectangular loop pair are still large enough that they produce magnetic fields that are approximately uniform over the region of the two ferrite cones. 

The entire experiment is controlled by a TTL pulse card (SpinCore PBESR-PRO-500) with an onboard ovenized crystal oscillator. The differential photodetector signal is digitized by the analog input of a data acquisition (DAQ) card (NI USB-6361). The DAQ sampling is synchronized to the overall pulse sequence through a trigger pulse from the TTL pulse card. External ovenized crystal oscillators are used to stabilize the internal clocks of the DAQ and RF signal generator. 

\begin{center}
\section{\label{sec:AppxPSN}} 
\setlength{\parskip}{-0.8em}{
\textbf {Diamond RF magnetometer sensitivity}}
\end{center}

\subsection{Photoelectron shot noise in balanced detection}
\label{sec:AppxPSN1}

\begin{figure}[htb]
    \centering
\includegraphics[width=0.7\textwidth]{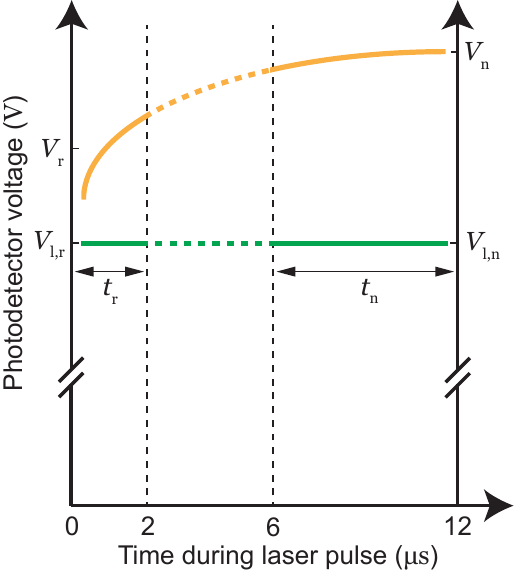}\hfill
\caption{\textbf{Definition of variables for balanced detection and normalization.} Schematic of photodetector voltage traces in the fluorescence channel (orange line) and laser channel (green line) during a single laser pulse. $V_{\rm r}$ and $V_{\rm l,r}$ are the average voltages during the readout window (duration: $t_{\rm r}$), and $V_{\rm n}$ and $V_{\rm l,n}$ are the average voltages during the normalization window (duration: $t_{\rm n}$).}
\label{fig:BlancPDvolts}
\end{figure}

The minimum detectable magnetic field of a magnetometer can be defined as:
\begin{equation}
\label{eq:BminSingle}
\delta B_{\rm min}=\frac{\delta S_{\rm min}}{\frac{\partial S}{\partial B}},
\end{equation}
where $S$ is the field-dependent signal measured by the magnetometer and $\delta S_{\rm min}$ is the minimum noise of a single measurement. In our diamond RF magnetometer, $S$ is the difference in balanced photodetector voltages averaged over two time windows--a ``readout window'' and a ``normalization window'', see Fig.~\ref{fig:BlancPDvolts}. It is given by:
\begin{equation}
\label{eq:PDsig}
\begin{split}
S=V_{\rm r}(B,t)-V_{\rm l,r}-(V_{\rm n}-V_{\rm l,n})~~~~~~~~& \\=G\bigl[\frac{N_{\rm r}(B,t)}{t_{\rm r}}-\frac{N_{\rm l,r}}{t_{\rm r}}-\frac{N_{\rm n}}{t_{\rm n}}+\frac{N_{\rm l,n}}{t_{\rm n}}\bigr].
\end{split}
\end{equation}
In Eq.~\eqref{eq:PDsig}, the quantity $V_{\rm r}(B,t)-V_{\rm l,r}$ is the average photodetector voltage during the readout window of duration $t_{\rm r}$, $N_{\rm r}$ is the equivalent number of photoelectrons detected on the fluorescence channel during the readout window, and $N_{\rm l,r}$ is the corresponding number on the laser channel. The quantity $V_{\rm n}-V_{\rm l,n}$ is the average photodetector voltage during the normalization window of duration $t_{\rm n}$, $N_{\rm n}$ is the equivalent number of photoelectrons detected on the fluorescence channel during the normalization window, and $N_{\rm l,n}$ is the corresponding number on the laser channel. To a good approximation, the only photodetector voltage that depends on magnetic field is $V_{\rm r}(B,t)$--the other voltages that comprise $S$ are independent of magnetic field. 

In the synchronized readout scheme, Fig.~\ref{fig:AC setup}(b), the signal is sampled at times $t=P\,\tau_{\rm sample}$, with $P$ being a non-negative integer, where $t$ is defined as the time at the beginning of the first $\pi/2$ pulse in an XY8-$N$ sequence. If an AC cosine magnetic field with amplitude $B$ and frequency $f$ is applied along the sensing axis in our NV-cones magnetometer, $V_{\rm r}(B,t)$ is modulated as:
\begin{equation}
\label{eq:NVFL}
\begin{split}
V_{\rm r}(B,t)~{=}~V_0\bigl[1\,&+\,C\,\sin{(\frac{4}{\sqrt{3}}\,\epsilon\,\gamma_{\rm nv}\,B\,\tau_{\rm tot})}\times \\
&\cos{(2\pi(f-f_{\rm ref})\,t+\phi_0)}\bigr].
\end{split}
\end{equation}
In Eq.~\eqref{eq:NVFL}, $V_0\approx30~{\rm V}$ is the mean fluorescence-channel voltage, $C$ is the effective fluorescence contrast of the XY8-$N$ sequence, $\epsilon$ is the flux-concentrator enhancement factor, $\tau_{\rm tot}$ is the total NV phase accumulation time during an XY8-$N$ sequence, $f_{\rm ref}$ is the reference frequency, and $\phi_0$ is the phase of the cosine AC field at $t=0$ (the first $\pi/2$ pulse of the first XY8-$N$ sequence).
 
For $f=f_{\rm ref}$ and $\phi_0=0$, a straightforward inspection of Eq.~\eqref{eq:NVFL} reveals that the magnetometer response to small amplitude RF fields ($\gamma_{\rm nv} B\tau_{\rm tot}\ll1$) is given by:
\begin{equation}
\label{eq:PDsigVsB}
\frac{\partial S}{\partial B}=\frac{\partial V_{\rm r}}{\partial B}\approx \frac{4}{\sqrt{3}} V_0\,C\,\epsilon\,\gamma_{\rm nv}\,\tau_{\rm tot}.
\end{equation}
Eq.~\eqref{eq:PDsigVsB} turns out to be valid for all values of $\phi_0$ and $f$ as long as $(f-f_{\rm ref})\,\tau_{\rm sample}<0.5$. Due to the constraints in the definition of $f_{\rm ref}$, this is equivalent to saying that Eq.~\eqref{eq:PDsigVsB} is valid for test field frequencies that fall within the frequency-domain filter-function response of the XY8-$N$ sequence~\cite{Gle2018,Smi2019}.

The noise in the processed signal $S$ is theoretically limited by photoelectron shot noise. Using the definition of $S$ in Eq.~\eqref{eq:PDsig}, the minimum detectable noise of a single readout, $\delta S_{\rm min}$, can be written in root-mean-squared (rms) voltage units as:
\begin{equation}
\label{eq:PDnois}
\delta S_{\rm min}=G\sqrt{\frac{N_{\rm r}}{t_{\rm r}^{2}}+\frac{N_{\rm l,r}}{t_{\rm r}^{2}}+\frac{N_{\rm n}}{t_{\rm n}^{2}}+\frac{N_{\rm l,n}}{t_{\rm n}^{2}}}.
\end{equation}
In the small contrast regime that our experiments operate in, $C\ll1$, the detected photoelectron rates in the fluorescence and laser channels are approximately the same for both readout and normalization windows:
\begin{equation}
\label{eq:PDrate} 
\frac{N_{\rm n}}{t_{\rm n}}\approx\frac{N_{\rm l,r}}{t_{\rm r}}\approx\frac{N_{\rm l,n}}{t_{\rm n}}\approx\frac{N_{\rm r}}{t_{\rm r}}=\frac{n_{\rm nv}V_{\rm sen}\phi}{t_{\rm r}}=\frac{V_0}{G},
\end{equation}
where $n_{\rm nv}$ is the NV concentration, $V_{\rm sen}$ is the illuminated NV sensor volume, and $\phi$ is the probability of detecting a photoelectron per NV center in a single readout. Inserting Eq.~\eqref{eq:PDrate} into Eq.~\eqref{eq:PDnois}, the noise becomes:
\begin{equation}
\label{eq:PDnois2}
\delta S_{\rm min}=\frac{G}{t_{\rm r}} \sqrt{2\,(1+\frac{t_{\rm r}}{t_{\rm n}})\,n_{\rm nv} V_{\rm sen} \phi}=\frac{G \xi}{t_{\rm r}} \sqrt{n_{\rm nv} V_{\rm sen} \phi},
\end{equation}
where $\xi=\sqrt{2\,(1+t_{\rm r}/t_{\rm n})}$ accounts for the extra photoelectron noise due to the balanced detection and normalization procedure. 

Inserting Eqs.~\eqref{eq:PDsigVsB},~\eqref{eq:PDrate}, and \eqref{eq:PDnois2} into Eq.~\eqref{eq:BminSingle}, the minimum detectable RF magnetic field from a single readout is:
\begin{equation}
\label{eq:BminSingle2}
\delta B_{\rm min}\approx\frac{\sqrt{3}\,\xi}{\epsilon}\frac{1}{4\,\gamma_{\rm nv}\,C\,\tau_{\rm tot}\sqrt{n_{\rm nv}\,V_{\rm sen}\,\phi}}.
\end{equation}
For successive NV readouts with repetition time $\tau_{\rm sample}$ and duty cycle $\delta=\tau_{\rm tot}/\tau_{\rm sample}$, the photoelectron-shot-noise limited sensitivity in the diamond RF magnetometer is given by:
\begin{equation}
\label{eq:PSNsen}
\eta_{\rm psn}=\delta B_{\rm min}\sqrt{\tau_{\rm sample}}\approx\frac{\sqrt{3}\,\xi}{\epsilon\,\sqrt{\delta}}\frac{1}{4\,\gamma_{\rm nv}\,C\sqrt{n_{\rm nv}\,V_{\rm sen}\,\phi\,\tau_{\rm tot}}},
\end{equation}
which is the expression given in Eq.~\eqref{eq:psn} of the main text.

\begin{figure}[t]
    \centering
\includegraphics[width=0.8\textwidth]{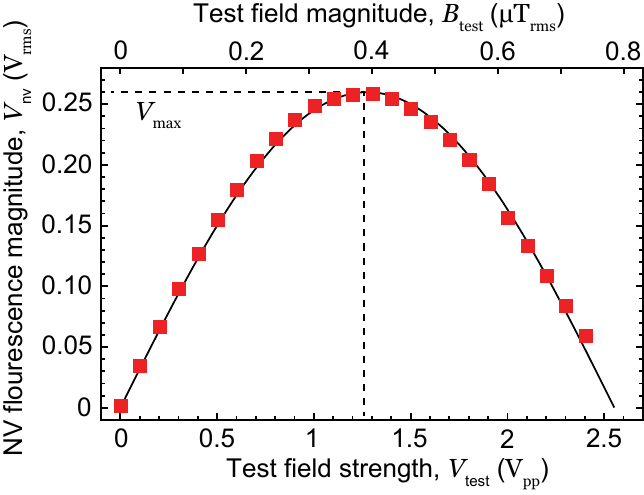}\hfill
       \caption{\textbf{NV test field saturation curve.}
NV test signal fluorescence magnitude $V_{\rm nv}$ without the cones as a function of the applied RF voltage to the test loop, $V_{\rm test}$. The RF test fields were applied at $0.35~{\rm MHz}$ frequency and their projections along the $z$-axis were detected by the NV centers in diamond. Here, an XY8-4 pulse sequence with $76\mbox{-}{\rm ns}$ MW $\pi$-pulse length and $\tau\,{=}\,684~{\rm ns}$ was used to detect the test signals. The vertical dashed line is the magnetic field amplitude that leads to an NV total phase accumulation of $\pi/2~{\rm radians}$ during a single XY8-4 sequence. The solid black line is a fit to an absolute-value-of-sine function, Eq.~\eqref{eq:NVsat}. From the fit, the scaling factor between the applied RF voltage and the NV detected magnetic field along the $z$-axis is found to be $\kappa\,{=}\,0.307\pm 0.004~{\rm \upmu T_{rms}/V_{pp}}$.}
\label{fig:calib}
\end{figure}

For sensitivity measurements in Fig.~\ref{fig:fT sensor}, the readout window duration was $t_{\rm r}=2~{\rm \upmu s}$ and the normalization window duration was $t_{\rm n}=6~{\rm \upmu s}$. For all NQR measurements, we used $t_{\rm r}=1.5~{\rm \upmu s}$ and $t_{\rm n}=5~{\rm \upmu s}$. In either case, $\xi\approx1.6$ and the calculated values for $\eta_{\rm psn}$ were similar.

\subsection{Comparing to experimental sensitivity}
\label{sec:AppxPSN2}
The photoelectron-shot-noise-limited sensitivity formulas given in Eqs.~\eqref{eq:psn} and \eqref{eq:PSNsen} describe the standard deviation of magnetometer signals in the time-domain for successive $1\mbox{-}{\rm s}$ measurements. In experiments, the magnetic field sensitivity was reported as the average noise floor of the NV fluorescence signal in the frequency-domain. It was obtained by taking the mean noise floor in spectra computed from the absolute value of the Fourier transform of $1\mbox{-}{\rm s}$ NV signals. For white noise, it can be shown that the standard deviation in the time domain (the theory method) is a factor of ${\sim}1.25$ smaller than the mean of the absolute value of the Fourier transform (experimental method). Incorporating this factor in Eq.~\eqref{eq:psn}, the photoelectron-shot-noise limited sensitivity of our ferrite-cones-diamond magnetometer would be ${\sim}38~{\rm fT_{rms}\,s^{1/2}}$ using the frequency-domain-analysis method.

\begin{table*}[t]
\scriptsize
\caption{\label{tab:calib} 
\textbf{Calibration factors and enhancement for different test-field frequencies.} The values for $\tau$, $N$, $\pi$-pulse length, and $\tau_{\rm sample}$ are for when the cones were assembled around the diamond. There are ${\sim}1\%$ fit uncertainty and ${\sim}8\%$ day-to-day systematic variation on the reported values for $\kappa$ and $\kappa_{\rm m}$. For each test frequency, the enhancement factor ($\epsilon$) was estimated by dividing the scaling factor with the cones by the value without them.}
\begin{tabular}{ |p{1.6cm}|p{0.8cm}|p{0.5cm}|p{1.5cm}|p{1.0cm}|p{2.1cm}|p{2.1cm}|p{1.7cm}|p{1.7cm}|p{1.2cm}| }
 \hline
 \textbf{Test frequ-ency} (MHz) & $\pmb{\tau}$ (ns) & $\pmb{N}$ & \textbf{${\pi}$-pulse length} (ns) & $\pmb{\tau_{\rm sample}}$ (${\rm \upmu s}$) & $\pmb{\kappa}$\,\textbf{with\,cones} (${\rm \upmu T_{rms,gap}/V_{pp}}$) & $\pmb{\kappa_{\rm m}}$\,\textbf{with\,cones} (${\rm \upmu T_{rms,gap}/V_{pp}}$) & $\pmb{\kappa}$\,\textbf{no\,cones} (${\rm \upmu T_{rms}/V_{pp}}$) & $\pmb{\kappa_{\rm m}}$\,\textbf{no\,cones} (${\rm \upmu T_{rms}/V_{pp}}$) & \textbf{Enhanc-ement, $\epsilon$}\\
 \hline
 0.07 & 3548 & 1 & 44 & 73.07 & 101 & 101 & 0.318 & 0.315 & ${319\pm3}$\\
 0.10 & 2478 & 1 & 44 & 60.61 & 84 & 84 & 0.310 & 0.306 & ${273\pm4}$\\
 0.20 & 1226 & 2 & 48 & 55.28 & 82 & 81 & 0.300 & 0.292 & ${275\pm4}$\\
 0.35 & 698 & 4 & 48 & 60.93 & 96 & 92 & 0.307 & 0.291 & ${314\pm3}$\\
 0.70 & 332 & 7 & 48 & 55.69 & 102 & 95 & 0.346 & 0.313 & ${299\pm9}$\\
 1.00 & 224 & 8 & 52 & 47.09 & 110 & 98 & 0.359 & 0.310 & ${311\pm10}$\\
 1.51 & 140 & 8 & 52 & 35.90 & 128 & 108 & 0.393 & 0.312 & ${336\pm20}$\\
 2.02 & 96 & 10 & 56 & 34.75 & 138 & 107 & 0.435 & 0.316 & ${328\pm21}$\\
 2.50 & 68 & 13 & 64 & 35.62 & 173 & 118 & 0.486 & 0.321 & ${362\pm12}$\\
 3.12 & 48 & 16 & 64 & 35.22 & 184 & 111 & 0.578 & 0.332 & ${326\pm16}$\\
 3.62 & 56 & 20 & 26 & 36.73 & ${\sim}130$ & ${\sim}106$ &  &  & \\
 \hline
\end{tabular}
\end{table*}

\begin{center}
\section{\label{sec:AppxCalib}} 
\setlength{\parskip}{-0.8em}{
\textbf {RF test field calibration and ferrite enhancement}}
\end{center}

\subsection{RF test field calibration}
\label{sec:AppxCalib1}
In the main text, the NV fluorescence processed signal, Eq.~\eqref{eq:PDsig}, was calibrated to a known applied AC magnetic field. We performed the test-field calibration with two methods: (i) using the theoretical signal response of the NV centers themselves, with and without the cones, and (ii) independent measurements of the frequency dependence of inductive pick-up in a wire loop.

\begin{figure}[b]
    \centering
\includegraphics[width=0.8\textwidth]{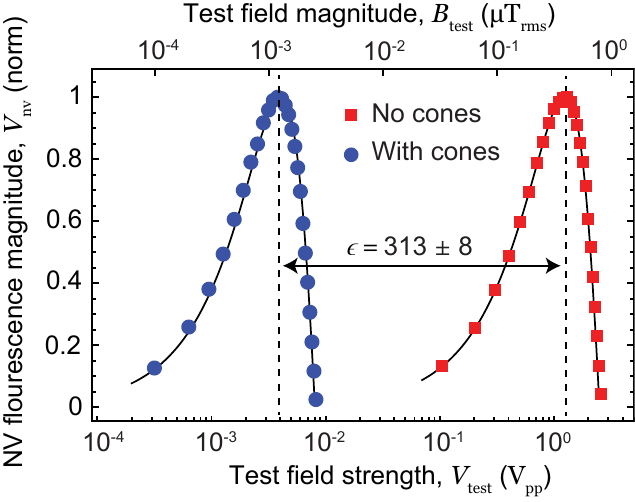}\hfill
\caption{\textbf{RF calibration with and without cones.} Normalized NV test signal fluorescence magnitude as a function of the applied RF voltage to the test loop with (blue circles) and without (red squares) the cones. RF test fields were applied at $0.35~{\rm MHz}$ frequency and XY8-4 readout sequences with $\tau\,{=}\,698~{\rm ns}$ and $684~{\rm ns}$ were used to detect their $z$-axis projections inside the diamond with and without the cones, respectively. The NV test signal magnitudes are fit to Eq.~\eqref{eq:NVsat} (solid black lines), revealing the scaling factor $\kappa\,{=}\,96\pm 2~{\rm \upmu T_{rms,gap}/V_{pp}}$ with the cones and $\kappa\,{=}\,0.307\pm 0.004~{\rm \upmu T_{rms}/V_{pp}}$ without them. From the ratio of these scaling factors, the RF field enhancement provided by the ferrite cones was estimated to be $\epsilon\,{=}\,313\pm 8$ at $0.35~{\rm MHz}$ frequency.}
\label{fig:RFenhanc}
\end{figure}

For method (i), an RF test signal, with a peak-to-peak voltage $V_{\rm test}$ and frequency $f_{\rm test}$, is applied to the rectangular wire loops placed around the diamond when the ferrite cones were absent ($\epsilon=1$). For each value of $f_{\rm test}$, the amplitude $V_{\rm test}$ is varied and the NV signal is recorded at each value of $V_{\rm test}$. From Eq.~\eqref{eq:NVFL}, the NV test signal fluorescence magnitude $V_{\rm nv}$ (in rms voltage units) is given by:
\begin{equation}
\label{eq:NVsat}
V_{\rm nv}~{=}~V_{\rm max}\,|\sin\,(4\,\sqrt{\frac{2}{3}}\,\kappa\,V_{\rm test}\,\gamma_{\rm nv}\,\tau_{\rm tot})|,
\end{equation}
where $V_{\rm max}=V_0\,C/\sqrt{2}$ is the maximum NV test signal magnitude, $\tau_{\rm tot}$ is the NV phase accumulation time, and $\kappa=B_{\rm test}/V_{\rm test}$ is the scaling factor which provides us with the calibration. Note that there are two plausible definitions of $\tau_{\rm tot}$ and each results in a slightly different value of $\kappa$. If we assume $\tau_{\rm tot}$ is the interval between $\pi/2$ pulses in an XY8-$N$ sequence, excluding the time for $\pi$-pulses~\cite{Smi2019}, then $\tau_{\rm tot}=16N\tau$, and we use the variable $\kappa$ for the scaling factor. If $\tau_{\rm tot}$ is the entire interval between $\pi/2$ pulses in an XY8-$N$ sequence~\cite{Gle2018}, then $\tau_{\rm tot}=8N(2\tau+t_{\rm \pi,mw})=4N/f_{\rm test}$, where $t_{\rm \pi,mw}$ is the MW $\pi$-pulse length, and we use $\kappa_{\rm m}$ for the scaling factor.

Figure~\ref{fig:calib} shows the NV fluorescence signal magnitude without the cones versus $V_{\rm test}$. Here, $f_{\rm test}=0.35~{\rm MHz}$, and an XY8-4 pulse sequence with $t_{\rm \pi,mw}=76~{\rm ns}$ and $\tau=684~{\rm ns}$ was used to detect the test signals. Fitting the data to Eq.~\eqref{eq:NVsat} reveals the scaling factor $\kappa\,{=}\,0.307\pm 0.004~{\rm \upmu T_{rms}/V_{pp}}$. The same procedure was applied to the setup with the cones in place ($\kappa=B_{\rm gap}/V_{\rm test}$), and the results with and without cones are shown in Fig.~\ref{fig:RFenhanc}. With the cones present, the maximum of the NV signal occurs at a ${\sim}313$ times lower test field amplitude. This is due to the flux-concentrator enhancement of the RF test field, and the ratio of the fitted response curves provides a measure of $\epsilon$, see~\ref{sec:Appxepsilon}.

\begin{figure}[b]
    \centering
\includegraphics[width=1.0\textwidth]{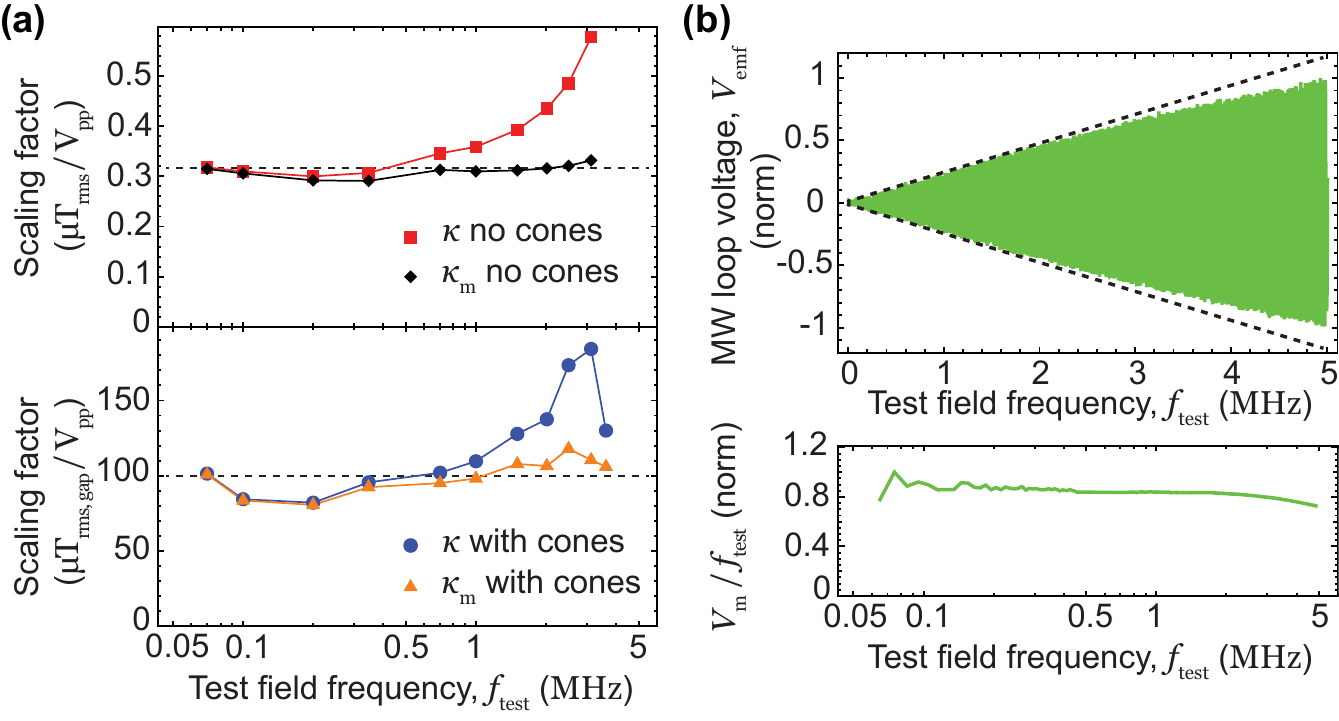}\hfill
\caption{\textbf{Frequency-dependent calibration of the test field.} (a) Scaling factor $\kappa$ and its modified version $\kappa_{\rm m}$ versus the test field frequency for with (bottom) and without (top) the cones. The dashed black lines indicate the scaling factors we applied to calibrate the diamond RF magnetometer for all frequencies throughout the main text. (b) Normalized induced voltage in the MW loop versus frequency of the RF carrier applied to the rectangular test loop. The induced voltage amplitude increases linearly with frequency consistent with Faraday's law (dashed black lines) up to ${\sim}3~{\rm MHz}$. This measurement was done with the diamond-cones assembly inside the test loop. The solid green line (bottom) shows the frequency-normalized induced voltage amplitude in the MW loop.}
\label{fig:scalefacVsfreq}
\end{figure}

We repeated the same process for different test frequencies. For each frequency, $\tau$, $t_{\rm \pi,mw}$, and $N$ were adjusted to maximize the NV test signal. Table~\ref{tab:calib} shows the resulting calibration factors, $\kappa$ and $\kappa_{\rm m}$, with and without the cones. Figure~\ref{fig:scalefacVsfreq}(a) shows a plot of the fitted values of $\kappa$ and $\kappa_{\rm m}$ with and without cones as a function of $f_{\rm test}$. For $f_{\rm test}\lesssim1~{\rm MHz}$, both definitions of calibration are approximately constant and consistent with $\epsilon\approx300$. For higher frequency, $\kappa$ begins to increase both with and without cones. An increase in field strength as a function of $f_{\rm test}$ was unexpected--if anything, we expected a decrease due to the finite inductance of the test-loop coil. While a full spin dynamics simulation is beyond the scope of this study, this behavior may imply that NV spin precession during the MW $\pi$-pulses should not be neglected when estimating the NV phase accumulation. To provide an independent check of the test-field frequency dependence, we turned to calibration method (ii).

For calibration method (ii), we studied the frequency response of the rectangular test loops using an inductive pickup loop. An RF carrier with $5~{\rm V_{pp}}$ amplitude was applied to the test loops, with the diamond-cones assembly in place, and the RF carrier frequency, $f_{\rm test}$, was swept linearly with rate $\alpha_{\rm s}\,{\approx}\,200~{\rm MHz/s}$ from $1~{\rm kHz}$ to $5~{\rm MHz}$. The oscillating pickup voltage in the small wire loop wrapped around one of the cones (that is usually used for MW delivery) was recorded by an oscilloscope (Yokogawa DL9140L). Figure~\ref{fig:scalefacVsfreq}(b, top) shows the MW loop voltage, $V_{\rm emf}$, as a function of $f_{\rm test}$. As expected from Faraday's law, $V_{\rm emf}$ increases linearly with frequency for $f_{\rm test}\lesssim3~{\rm MHz}$. At higher frequency, the slope gradually reduces, presumably due to a rise in impedance of the inductive test-field loops.

\begin{table*}[t]
\scriptsize
\caption{\label{tab:sens} 
\textbf{XY8-$\pmb{N}$ sensitivity vs test field frequency.} The sensitivities with and without the cones were obtained by applying a uniform RF test field with magnitude $B_{\rm test}\,{\approx}\,100~{\rm pT_{rms}}$ along the $z$-axis. The effective fluorescence contrast $C$ with the cones was obtained from Eq.~\eqref{eq:NVcontrst} assuming $\epsilon\,{=}\,316$ for all test frequencies. For each test frequency, the average noise floor of the absolute Fourier transform spectrum was extracted from a  ${\sim}100\mbox{-}{\rm Hz}$ wide region of the spectrum that did not contain spurious noise spikes.}
\begin{tabular}{ |p{1.6cm}|p{0.8cm}|p{0.5cm}|p{1.5cm}|p{1.0cm}|p{0.8cm}|p{2.3cm}|p{2.3cm}| }
 \hline
 \textbf{Test frequ-ency} (MHz) & $\pmb{\tau}$ (ns) & $\pmb{N}$ & \textbf{${\pi}$-pulse length} (ns) & $\pmb{\tau_{\rm sample}}$ (${\rm \upmu s}$) & $\pmb{C}$ & \textbf{Sensitivity\,with cones}\,(${\rm fT_{rms}\,s^{1/2}}$) & \textbf{Sensitivity\,no cones}\,(${\rm pT_{rms}\,s^{1/2}}$)\\
 \hline
 0.07 & 3548 & 1 & 44 & 73.07 & 0.002 & ${200\pm13}$ & ${60.4\pm3.6}$ \\
 0.10 & 2478 & 1 & 44 & 60.61 & 0.004 & ${150\pm8}$ & ${28.6\pm1.6}$ \\
 0.20 & 1226 & 2 & 48 & 55.28 & 0.006 & ${89\pm4}$ & ${19.3\pm1.2}$ \\
 0.35 & 698 & 4 & 48 & 60.93 & 0.009 & ${73\pm4}$ & ${18.5\pm0.9}$ \\
 0.70 & 332 & 7 & 48 & 55.69 & 0.007 & ${106\pm6}$ & ${21.7\pm1.4}$ \\
 1.00 & 224 & 8 & 52 & 47.09 & 0.008 & ${105\pm5}$ & ${20.6\pm1.1}$ \\
 1.51 & 140 & 8 & 52 & 35.90 & 0.010 & ${119\pm8}$ & ${22.6\pm1.2}$ \\
 2.02 & 96 & 10 & 56 & 34.75 & 0.009 & ${128\pm6}$ & ${25.6\pm1.4}$ \\
 2.52 & 86 & 13 & 26 & 35.27 & 0.013 & ${172\pm8}$ & ${26.7\pm1.5}$ \\
 3.09 & 68 & 16 & 26 & 35.34 & 0.009 & ${229\pm8}$ & ${49.3\pm3.2}$ \\
 3.62 & 56 & 20 & 26 & 36.73 & 0.010 & ${202\pm8}$ & \\
 \hline
\end{tabular}
\end{table*}

The data set in Fig.~\ref{fig:scalefacVsfreq}(b, top) was divided into $10~{\rm kHz}$ segments, and each segment was fit to a function $V_{\rm emf}\,{=}\,V_{\rm m}\sin{(2\pi\,f_{\rm test}^{2}/\alpha_{\rm s}+\phi_{\rm i})}$ to extract the amplitude $V_{\rm m}$ and initial phase $\phi_{\rm i}$ of the oscillation. The induced voltage's amplitude divided by frequency, $V_{\rm m}/f_{\rm test}$, versus the test field frequency is plotted in Fig.~\ref{fig:scalefacVsfreq}(b, bottom). The response is approximately flat in the $0.07\mbox{-}3.6~{\rm MHz}$ frequency range studied in our experiments.

The combined observation from both calibration methods is that, to a decent approximation, the RF test field amplitude is independent of frequency in the $0.07\mbox{-}3.6~{\rm MHz}$ range studied here. Thus, we applied constant scaling factors,  $\kappa_{\rm m}\,{=}\,0.316~{\rm \upmu T_{rms}/V_{pp}}$ for data without the cones and $\kappa_{\rm m}\,{=}\,100~{\rm \upmu T_{rms,gap}/V_{pp}}$ for data with cones, for all frequencies $f_{\rm test}$ throughout the main text. These values are denoted as dashed black lines in Fig.~\ref{fig:scalefacVsfreq}(a). The only exception is the data in Fig.~\ref{fig:fT sensor}(d), where we use the measured values in Tab.~\ref{tab:calib} to explicitly show the small fluctuations ($\lesssim10\%$) of enhancement factor as a function of frequency.

\subsection{Ferrite RF field enhancement measurement}
\label{sec:Appxepsilon}
For each test frequency, the RF enhancement provided by the ferrite cones is estimated as the ratio of the scaling factor with the cones to the value without them. Two enhancement factors were obtained for each frequency, one from the $\kappa$-ratio and another from the $\kappa_{\rm m}$-ratio, and their mean value as well as deviation from the mean are reported in Tab.~\ref{tab:calib} and plotted in Fig.~\ref{fig:fT sensor}(d).

\begin{center}
\section{\label{sec:AppxContrst}} 
\setlength{\parskip}{-0.8em}{
\textbf {Frequency-dependence of magnetometer sensitivity}}
\end{center}

In Fig.~\ref{fig:fT sensor}(e) of the main text, we present the sensitivity of the diamond RF magnetometer as a function of test frequency with and without the ferrite cones. For these measurements, a sinusoidal RF carrier with a small voltage amplitude $V_{\rm test}\,{\approx}\,316~{\rm \upmu V_{pp}}$ was applied to the test field loops. This small voltage was provided by inserting a 30-dB attenuator in the output of the RF signal generator. Based on calibration measurements (\ref{sec:AppxCalib}), this produced a uniform magnetic test field with magnitude $B_{\rm test}\,{\approx}\,100~{\rm pT_{rms}}$ along the $z$-axis for the $0.07\mbox{-}3.6~{\rm MHz}$ frequency range.

Table~\ref{tab:sens} summarizes the XY8-$N$ setting parameters for each test frequency when the cones were arranged around the diamond, along with the experimentally-determined sensitivities with and without the cones. At each frequency setting, $\tau$, $N$, and the microwave $\pi$-pulse length were optimized to give the best sensitivity. Then, the magnetometer signal was recorded for $100~{\rm s}$ with and without the ferrite cones. Each $100\mbox{-}\rm{s}$ data set was divided into one hundred $1\mbox{-}\rm{s}$ segments, and a spectrum was obtained for each segment by taking the absolute value of the Fourier transform. Then the NV test signal's fluorescence magnitude ($V_{\rm nv}$) as well as the mean noise floor were extracted for each spectrum. The noise band was selected within a spike-free $100~{\rm Hz}$ wide band. Typically, we chose the band at a relatively high alias frequency to avoid picking up any low-frequency drifts of the NV fluorescence signal. The sensitivity for each frequency is reported as the mean value of the average noise floors for one hundred segments, see Fig.~\ref{fig:fT sensor}(e) and Table~\ref{tab:sens}. The error bar on the reported sensitivity is the standard deviation of the 100 sensitivity measurements.

In Fig.~\ref{fig:fT sensor}(b) of the main text, the frequency band used to calculate noise was $3.0{-}3.7~{\rm kHz}$ for the magnetic spectrum without the cones (red), $2.85{-}3.10~{\rm kHz}$ for the spectrum with the cones (blue), and $2.8{-}3.0~{\rm kHz}$ for the spectrum with the cones when the MW frequency was detuned (green). The bands used to calculate the noise are different in each data set to avoid including the small spikes that appear in different regions of the spectrum (presumably due to RF interference) from time to time.

In a small field approximation, the contrast can be derived from Eq.~\eqref{eq:NVFL} as:
\begin{equation}
\label{eq:NVcontrst}
C~{\approx}~\frac{\sqrt{3}\,V_{\rm nv}}{4\,\epsilon\,B_{\rm test}\,\gamma_{\rm nv}\,\tau_{\rm tot}\,V_0}.
\end{equation}
For each test frequency, we used the observed value of $V_{\rm nv}$ and Eq.~\eqref{eq:NVcontrst} to estimate $C$ for each test frequency, Table~\ref{tab:sens}, assuming $B_{\rm test}\,{=}\,100~{\rm pT_{rms}}$ and $\epsilon\,{=}\,316$. A typical value of the contrast obtained in our setup for $f_{\rm test}=0.35~{\rm MHz}$ is $C\,{\approx}\,0.01$.

\begin{figure}[t]
    \centering
\includegraphics[width=0.99\textwidth]{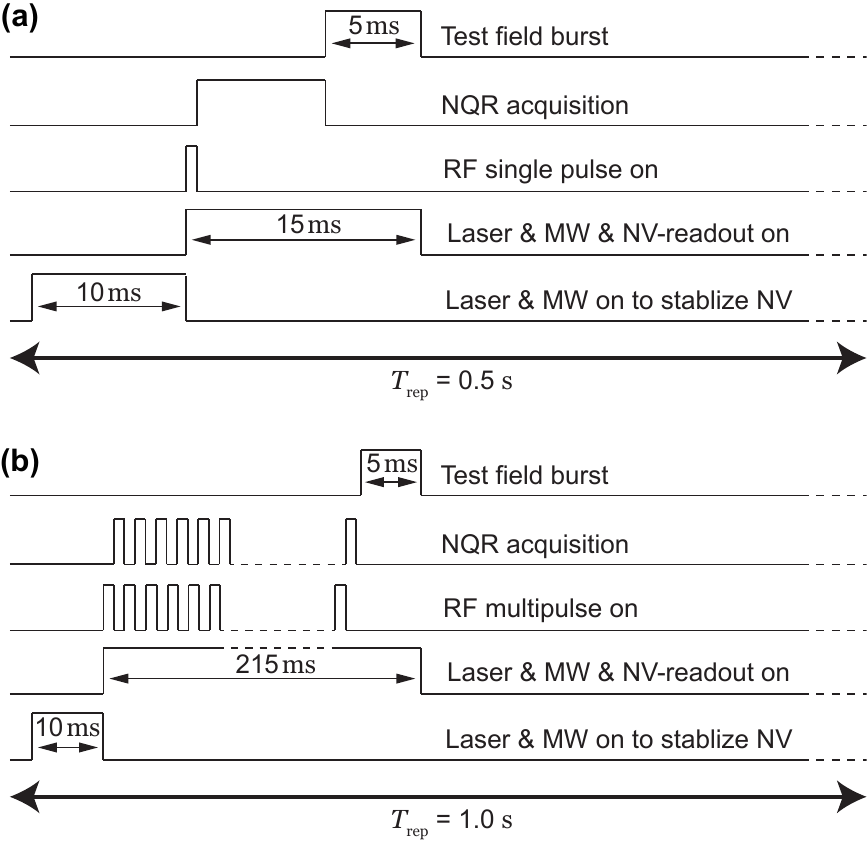}\hfill
       \caption{
\textbf{Timing protocol for NV NQR measurements.} NV NQR measurement protocol used for (a) single-RF-pulse NQR experiments and (b) SLSE NQR experiments.}
\label{fig:NVNQRmeasPuls}
\end{figure}

\begin{center}
\section{\label{sec:AppxNQRsetup}} 
\setlength{\parskip}{-0.8em}{
\textbf {NQR setup}}
\end{center}

\subsection{Timing diagram}
\label{appxnqrtime}
Figure~\ref{fig:NVNQRmeasPuls}(a) shows the NV NQR measurement protocol used for the single-RF-pulse experiments (Figs.~\ref{fig:21gr NQR},~\ref{fig:NV_NQR deadtime} in the main text). To stabilize the NV fluorescence response, the laser and MW pulses are turned on for $10~{\rm ms}$ prior to acquiring NQR signals. Then, an RF excitation pulse is applied, and the synchronized XY8-20 NV signals are acquired for $15~{\rm ms}$. The MW and laser pulses are then turned off for the remaining $475~{\rm ms}$ of the sequence. We do this because we found that keeping the MW and laser pulses on for the entire repetition time, $T_{\rm rep}=500~{\rm ms}$, resulted in a broadening, shift, and eventual loss of the NQR signal after several minutes. This is likely due to a rise in temperature and temperature gradients across the NQR sample generated by heat dissipation from the laser and MW pulses. Fortunately, turning the pulses on for $25~{\rm ms}$ out of the total $500~{\rm ms}$ sequence time was a low enough duty cycle to eliminate this effect, while still allowing for acquisition of the full NQR transient signal.

A $100~{\rm pT_{rms}}$ RF test field with a frequency a-few-${\rm kHz}$ above the NQR frequency is applied for the last $5~{\rm ms}$ of the NV readout in each repetition, after the NQR signals had decayed. This ``test-field burst'' is added to monitor the magnetometer's sensitivity over hours of averaging and make sure the NV test signal isn't dropping below a threshold value. The test-field data were dropped when analyzing NQR signals. For NQR experiments with the  the $4\mbox{-}\rm g$ sample (Fig.~\ref{fig:NV_NQR deadtime}), the average NV test signal fluorescence was $V_{\rm nv}\approx7~{\rm mV_{rms}}$, and the signal remained within a factor of $1.3$ of this level throughout the measurements. For NQR experiments with the $21\mbox{-}\rm g$ sample (Figs.~\ref{fig:21gr NQR} and~\ref{fig:NQR SLSE}), the average NV test signal fluorescence was $V_{\rm nv}\approx6~{\rm mV_{rms}}$, and the signal remained within a factor of $1.15$ of this level throughout the measurements.

Figure~\ref{fig:NVNQRmeasPuls}(b) shows the NV NQR measurement protocol used for the RF multipulse SLSE experiment (Fig.~\ref{fig:NQR SLSE} in the main text). The NV XY8-20 sequences are synchronized with the SLSE RF excitation pulses, and both pulse trains are applied for ${\sim}215~{\rm ms}$ in each repetition. In order to avoid heating of the sample, the repetition time in the SLSE experiment was set to be twice as long as in single-RF-pulse experiments, $T_{\rm rep}\,{=}\,1~{\rm s}$.

In order to alternate the phase of the RF excitation pulses, we combine the outputs of two Teledyne LeCroy waveform signal generators. The $10~{\rm MHz}$ frequency reference output of one generator is connected to the frequency reference input of the other to synchronize their internal clocks. The two RF sources are set to the same frequency and $90^{\degree}$ out of phase with each other. Each source is triggered by its own channel from the TTL pulse card; one trigger is repeated every $T_{\rm rep}=1~{\rm s}$ and the other is repeated every ${\sim}2~{\rm ms}$ (with an initial $1~{\rm ms}$ delay with respect to the first source), corresponding to the time delay between the echo pulses.

\subsection{Sodium nitrite samples}
\label{sec:appxNQRNaNO2}
Sodium nitrite (${\rm NaNO_2}$) powder sample was purchased from Sigma-Aldrich (Lot${\#}\,$MKBX1577V). Plastic cylinder containers with ${\sim}1~{\rm mm}$ wall thickness were assembled to hold the powder samples in the setup.

\begin{table*}[t]
\caption{\label{tab:coils} \textbf{Parameters of the RF coils used in the NQR experiments.} The inductance was measured by an RLC-meter device. The quality factor, $Q$, was measured by the method discussed in~\ref{sec:AppxNQRcoil}.}
\begin{tabular}{ |p{1.0cm}|p{1.3cm}|p{1.0cm}|p{1.3cm}|p{1.3cm}|p{1.3cm}|p{2.3cm}|p{0.8cm}| }
 \hline
 \textbf{Coil} & \textbf{Gauge} & \textbf{Turns} & $\pmb{D}$ & $\pmb{H}$ & $\pmb{V_{\rm c}}$ & \textbf{Inductance}, $\pmb{L}$ & $\pmb{Q}$\\
 \hline
 Small & 21 AWG & 20 & 23.1 mm & 14.0 mm & 5.9 ${\rm cm^{3}}$ & 8.3 ${\rm \upmu H}$ & ${\sim}8$\\
 Big & 20 AWG & 58 & 21.6 mm & 53.2 mm & 19.5 ${\rm cm^{3}}$ & 23.5 ${\rm \upmu H}$ & ${\sim}23$\\
 \hline
\end{tabular}
\end{table*}

\subsection{Thermal housing}
\label{sec:appxNQRtemp}
To improve temperature stability during NQR measurements, a $0.5\mbox{-}{\rm inch}$-thick thermal insulation sheet was glued to the exterior walls of the aluminum-shield housing. A temperature control device (Thorlabs ITC4005) is used to control the temperature of a Peltier element which was thermally attached to the Al-shield with the help of silicone thermal paste.

\subsection{RF pulse amplification}
\label{sec:AppxRFamplify}
\begin{figure}[b]
    \centering
\includegraphics[width=0.83\textwidth]{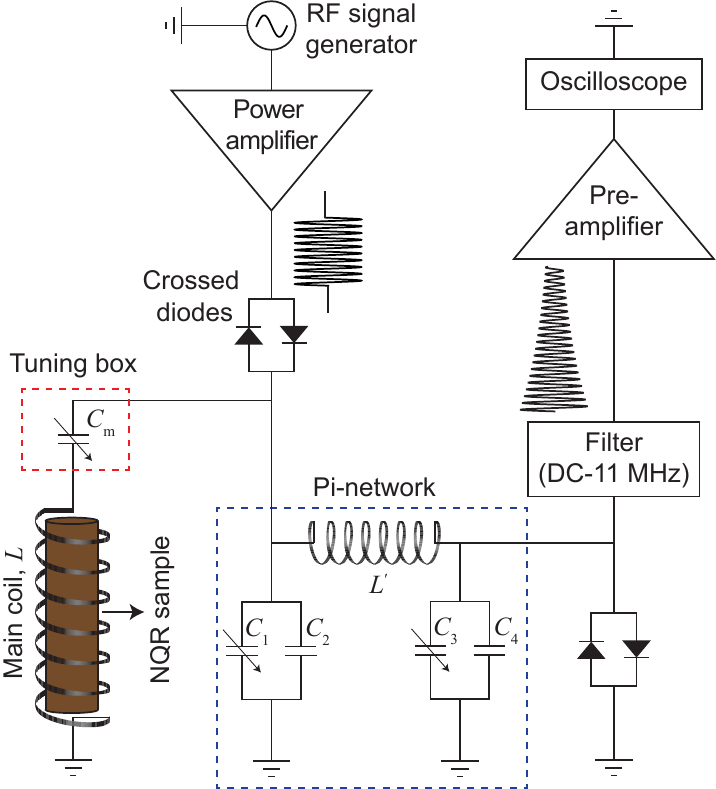}\hfill
       \caption{
\textbf{Block diagram of the setup used for NQR coil excitation and detection.}
$C_{\rm m}$, $C_1$, and $C_3$: variable ($9\mbox{-}200~{\rm pF}$) capacitors;
$L$: inductance of the main coil (see Table~\ref{tab:coils});
$C_2=0.9~{\rm nF}$, $C_4=0.22~{\rm nF}$, $L^{'}\approx 1.7~{\rm \upmu H}$.}
\label{fig:RLC}
\end{figure}

RF pulses are amplified by a 250-W RF power amplifier (Tomco BT00250-AlphaS) before entering a resonant multi-turn coil wrapped around the plastic cylinder container, see Fig.~\ref{fig:RLC}. In order to suppress the amplifier noise, two pairs of crossed-diodes with a minimum 100-V breakdown voltage (onsemi 1N4446) are connected in series to the amplifier's output. Also, the amplifier is blanked before and after each RF pulse by applying a gated DC voltage to the amplifier's TTL blanking port. From the fit in Fig.~\ref{fig:21gr NQR}(c), the conversion factor between the applied RF pulse voltage to the amplifier and generated RF magnetic field amplitude inside the coil was found to be $K=B_{\rm rf}/V_{\rm rf}=5.01\pm 0.05~{\rm mT/V_{pp}}$.

\subsection{Resonant RF coil tuning circuit and pi-network}
\label{sec:AppxNQRcoil}
Parameters of the resonant RF coils wrapped around the sodium nitrite powder samples and used for exciting the $3.6~{\rm MHz}$ NQR transition are shown in Table~\ref{tab:coils}. The small coil was used for the $4\mbox{-}\rm g$ sample (Fig.~\ref{fig:NV_NQR deadtime} in the main text) and the big coil was used for the $21\mbox{-}\rm g$ sample (Figs.~\ref{fig:21gr NQR} and \ref{fig:NQR SLSE} in the main text). Both coils were only partially filled with powder, so the coil volumes are larger than the actual sample volumes. A variable capacitor (Sprague-Goodman GZN20100) with capacitance $C_{\rm m}=9\mbox{-}200~{\rm pF}$ is connected in series with the RF coils (Fig.~\ref{fig:RLC}). For both RF coils, the combination of the variable capacitor and the coil's parasitic capacitance provided the ability to tune the circuit to the $3.6~{\rm MHz}$ NQR transition. Adding the usual parallel tuning capacitor prevented us from tuning the circuit to $3.6~{\rm MHz}$, which implies that the self-resonance of each coil was already near $3.6~{\rm MHz}$. Even without the parallel tuning capacitor, the impedance matching was sufficient to excite and inductively detect NQR signals efficiently.

The same resonant RF coil is used for both exciting and detecting NQR signals. To do this, the signal voltage picked-up by the resonant RF coil is passed through a custom-built pi-network and a low-pass filter (MiniCircuits BLP-10.7), amplified by a low-noise pre-amplifier (NF CMP61665-2), and recorded by an oscilloscope.

The custom-built pi-network, Fig.~\ref{fig:RLC} (dashed blue box), acts as a limiter that prevents the high-power RF excitation pulses from saturating the pre-amplifier, while still passing the weak inductive NQR signals~\cite{Hib2008}. A requirement is that the pi-network satisfies an ``anti-resonance'' condition $f_0\,{=}\,1/(2\pi\sqrt{L^{'}(C_1+C_2)})$, where $L^{'}$, $C_1$, $C_2$ are pi-network elements defined in Fig.~\ref{fig:RLC}. Three pairs of crossed-diodes are inserted after the pi-network to complete the circuit. We tuned $C_1$ and $C_3$ to maximize the input impedance when the three diode pairs were shorted and minimize the reflection when there was no short. 
 
\begin{figure}[hbt]
    \centering
\includegraphics[width=0.85\textwidth]{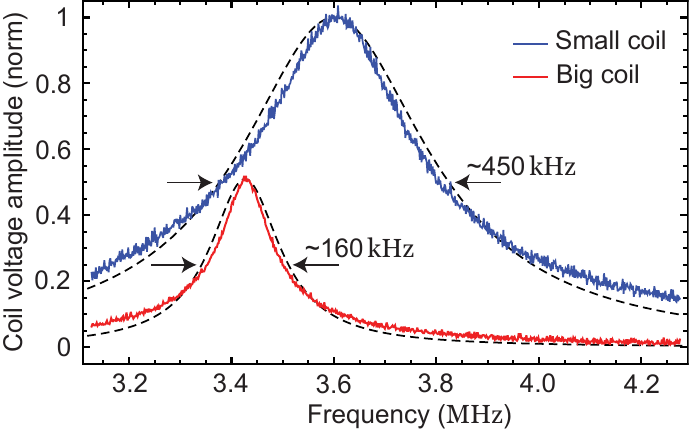}\hfill
       \caption{
\textbf{Coil quality factor.} The loaded quality factor of the resonant RF coil was measured by sweeping the frequency of an RF signal applied to a single-turn loop and picking up the oscillating induced voltage by the resonant RF coil. Similar to Fig.~\ref{fig:scalefacVsfreq}(b), the amplitude of the oscillating voltage is extracted for each $1.66~{\rm kHz}$ segment, and plotted as a function of frequency for both resonant RF coils. The ``Small coil'' is the coil used for the 4-g sample, and the ``Big coil'' is the coil used for the 21-g sample. The spectra are fit to Lorentzian functions (dashed black lines) to extract the center frequency, $f_0$, and FWHM, $\Delta f$. The quality factor is estimated as $Q\,{=}\,f_{0}/\Delta f$ and is given in Table~\ref{tab:coils}. Although the big coil's resonance frequency is detuned from NQR resonance in this figure, it was always tuned to the $3.6~{\rm MHz}$ NQR frequency prior to performing NQR spectroscopy.}
\label{fig:Qfac}
\end{figure}

To characterize the resonant RF coil's frequency response, a sinusoidal RF carrier was applied to a ${\sim}5.8~{\rm cm}$ diameter single-turn loop that was placed above the resonant RF coil. The carrier frequency was swept, and the oscillating induced voltage in the coil was recorded by the oscilloscope using the setup shown in Fig.~\ref{fig:RLC}. By adjusting the capacitance $C_{\rm m}$, the coil resonance frequency was tuned to the NQR frequency ($f_0=3.6~{\rm MHz}$). Figure~\ref{fig:Qfac} shows the amplitude of the induced voltage as a function of frequency for both coils. The amplitude was extracted from the oscillating signal using a similar procedure as described for Fig.~\ref{fig:scalefacVsfreq}(b). From these data, we estimate the loaded quality factor is $Q\approx8$ for the small coil used for the 4-g sample and $Q\approx23$ for the large coil used for the 21-g sample.

\subsection{NQR Fourier transform spectra in the main text}
For the NQR experiments, both the NV and coil readouts were synchronized by trigger pulses from the TTL pulse card. In Figs.~\ref{fig:21gr NQR}(b,c,d) of the main text, the imaginary part of the Fourier transform of the time-domain NQR signals were taken because the symmetric lineshape of the spectra allowed us to determine the signals' phases with sufficient precision. In Fig.~\ref{fig:NV_NQR deadtime}(b) and Fig.~\ref{fig:NQR SLSE}(c) of the main text, the absolute value of the Fourier transform of the time-domain NQR signals were taken because the asymmetric lineshape prevented us from determining an exact phase.

\begin{center}
\section{\label{sec:AppxNQRmagn}} 
\setlength{\parskip}{-0.8em}{
\textbf {Spin dynamics under NQR and RF excitation Hamiltonians}}
\end{center}

An NQR transition can be excited by applying a resonant RF magnetic field. In powder samples, an RF pulse applied along the $z$-axis (in the lab frame) can induce a magnetization oscillating at the NQR frequency, along the $z$-axis, given by:
\begin{equation}
\label{eq:NQRmagnet}
M_{\rm z}^{\rm lab}~{=}~n_{\rm n}\,\gamma_{\rm n}\,h\,\langle I_{\rm z}^{\rm lab}\rangle,
\end{equation}
where $n_{\rm n}$ is the nuclear-spin concentration in the sample, $\gamma_{\rm n}$ is the nuclear-spin gyromagnetic ratio, $h$ is the Planck constant, and $\langle I_{\rm z}^{\rm lab}\rangle$ is the expectation value of the nuclear-spin projection along the $z$-axis. In thermal equilibrium, the density operator populated under the nuclear quadrupole Hamiltonian $H_{\rm Q}$, Eq.~\eqref{eq:Hnqr}, is:
\begin{equation}
\label{eq:IniDensOp}\rho_{\rm 0}~{=}~\frac{e^{-h H_{\rm Q}/(k_{\rm B}T_{\rm s})}}{{\rm Tr}[e^{-h H_{\rm Q}/(k_{\rm B}T_{\rm s})}]},
\end{equation}
where ${\rm Tr}$ is trace of the operator, $k_{\rm B}$ is the Boltzmann constant, and $T_{\rm s}$ is the sample temperature. In the high temperature approximation, Eq.~\eqref{eq:IniDensOp} becomes:
\begin{equation}
\label{eq:IniDensOpApprox}\rho_{\rm 0}~{\approx}~\frac{\mathbbm{1}-h H_{\rm Q}/(k_{\rm B}T_{\rm s})}{2I+1},
\end{equation}
where $\mathbbm{1}$ is the identity operator and $2I+1$ is the number of nuclear-spin energy levels. 

The Hamiltonian describing the interaction of nuclear spins with an RF excitation field applied along the $z$-axis can be written (in frequency units) as:
\begin{equation}
\label{eq:rfHam}
\begin{split}
H_{\rm rf}&(t)~{=}~-\gamma_{\rm n}B_{\rm rf}I_{\rm z}^{\rm lab}\cos{(2\pi\,f_{\rm rf}\,t)} \\
&\,~~~~{=}~-\gamma_{\rm n}B_{\rm rf}\cos{(2\pi\,f_{\rm rf}\,t)}\times \\
&~~~\bigl[I_{\rm x^{'}}\sin{\theta}\cos{\varphi}+I_{\rm y^{'}}\sin{\theta}\sin{\varphi}+I_{\rm z^{'}}\cos{\theta}\bigr],
\end{split}
\end{equation}
where $B_{\rm rf}$ and $f_{\rm rf}$ are the amplitude and frequency of the RF magnetic field, respectively. In Eq.~\eqref{eq:rfHam}, $\{I_{\rm x^{'}},I_{\rm y^{'}},I_{\rm z^{'}}\}$ are the unitless nuclear-spin operators along each principle axis of a given crystallite, $\theta$ is the polar angle and $\varphi$ is the azimuthal angle that the applied RF field vector makes with respect to the principle axes. For a spin-1 system ($I=1$, $\eta\,{\neq}\,0$), there are three NQR transitions with frequencies: $f_{\rm x^{'}}=f_{\rm Q}(1\,{+}\,\eta/3)$ for the $E_{\rm z^{'}}\leftrightarrow E_{\rm y^{'}}$ transition, $f_{\rm y^{'}}=f_{\rm Q}(1\,{-}\,\eta/3)$ for the $E_{\rm z^{'}}\leftrightarrow E_{\rm x^{'}}$ transition, and $f_{\rm z^{'}}=2\,f_{\rm Q}\,\eta\,/3$ for the $E_{\rm x^{'}}\leftrightarrow E_{\rm y^{'}}$ transition, see Fig.~\ref{fig:NQR setup}(b). In the rotating frame of the quadrupolar Hamiltonian, the time-averaged RF Hamiltonian in resonance with one of the NQR transitions is reduced to:
\begin{equation}
\label{eq:rfHamInt}
\begin{split}
\overline{\widehat{H}}_{\rm rf}~{=}~-\frac{\gamma_{\rm n}B_{\rm rf}}{2}&\times \\
&(I_{\rm x^{'}}\sin{\theta}\cos{\varphi})~~{\rm if}~~f_{\rm rf}=f_{\rm x^{'}} \\
&(I_{\rm y^{'}}\sin{\theta}\sin{\varphi})~~{\rm if}~~f_{\rm rf}=f_{\rm y^{'}} \\
&(I_{\rm z^{'}}\cos{\theta})~~~~~~~~\,{\rm if}~~f_{\rm rf}=f_{\rm z^{'}}.
\end{split}
\end{equation}

The density operator's time evolution can be described by the Liouville equation. Following an RF-excitation pulse of length $t_{\rm rf}$, the density operator in the rotating frame becomes:
\begin{equation}
\label{eq:DensOpAftr}
\hat{\rho_1}~{=}~e^{-i\,2\pi\overline{\widehat{H}}_{\rm rf}t_{\rm rf}}\hat{\rho_{\rm 0}}e^{i\,2\pi\overline{\widehat{H}}_{\rm rf}t_{\rm rf}}.
\end{equation}
An RF excitation pulse applied at the $f_{\rm y^{'}}$ resonance frequency induces an oscillating magnetization along the $y^{\prime}$-axis of each crystallite. The expectation value of the spin projection along the $z$-axis is:
\begin{equation}
\label{eq:PolIzLab}
\begin{split}
\langle I_{\rm z}^{\rm lab}(t_0, \alpha)\rangle~{=}~&{\rm Tr}[I_{\rm z}^{\rm lab}e^{-i\,2\pi H_{\rm Q} t_0} \hat{\rho_1} e^{i 2\pi H_{\rm Q} t_0}] \\
{=}~&{\rm Tr}[(I_{\rm y^{'}} \sin{\theta}\sin{\varphi})e^{-i 2\pi H_{\rm Q} t_0} \hat{\rho_1} e^{i 2\pi H_{\rm Q} t_0}] \\
{=}~\frac{h\,f_{\rm y^{'}}}{3\,k_{\rm B}\,T_{\rm s}}&\sin{(2\pi f_{\rm y^{'}} t_0)}\sin{(\alpha\sin{\theta}\sin{\varphi})}\sin{\theta}\sin{\varphi}.
\end{split}
\end{equation}
In Eq.~\eqref{eq:PolIzLab}, $t_0$ is the time after the RF pulse and $\alpha=2\pi\gamma_{\rm n}B_{\rm rf}t_{\rm rf}$ is the RF nutation angle in radians. Inserting Eq.~\eqref{eq:PolIzLab} into Eq.~\eqref{eq:NQRmagnet} and averaging over all $\theta$ and $\varphi$ to take the limit of many randomly-oriented crystallites, the net NQR magnetization in the lab frame becomes:
\begin{equation}
\label{eq:NQRmagnetPwdr}
\begin{split}
M_{\rm z}^{\rm lab}({t_0, \alpha})&~{=}~n_{\rm n}\,\gamma_{\rm n}\,\frac{h^{2} f_{\rm y^{'}}}{3\,k_{\rm B}\,T_{\rm s}}\sin{(2\pi f_{\rm y^{'}} t_0)}\times \\
\int_{0}^{\pi}\int_{0}^{2\pi}&\frac{1}{4\pi}\biggl[\sin{(\alpha\sin{\theta}\sin{\varphi})}\sin{\theta}\sin{\varphi}\biggr]\sin{\theta}\,d\theta\,d\varphi \\
~{=}~&n_{\rm n}\,\gamma_{\rm n}\,\frac{h^{2} f_{\rm y^{'}}}{3\,k_{\rm B}\,T_{\rm s}}\sin{(2\pi f_{\rm y^{'}} t_0)}\sqrt{\frac{\pi}{2\alpha}}\,J_{3/2}(\alpha),
\end{split}
\end{equation}
where $J_{3/2}$ is the Bessel function of first kind and order 3/2. In Eq.~\eqref{eq:NQRmagnetPwdr}, the nutation function, $\sqrt{\pi/(2\alpha)}\,J_{3/2}(\alpha)$, has a maximum of ${\sim}0.436$ for an RF excitation pulse with $\alpha\,{\approx}\,2.08\,{\rm rad}\,{=}\,119^{\degree}$. Therefore, the amplitude of the NQR magnetization immediately after an optimal RF excitation pulse can be written as~\cite{Gar2001}:
\begin{equation}
\label{eq:NQRmagnetPwdr119}
M_{\rm z}^{\rm lab}~{\approx}~0.436\,n_{\rm n}\,\gamma_{\rm n}\,\frac{h^{2} f_{\rm y^{'}}}{3\,k_{\rm B}\,T_{\rm s}}.
\end{equation}
Using the same steps as above, it can be shown that in the case of a powder, the NQR magnetization components on a plane perpendicular to the RF field direction average to zero, i.e. $M_{\rm x}^{\rm lab}=M_{\rm y}^{\rm lab}=0$. Using Eq.~\ref{eq:NQRmagnetPwdr119} and for the case of $^{14}{\rm N}$ nuclear spins in a perfectly packed sodium nitrite powder at room temperature ($n_{\rm n}\,{=}\,1.89 \times 10^{22}~{\rm cm^{-3}}$, $\gamma_{\rm n}\,{=}\,3.077~{\rm MHz/T}$, $f_{\rm y^{'}}\,{=}\,3.608~{\rm MHz}$, $T_{\rm s}\,{=}\,293~{\rm K}$), the nuclear-spin projection induced by an optimal RF pulse is calculated to be $\langle I_{\rm z}^{\rm lab}\rangle\,{=}\,8.6\times10^{-8}$ which results in an NQR magnetization $M_{\rm z}^{\rm lab}\,{=}\,M_0\,{=}\,3.3~{\rm \upmu A/m}$ in the sample.

\begin{figure}[b]
    \centering
\includegraphics[width=0.95\textwidth]{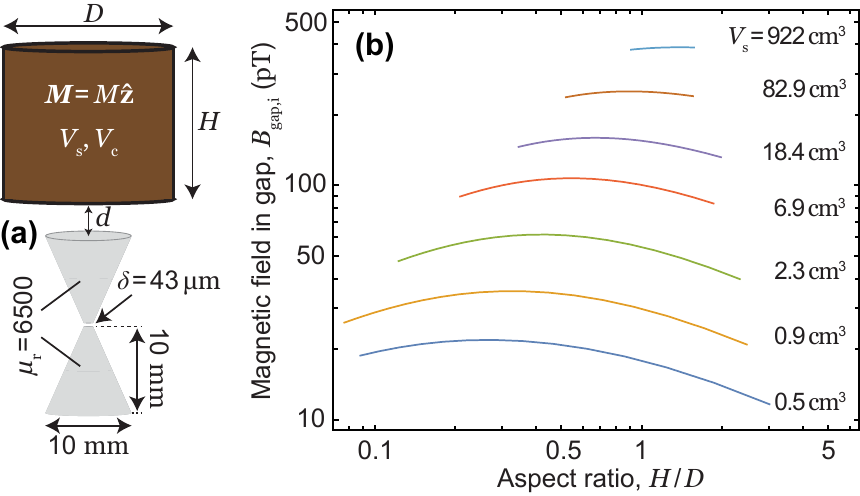}\hfill
       \caption{
\textbf{Magnetic field in the gap vs sample aspect ratio.} (a) Diagram of the model used for the finite-element simulations. The sample was modeled as a cylindrical magnet with height $H$, diameter $D$, and a uniform magnetization $M$ along the $z$-axis. The distance from the bottom of the sample to the top of the cones is $d$. Two ferrite cones with relative permeability $\mu_{\rm r}\,{=}\,6500$, $10~{\rm mm}$ length, $10~{\rm mm}$ base diameter, and $370~{\rm \upmu m}$ truncated-tip-diameter  are arranged in a bowtie configuration with the gap length ${\delta}\,{=}\,43~{\rm \upmu m}$. (b) Average magnetic field within the gap calculated as a function of the cylinder's aspect ratio $H/D$ for different sample volumes $V_{\rm s}$. Here, $d\,{=}\,4~{\rm mm}$ and the sample magnetization was set to $M\,{=}\,M_{0}\,{=}\,3.3~{\rm \upmu A/m}$.}
\label{fig:simulat}
\end{figure}

\begin{center}
\section{\label{sec:AppxSimuMagnCyl}} 
\setlength{\parskip}{-0.8em}{
\textbf {Simulation of ferrite cones and a magnetized cylinder}}
\end{center}

Figure~\ref{fig:simulat}(a) describes the model we used for the initial finite-element simulations presented in Fig.~\ref{fig:NQR setup}(c) in the main text. The sample was modeled as a cylindrical magnet, with volume $V_{\rm s}$ and uniform magnetization $M\,{=}\,M_0\,{=}\,3.3~{\rm \upmu A/m}$ along the $z$-axis, located a distance $d=4~{\rm mm}$ above the cones. Figure~\ref{fig:simulat}(b) shows the initial nuclear AC magnetic field amplitude in the gap, $B_{\rm gap,i}$, calculated for different sample volumes and aspect ratios. As $V_s$ increases, the optimal aspect ratio shifts to larger values and the peak value of $B_{\rm gap,i}$ increases. However, for $V_{\rm s}\,{\gtrsim}\,100~{\rm cm^{3}}$, $B_{\rm gap,i}$ begins to saturate to a practically-achievable maximum value of ${\sim}500~{\rm pT}$. The peak value of $B_{\rm gap,i}$ for each value of $V_s$ were plotted in Fig.~\ref{fig:NQR setup}(c) in the main text.

Several parameters in the experiment turned out to be different from the optimal geometry we initially modeled for. Taking into account the sodium nitrite powder filling fraction inside the coils (Table~\ref{tab:coils}) and the magnetization amplification due to the resonant RF coil, Eq.~\eqref{eq:EffMag}, the effective NQR magnetization for the $4\mbox{-}\rm g$ and $21\mbox{-}\rm g$ samples are estimated to be $M_{\rm eff}\,{\approx}\,8~{\rm \upmu A/m}$ and $M_{\rm eff}\,{\approx}\,37~{\rm \upmu A/m}$, respectively. Also, instead of the $d=4~{\rm mm}$ standoff we initially assumed in the model, the standoffs from the bottom of the coils to the top of the cones were measured to be ${\sim}\,5~{\rm mm}$ in the $4\mbox{-}\rm g$ NQR experiment and ${\sim}\,9~{\rm mm}$ in the $21\mbox{-}\rm g$ experiment. Finally, the gap length was corrected to $\delta\,{=}\,40~{\rm \upmu m}$ (due to a small glue layer), which resulted in a simulated DC enhancement of $\epsilon\,{=}\,300$.

Using these experimental conditions, we carried out new simulations for the 4-g and 21-g samples. The ``samples'', in this case, had the dimensions of the respective coils (Tab.~\ref{tab:coils}), and the magnetization was set as the calculated values of $M_{\rm eff}$. After these adjustments, the calculated initial magnetic field amplitude in the gap was $B_{\rm gap,i}\,{\approx}\,230~{\rm pT}$ for the 4-g sample and $B_{\rm gap,i}\,{\approx}\,930~{\rm pT}$ for the 21-g sample. Converting these values to equivalent uniform magnetic fields ($B_{\rm equiv,i}\,{=}\,B_{\rm gap,i}/\epsilon$ with $\epsilon\,{=}\,300$) gives $B_{\rm equiv,i}\,{\approx}\,760~{\rm fT}$  for the $4\mbox{-}\rm g$ sample and $B_{\rm equiv,i}\,{\approx}\,3100~{\rm fT}$ for the $21\mbox{-}\rm g$ sample. These estimates were much closer to the experimentally observed values, well within a factor of 2.

\begin{center}
\section{\label{sec:AppxSLSE}} 
\setlength{\parskip}{-0.8em}{
\textbf {Signal-to-noise ratio improvement in SLSE}}
\end{center}

In the SLSE experiments, the NQR signals following the echo pulses decay exponentially with an effective relaxation time $T_{\rm 2}^{\rm SLSE}$. $T_{\rm 2}^{\rm SLSE}$ is typically much longer than the decay of an NQR transient following a single RF pulse, and it is typically limited only by homonuclear dipolar and spin-lattice couplings in the sample~\cite{Gre2008}. For this reason, SLSE is often applied to increase the SNR of NQR spectra.

In order to find the NQR SNR for each acquisition time in the SLSE experiment, shown in Fig.~\ref{fig:NQR SLSE}(e), the first-25 NV readouts with ${\sim}1.65~{\rm ms}$ length were averaged together in the time-domain. A digital high-pass filter with cutoff frequency $2.6~{\rm kHz}$ and a Tukey window function centered to the middle of the time-averaged data were applied. Then, the absolute value of the Fourier transform was taken to obtain the NQR spectrum for each echo burst. The SLSE SNR was found by dividing the signal level (contained within a single frequency point, see Fig.~\ref{fig:NQR SLSE}) by the average noise within the $9{-}12~{\rm kHz}$ alias-frequency band. The noise floor in the SLSE experiment was found to be ${\sim}160~{\rm fT_{rms}\,s^{1/2}}$.

To find the SNR in the single-RF-pulse experiments, the initial ${\sim}3.85~{\rm ms}$ of the NV readout obtained at $21.8\,{\rm ^{\circ}C}$ temperature (green spectrum in Fig.~\ref{fig:21gr NQR}(d)) was selected. A Lorentz-to-Gauss window function $W(t)\,{=}\,\exp(0.085\,t)\exp(-(0.04\,t)^{2})$ was applied, and the absolute value of the Fourier transform was taken to obtain an NQR spectrum. The SNR was found by dividing the signal level to the average noise within the $8{-}11~{\rm kHz}$ alias-frequency band. The noise floor in the single-RF-pulse experiment was found to be ${\sim}400~{\rm fT_{rms}\,s^{1/2}}$.

In Fig.~\ref{fig:NQR SLSE}(b), the SLSE signal amplitude for the $21\mbox{-}\rm g$ sample is $B_{\rm equiv,i}\approx0.5~{\rm pT}$, which is ${\sim}3$-times less than the one obtained in the single-RF-pulse experiments (Fig.~\ref{fig:21gr NQR}(a), bottom). This can be due to variation in sample temperature and thermal gradients which led to a detuning of the RF pulses and reduced their fidelity~\cite{Mal2011}. Nevertheless, for a given experimental acquisition time, we found that the SLSE SNR exhibits a ${\sim}3$-fold improvement over the SNR in the single-RF-pulse experiments, see Fig.~\ref{fig:NQR SLSE}(e). This improvement is due to an order-of-magnitude higher measurement duty cycle and the ${\sim}2.5$-times lower noise floor in the SLSE experiment. The latter effect was not expected but may have been due to RF interference being worse during the single-RF-pulse experiment.

\begin{figure}[t]
    \centering
\includegraphics[width=0.55\textwidth]{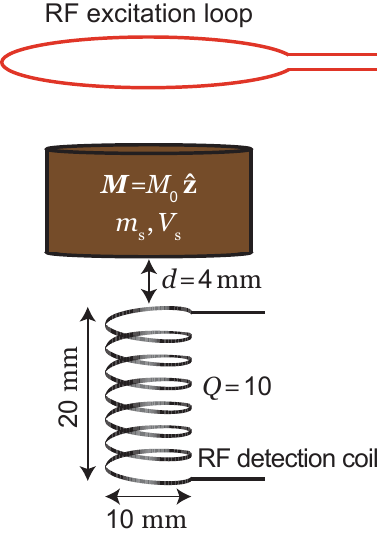}\hfill
\caption{
\textbf{Remote NQR detection.} Schematic of a setup for remotely-detected NQR spectroscopy. The sample has mass $m_{\rm s}$, volume $V_{\rm s}$, and magnetization $M_{0}$. The NQR transition is excited by a non-resonant wire loop located far enough from the sample to neglect resonant signal amplification. The detection coil has a loaded quality factor $Q\,{=}\,10$ and volume $V_{\rm c}=1.5~{\rm cm^{3}}$ (similar to the volume of our ferrite-cones-diamond magnetometer).}
\label{fig:NqrCoilRemot}
\end{figure}

\begin{center}
\section{\label{sec:AppxCoilEnhanc}} 
\setlength{\parskip}{-0.8em}{
\textbf {Calculations of resonant-induction amplification}}
\end{center}

\subsection{Johnson-noise-limited SNR with a resonant-RF-coil and comparison to experiment}
\label{sec:AppxCoilEnhanc1}
In Sec.~\ref{sec:Discussions} of the main text, we calculated the Johnson-noise-limited SNR (${\rm SNR_J}$) of NQR detection for the case when a resonant RF coil is used for signal amplification. Note that Eq.~\eqref{eq:JohLimSNR} neglects any deadtimes due to re-thermalization, so it is valid only for the interval of a measurement where the NQR signal is large. Realistically, for a single-RF-pulse experiment where the signal is averaged over many repetitions, the SNR would be a factor of ${\sim}(T_{\rm 1,nuc}/T_{\rm 2,nuc}^{\ast})^{1/2}$ lower than in Eq.~\eqref{eq:JohLimSNR}. Note also that in Eq.~\eqref{eq:JohNois} of the main text, the Johnson noise was derived using magnetic units of ${\rm T_{rms}}$. In order to derive the Johnson-noise-limited SNR, Eq.~\eqref{eq:JohLimSNR}, we converted the Johnson noise to magnetic amplitude units ($\rm T$) by multiplying by $\sqrt{2}$.

We also estimated that ${\rm SNR_J}$ is ${\sim}2$ orders of magnitude higher than the NQR SNR in our ferrite-cones diamond RF magnetometer. This assumes that the diamond RF magnetometer's sensitivity is ${\sim}200~{\rm fT_{rms}\,s^{1/2}}$, see Fig.~\ref{fig:fT sensor}(e), and that the signal amplitudes are $500~{\rm fT}$ for the 4-g sample and $2300~{\rm fT}$ for the 21-g sample, see Fig.~\ref{fig:NQR setup}(c). Here we neglect a possible factor of $\sqrt{2}$ that comes from converting noise in units of $\rm T\,s^{1/2}$ to $\rm T\,Hz^{-1/2}$~\cite{Fes2020}. 

\subsection{Johnson magnetic noise in our experiment}
\label{sec:AppxCoilEnhanc2}
It is worthwhile to directly calculate the effect of the resonant RF coil's Johnson noise on the diamond RF magnetometer to verify it was not limiting our sensitivity. Using Eq.~\eqref{eq:JohNois} and parameters of our RF coil assemblies (Table~\ref{tab:coils}), the rms Johnson magnetic noise inside the resonant RF coils at room temperature is calculated to be $\eta_{\rm J}\approx35~{\rm fT_{rms}\,Hz^{-1/2}}$ for the small coil and $\eta_{\rm J}\approx32~{\rm fT_{rms}\,Hz^{-1/2}}$ for the big coil. However, our diamond RF magnetometer is located outside of the coil in a remote-detection configuration. In remotely-detected NQR, the magnetic field produced by the sample decays with increasing standoff, resulting in a loss of signal amplitude. The loss factor can be defined as $\alpha_{\rm loss}=B_{\rm equiv,i}/(\mu_0 M_{\rm eff})$. Based on the calculations in \ref{sec:AppxSimuMagnCyl}, we estimate $\alpha_{\rm loss}\approx 0.08$ for the $4\mbox{-}\rm g$ NQR experiment and $\alpha_{\rm loss}\approx 0.07$ for the $21\mbox{-}\rm g$ experiment. Using these factors, the equivalent Johnson noise that would be detected by our diamond RF magnetometer is ${\sim}2.8~{\rm fT_{\rm rms}\,Hz^{-1/2}}$ for the small coil and ${\sim}2.2~{\rm fT_{\rm rms}\,Hz^{-1/2}}$ for the big coil. These noise levels are much lower than the photoelectron-shot-noise-limited sensitivity of our diamond RF magnetometer and thus are negligible in the present experiments.

\subsection{Remote NQR detection}
\label{sec:AppxCoilEnhanc3}
The remote-detected NQR sensitivity calculations presented in Sec.~\ref{sec:Discussions} of the main text assumed the geometry depicted in Fig.~\ref{fig:NqrCoilRemot}. A non-resonant single-turn loop far from a sodium nitrite powder sample is used to excite its $3.6~{\rm MHz}$ NQR transition. The equivalent sensitivity of a Johnson-noise-limited RF coil to external magnetic fields is $\eta_{\rm J}/Q$, where $\eta_{\rm J}$ is given in Eq.~\eqref{eq:JohNois}. The factor of $1/Q$ comes because any external fields would be amplified by a factor of ${\sim}Q$ due to resonant induction. For an RF coil with volume $V_{\rm c}=1.5~{\rm cm^3}$ (equivalent to our ferrite-cones diamond RF magnetometer volume) and loaded quality factor $Q=10$, the sensitivity is calculated to be $\eta_{\rm J}/Q\,{\approx}\,8~{\rm fT_{rms}\,Hz^{-1/2}}$. This sensitivity is ${\sim}9$ times better than our ferrite-cones diamond magnetometer's sensitivity if we assume the measured sensitivity of ${\sim}70~{\rm fT_{rms}\,s^{1/2}}$ at $0.35~{\rm MHz}$. 


\begin{thebibliography}{75}%
\makeatletter
\providecommand \@ifxundefined [1]{%
 \@ifx{#1\undefined}
}%
\providecommand \@ifnum [1]{%
 \ifnum #1\expandafter \@firstoftwo
 \else \expandafter \@secondoftwo
 \fi
}%
\providecommand \@ifx [1]{%
 \ifx #1\expandafter \@firstoftwo
 \else \expandafter \@secondoftwo
 \fi
}%
\providecommand \natexlab [1]{#1}%
\providecommand \enquote  [1]{``#1''}%
\providecommand \bibnamefont  [1]{#1}%
\providecommand \bibfnamefont [1]{#1}%
\providecommand \citenamefont [1]{#1}%
\providecommand \href@noop [0]{\@secondoftwo}%
\providecommand \href [0]{\begingroup \@sanitize@url \@href}%
\providecommand \@href[1]{\@@startlink{#1}\@@href}%
\providecommand \@@href[1]{\endgroup#1\@@endlink}%
\providecommand \@sanitize@url [0]{\catcode `\\12\catcode `\$12\catcode
  `\&12\catcode `\#12\catcode `\^12\catcode `\_12\catcode `\%12\relax}%
\providecommand \@@startlink[1]{}%
\providecommand \@@endlink[0]{}%
\providecommand \url  [0]{\begingroup\@sanitize@url \@url }%
\providecommand \@url [1]{\endgroup\@href {#1}{\urlprefix }}%
\providecommand \urlprefix  [0]{URL }%
\providecommand \Eprint [0]{\href }%
\providecommand \doibase [0]{http://dx.doi.org/}%
\providecommand \selectlanguage [0]{\@gobble}%
\providecommand \bibinfo  [0]{\@secondoftwo}%
\providecommand \bibfield  [0]{\@secondoftwo}%
\providecommand \translation [1]{[#1]}%
\providecommand \BibitemOpen [0]{}%
\providecommand \bibitemStop [0]{}%
\providecommand \bibitemNoStop [0]{.\EOS\space}%
\providecommand \EOS [0]{\spacefactor3000\relax}%
\providecommand \BibitemShut  [1]{\csname bibitem#1\endcsname}%
\let\auto@bib@innerbib\@empty
\bibitem [{\citenamefont {Barry}\ \emph {et~al.}(2020)\citenamefont {Barry},
  \citenamefont {Schloss}, \citenamefont {Bauch}, \citenamefont {Turner},
  \citenamefont {Hart}, \citenamefont {Pham},\ and\ \citenamefont
  {Walsworth}}]{Bar2020}%
  \BibitemOpen
  \bibfield  {author} {\bibinfo {author} {\bibfnamefont {J.~F.}\ \bibnamefont
  {Barry}}, \bibinfo {author} {\bibfnamefont {J.~M.}\ \bibnamefont {Schloss}},
  \bibinfo {author} {\bibfnamefont {E.}~\bibnamefont {Bauch}}, \bibinfo
  {author} {\bibfnamefont {M.~J.}\ \bibnamefont {Turner}}, \bibinfo {author}
  {\bibfnamefont {C.~A.}\ \bibnamefont {Hart}}, \bibinfo {author}
  {\bibfnamefont {L.~M.}\ \bibnamefont {Pham}}, \ and\ \bibinfo {author}
  {\bibfnamefont {R.~L.}\ \bibnamefont {Walsworth}},\ }\bibfield  {title}
  {\enquote {\bibinfo {title} {{Sensitivity optimization for NV-diamond
  magnetometry}},}\ }\href {\doibase 10.1103/RevModPhys.92.015004} {\bibfield
  {journal} {\bibinfo  {journal} {Reviews of Modern Physics}\ }\textbf
  {\bibinfo {volume} {92}},\ \bibinfo {pages} {15004} (\bibinfo {year}
  {2020})}\BibitemShut {NoStop}%
\bibitem [{\citenamefont {Barry}\ \emph {et~al.}(2016)\citenamefont {Barry},
  \citenamefont {Turner}, \citenamefont {Schloss}, \citenamefont {Glenn},
  \citenamefont {Song}, \citenamefont {Lukin}, \citenamefont {Park},\ and\
  \citenamefont {Walsworth}}]{Bar2016}%
  \BibitemOpen
  \bibfield  {author} {\bibinfo {author} {\bibfnamefont {J.~F.}\ \bibnamefont
  {Barry}}, \bibinfo {author} {\bibfnamefont {M.~J.}\ \bibnamefont {Turner}},
  \bibinfo {author} {\bibfnamefont {J.~M.}\ \bibnamefont {Schloss}}, \bibinfo
  {author} {\bibfnamefont {D.~R.}\ \bibnamefont {Glenn}}, \bibinfo {author}
  {\bibfnamefont {Y.}~\bibnamefont {Song}}, \bibinfo {author} {\bibfnamefont
  {M.~D.}\ \bibnamefont {Lukin}}, \bibinfo {author} {\bibfnamefont
  {H.}~\bibnamefont {Park}}, \ and\ \bibinfo {author} {\bibfnamefont {R.~L.}\
  \bibnamefont {Walsworth}},\ }\bibfield  {title} {\enquote {\bibinfo {title}
  {{Optical magnetic detection of single-neuron action potentials using quantum
  defects in diamond}},}\ }\href {\doibase 10.1073/pnas.1601513113} {\bibfield
  {journal} {\bibinfo  {journal} {Proceedings of the National Academy of
  Sciences of the United States of America}\ }\textbf {\bibinfo {volume}
  {113}},\ \bibinfo {pages} {14133} (\bibinfo {year} {2016})}\BibitemShut
  {NoStop}%
\bibitem [{\citenamefont {Fescenko}\ \emph {et~al.}(2020)\citenamefont
  {Fescenko}, \citenamefont {Jarmola}, \citenamefont {Savukov}, \citenamefont
  {Kehayias}, \citenamefont {Smits}, \citenamefont {Damron}, \citenamefont
  {Ristoff}, \citenamefont {Mosavian},\ and\ \citenamefont {Acosta}}]{Fes2020}%
  \BibitemOpen
  \bibfield  {author} {\bibinfo {author} {\bibfnamefont {I.}~\bibnamefont
  {Fescenko}}, \bibinfo {author} {\bibfnamefont {A.}~\bibnamefont {Jarmola}},
  \bibinfo {author} {\bibfnamefont {I.}~\bibnamefont {Savukov}}, \bibinfo
  {author} {\bibfnamefont {P.}~\bibnamefont {Kehayias}}, \bibinfo {author}
  {\bibfnamefont {J.}~\bibnamefont {Smits}}, \bibinfo {author} {\bibfnamefont
  {J.}~\bibnamefont {Damron}}, \bibinfo {author} {\bibfnamefont
  {N.}~\bibnamefont {Ristoff}}, \bibinfo {author} {\bibfnamefont
  {N.}~\bibnamefont {Mosavian}}, \ and\ \bibinfo {author} {\bibfnamefont
  {V.~M.}\ \bibnamefont {Acosta}},\ }\bibfield  {title} {\enquote {\bibinfo
  {title} {{Diamond magnetometer enhanced by ferrite flux concentrators}},}\
  }\href {\doibase 10.1103/PhysRevResearch.2.023394} {\bibfield  {journal}
  {\bibinfo  {journal} {Physical Review Research}\ }\textbf {\bibinfo {volume}
  {2}},\ \bibinfo {pages} {23394} (\bibinfo {year} {2020})}\BibitemShut
  {NoStop}%
\bibitem [{\citenamefont {Eisenach}\ \emph {et~al.}(2021)\citenamefont
  {Eisenach}, \citenamefont {Barry}, \citenamefont {O'Keeffe}, \citenamefont
  {Schloss}, \citenamefont {Steinecker}, \citenamefont {Englund},\ and\
  \citenamefont {Braje}}]{Eis2021}%
  \BibitemOpen
  \bibfield  {author} {\bibinfo {author} {\bibfnamefont {E.~R.}\ \bibnamefont
  {Eisenach}}, \bibinfo {author} {\bibfnamefont {J.~F.}\ \bibnamefont {Barry}},
  \bibinfo {author} {\bibfnamefont {M.~F.}\ \bibnamefont {O'Keeffe}}, \bibinfo
  {author} {\bibfnamefont {J.~M.}\ \bibnamefont {Schloss}}, \bibinfo {author}
  {\bibfnamefont {M.~H.}\ \bibnamefont {Steinecker}}, \bibinfo {author}
  {\bibfnamefont {D.~R.}\ \bibnamefont {Englund}}, \ and\ \bibinfo {author}
  {\bibfnamefont {D.~A.}\ \bibnamefont {Braje}},\ }\bibfield  {title} {\enquote
  {\bibinfo {title} {{Cavity-enhanced microwave readout of a solid-state spin
  sensor}},}\ }\href {\doibase 10.1038/s41467-021-21256-7} {\bibfield
  {journal} {\bibinfo  {journal} {Nature Communications}\ }\textbf {\bibinfo
  {volume} {12}},\ \bibinfo {pages} {1} (\bibinfo {year} {2021})}\BibitemShut
  {NoStop}%
\bibitem [{\citenamefont {Zhang}\ \emph {et~al.}(2021)\citenamefont {Zhang},
  \citenamefont {Shagieva}, \citenamefont {Widmann}, \citenamefont
  {K{\"{u}}bler}, \citenamefont {Vorobyov}, \citenamefont {Kapitanova},
  \citenamefont {Nenasheva}, \citenamefont {Corkill}, \citenamefont {Rhrle},
  \citenamefont {Nakamura}, \citenamefont {Sumiya}, \citenamefont {Onoda},
  \citenamefont {Isoya},\ and\ \citenamefont {Wrachtrup}}]{Zha2021}%
  \BibitemOpen
  \bibfield  {author} {\bibinfo {author} {\bibfnamefont {C.}~\bibnamefont
  {Zhang}}, \bibinfo {author} {\bibfnamefont {F.}~\bibnamefont {Shagieva}},
  \bibinfo {author} {\bibfnamefont {M.}~\bibnamefont {Widmann}}, \bibinfo
  {author} {\bibfnamefont {M.}~\bibnamefont {K{\"{u}}bler}}, \bibinfo {author}
  {\bibfnamefont {V.}~\bibnamefont {Vorobyov}}, \bibinfo {author}
  {\bibfnamefont {P.}~\bibnamefont {Kapitanova}}, \bibinfo {author}
  {\bibfnamefont {E.}~\bibnamefont {Nenasheva}}, \bibinfo {author}
  {\bibfnamefont {R.}~\bibnamefont {Corkill}}, \bibinfo {author} {\bibfnamefont
  {O.}~\bibnamefont {Rhrle}}, \bibinfo {author} {\bibfnamefont
  {K.}~\bibnamefont {Nakamura}}, \bibinfo {author} {\bibfnamefont
  {H.}~\bibnamefont {Sumiya}}, \bibinfo {author} {\bibfnamefont
  {S.}~\bibnamefont {Onoda}}, \bibinfo {author} {\bibfnamefont
  {J.}~\bibnamefont {Isoya}}, \ and\ \bibinfo {author} {\bibfnamefont
  {J.}~\bibnamefont {Wrachtrup}},\ }\bibfield  {title} {\enquote {\bibinfo
  {title} {{Diamond Magnetometry and Gradiometry towards Subpicotesla dc Field
  Measurement}},}\ }\href {\doibase 10.1103/PhysRevApplied.15.064075}
  {\bibfield  {journal} {\bibinfo  {journal} {Physical Review Applied}\
  }\textbf {\bibinfo {volume} {15}},\ \bibinfo {pages} {1} (\bibinfo {year}
  {2021})}\BibitemShut {NoStop}%
\bibitem [{\citenamefont {Masuyama}\ \emph {et~al.}(2018)\citenamefont
  {Masuyama}, \citenamefont {Mizuno}, \citenamefont {Ozawa}, \citenamefont
  {Ishiwata}, \citenamefont {Hatano}, \citenamefont {Ohshima}, \citenamefont
  {Iwasaki},\ and\ \citenamefont {Hatano}}]{Mas2018}%
  \BibitemOpen
  \bibfield  {author} {\bibinfo {author} {\bibfnamefont {Y.}~\bibnamefont
  {Masuyama}}, \bibinfo {author} {\bibfnamefont {K.}~\bibnamefont {Mizuno}},
  \bibinfo {author} {\bibfnamefont {H.}~\bibnamefont {Ozawa}}, \bibinfo
  {author} {\bibfnamefont {H.}~\bibnamefont {Ishiwata}}, \bibinfo {author}
  {\bibfnamefont {Y.}~\bibnamefont {Hatano}}, \bibinfo {author} {\bibfnamefont
  {T.}~\bibnamefont {Ohshima}}, \bibinfo {author} {\bibfnamefont
  {T.}~\bibnamefont {Iwasaki}}, \ and\ \bibinfo {author} {\bibfnamefont
  {M.}~\bibnamefont {Hatano}},\ }\bibfield  {title} {\enquote {\bibinfo {title}
  {{Extending coherence time of macro-scale diamond magnetometer by dynamical
  decoupling with coplanar waveguide resonator}},}\ }\href {\doibase
  10.1063/1.5047078} {\bibfield  {journal} {\bibinfo  {journal} {Review of
  Scientific Instruments}\ }\textbf {\bibinfo {volume} {89}},\ \bibinfo {pages}
  {125007} (\bibinfo {year} {2018})}\BibitemShut {NoStop}%
\bibitem [{\citenamefont {Glenn}\ \emph {et~al.}(2018)\citenamefont {Glenn},
  \citenamefont {Bucher}, \citenamefont {Lee}, \citenamefont {Lukin},
  \citenamefont {Park},\ and\ \citenamefont {Walsworth}}]{Gle2018}%
  \BibitemOpen
  \bibfield  {author} {\bibinfo {author} {\bibfnamefont {D.~R.}\ \bibnamefont
  {Glenn}}, \bibinfo {author} {\bibfnamefont {D.~B.}\ \bibnamefont {Bucher}},
  \bibinfo {author} {\bibfnamefont {J.}~\bibnamefont {Lee}}, \bibinfo {author}
  {\bibfnamefont {M.~D.}\ \bibnamefont {Lukin}}, \bibinfo {author}
  {\bibfnamefont {H.}~\bibnamefont {Park}}, \ and\ \bibinfo {author}
  {\bibfnamefont {R.~L.}\ \bibnamefont {Walsworth}},\ }\bibfield  {title}
  {\enquote {\bibinfo {title} {{High-resolution magnetic resonance spectroscopy
  using a Solid-State spin sensor}},}\ }\href {\doibase 10.1038/nature25781}
  {\bibfield  {journal} {\bibinfo  {journal} {Nature}\ }\textbf {\bibinfo
  {volume} {555}},\ \bibinfo {pages} {351} (\bibinfo {year}
  {2018})}\BibitemShut {NoStop}%
\bibitem [{\citenamefont {Smits}\ \emph {et~al.}(2019)\citenamefont {Smits},
  \citenamefont {Damron}, \citenamefont {Kehayias}, \citenamefont {McDowell},
  \citenamefont {Mosavian}, \citenamefont {Fescenko}, \citenamefont {Ristoff},
  \citenamefont {Laraoui}, \citenamefont {Jarmola},\ and\ \citenamefont
  {Acosta}}]{Smi2019}%
  \BibitemOpen
  \bibfield  {author} {\bibinfo {author} {\bibfnamefont {J.}~\bibnamefont
  {Smits}}, \bibinfo {author} {\bibfnamefont {J.~T.}\ \bibnamefont {Damron}},
  \bibinfo {author} {\bibfnamefont {P.}~\bibnamefont {Kehayias}}, \bibinfo
  {author} {\bibfnamefont {A.~F.}\ \bibnamefont {McDowell}}, \bibinfo {author}
  {\bibfnamefont {N.}~\bibnamefont {Mosavian}}, \bibinfo {author}
  {\bibfnamefont {I.}~\bibnamefont {Fescenko}}, \bibinfo {author}
  {\bibfnamefont {N.}~\bibnamefont {Ristoff}}, \bibinfo {author} {\bibfnamefont
  {A.}~\bibnamefont {Laraoui}}, \bibinfo {author} {\bibfnamefont
  {A.}~\bibnamefont {Jarmola}}, \ and\ \bibinfo {author} {\bibfnamefont
  {V.~M.}\ \bibnamefont {Acosta}},\ }\bibfield  {title} {\enquote {\bibinfo
  {title} {{Two-dimensional nuclear magnetic resonance spectroscopy with a
  microfluidic diamond quantum sensor}},}\ }\href {\doibase
  10.1126/sciadv.aaw7895} {\bibfield  {journal} {\bibinfo  {journal} {Science
  Advances}\ }\textbf {\bibinfo {volume} {5}},\ \bibinfo {pages} {eaaw7895}
  (\bibinfo {year} {2019})}\BibitemShut {NoStop}%
\bibitem [{\citenamefont {Wang}\ \emph {et~al.}(2022)\citenamefont {Wang},
  \citenamefont {Kong}, \citenamefont {Zhao}, \citenamefont {Huang},
  \citenamefont {Yu}, \citenamefont {Wang}, \citenamefont {Shi},\ and\
  \citenamefont {Du}}]{Wan2022}%
  \BibitemOpen
  \bibfield  {author} {\bibinfo {author} {\bibfnamefont {Z.}~\bibnamefont
  {Wang}}, \bibinfo {author} {\bibfnamefont {F.}~\bibnamefont {Kong}}, \bibinfo
  {author} {\bibfnamefont {P.}~\bibnamefont {Zhao}}, \bibinfo {author}
  {\bibfnamefont {Z.}~\bibnamefont {Huang}}, \bibinfo {author} {\bibfnamefont
  {P.}~\bibnamefont {Yu}}, \bibinfo {author} {\bibfnamefont {Y.}~\bibnamefont
  {Wang}}, \bibinfo {author} {\bibfnamefont {F.}~\bibnamefont {Shi}}, \ and\
  \bibinfo {author} {\bibfnamefont {J.}~\bibnamefont {Du}},\ }\bibfield
  {title} {\enquote {\bibinfo {title} {{Picotesla magnetometry of microwave
  fields with diamond sensors}},}\ }\href
  {https://www.science.org/doi/10.1126/sciadv.abq8158} {\bibfield  {journal}
  {\bibinfo  {journal} {Science Advances}\ }\textbf {\bibinfo {volume} {8}}
  (\bibinfo {year} {2022})}\BibitemShut {NoStop}%
\bibitem [{\citenamefont {Alsid}\ \emph {et~al.}(2022)\citenamefont {Alsid},
  \citenamefont {Schloss}, \citenamefont {Steinecker}, \citenamefont {Barry},
  \citenamefont {Maccabe}, \citenamefont {Wang}, \citenamefont {Cappellaro},\
  and\ \citenamefont {Braje}}]{Als2022}%
  \BibitemOpen
  \bibfield  {author} {\bibinfo {author} {\bibfnamefont {S.~T.}\ \bibnamefont
  {Alsid}}, \bibinfo {author} {\bibfnamefont {J.~M.}\ \bibnamefont {Schloss}},
  \bibinfo {author} {\bibfnamefont {M.~H.}\ \bibnamefont {Steinecker}},
  \bibinfo {author} {\bibfnamefont {J.~F.}\ \bibnamefont {Barry}}, \bibinfo
  {author} {\bibfnamefont {A.~C.}\ \bibnamefont {Maccabe}}, \bibinfo {author}
  {\bibfnamefont {G.}~\bibnamefont {Wang}}, \bibinfo {author} {\bibfnamefont
  {P.}~\bibnamefont {Cappellaro}}, \ and\ \bibinfo {author} {\bibfnamefont
  {D.~A.}\ \bibnamefont {Braje}},\ }\bibfield  {title} {\enquote {\bibinfo
  {title} {{A Solid-State Microwave Magnetometer with Picotesla-Level
  Sensitivity}},}\ }\href {http://arxiv.org/abs/2206.15440} {\bibfield
  {journal} {\bibinfo  {journal} {arXiv:2206.15440}\ } (\bibinfo {year}
  {2022})}\BibitemShut {NoStop}%
\bibitem [{\citenamefont {Savukov}\ \emph {et~al.}(2005)\citenamefont
  {Savukov}, \citenamefont {Seltzer}, \citenamefont {Romalis},\ and\
  \citenamefont {Sauer}}]{Sav2005}%
  \BibitemOpen
  \bibfield  {author} {\bibinfo {author} {\bibfnamefont {I.~M.}\ \bibnamefont
  {Savukov}}, \bibinfo {author} {\bibfnamefont {S.~J.}\ \bibnamefont
  {Seltzer}}, \bibinfo {author} {\bibfnamefont {M.~V.}\ \bibnamefont
  {Romalis}}, \ and\ \bibinfo {author} {\bibfnamefont {K.~L.}\ \bibnamefont
  {Sauer}},\ }\bibfield  {title} {\enquote {\bibinfo {title} {{Tunable atomic
  magnetometer for detection of radio-frequency magnetic fields}},}\ }\href
  {\doibase 10.1103/PhysRevLett.95.063004} {\bibfield  {journal} {\bibinfo
  {journal} {Physical Review Letters}\ }\textbf {\bibinfo {volume} {95}},\
  \bibinfo {pages} {3} (\bibinfo {year} {2005})}\BibitemShut {NoStop}%
\bibitem [{\citenamefont {Ledbetter}\ \emph {et~al.}(2007)\citenamefont
  {Ledbetter}, \citenamefont {Acosta}, \citenamefont {Rochester}, \citenamefont
  {Budker}, \citenamefont {Pustelny},\ and\ \citenamefont
  {Yashchuk}}]{Led2007}%
  \BibitemOpen
  \bibfield  {author} {\bibinfo {author} {\bibfnamefont {M.~P.}\ \bibnamefont
  {Ledbetter}}, \bibinfo {author} {\bibfnamefont {V.~M.}\ \bibnamefont
  {Acosta}}, \bibinfo {author} {\bibfnamefont {S.~M.}\ \bibnamefont
  {Rochester}}, \bibinfo {author} {\bibfnamefont {D.}~\bibnamefont {Budker}},
  \bibinfo {author} {\bibfnamefont {S.}~\bibnamefont {Pustelny}}, \ and\
  \bibinfo {author} {\bibfnamefont {V.~V.}\ \bibnamefont {Yashchuk}},\
  }\bibfield  {title} {\enquote {\bibinfo {title} {{Detection of
  radio-frequency magnetic fields using nonlinear magneto-optical rotation}},}\
  }\href {\doibase 10.1103/PhysRevA.75.023405} {\bibfield  {journal} {\bibinfo
  {journal} {Physical Review A}\ }\textbf {\bibinfo {volume} {75}},\ \bibinfo
  {pages} {023405} (\bibinfo {year} {2007})}\BibitemShut {NoStop}%
\bibitem [{\citenamefont {Griffith}\ \emph {et~al.}(2010)\citenamefont
  {Griffith}, \citenamefont {Knappe},\ and\ \citenamefont
  {Kitching}}]{Gri2010}%
  \BibitemOpen
  \bibfield  {author} {\bibinfo {author} {\bibfnamefont {W.~C.}\ \bibnamefont
  {Griffith}}, \bibinfo {author} {\bibfnamefont {S.}~\bibnamefont {Knappe}}, \
  and\ \bibinfo {author} {\bibfnamefont {J.}~\bibnamefont {Kitching}},\
  }\bibfield  {title} {\enquote {\bibinfo {title} {{Femtotesla atomic
  magnetometry in a microfabricated vapor cell}},}\ }\href {\doibase
  10.1364/oe.18.027167} {\bibfield  {journal} {\bibinfo  {journal} {Optics
  Express}\ }\textbf {\bibinfo {volume} {18}},\ \bibinfo {pages} {27167}
  (\bibinfo {year} {2010})}\BibitemShut {NoStop}%
\bibitem [{\citenamefont {Wasilewski}\ \emph {et~al.}(2010)\citenamefont
  {Wasilewski}, \citenamefont {Jensen}, \citenamefont {Krauter}, \citenamefont
  {Renema}, \citenamefont {Balabas},\ and\ \citenamefont {Polzik}}]{Was2010}%
  \BibitemOpen
  \bibfield  {author} {\bibinfo {author} {\bibfnamefont {W.}~\bibnamefont
  {Wasilewski}}, \bibinfo {author} {\bibfnamefont {K.}~\bibnamefont {Jensen}},
  \bibinfo {author} {\bibfnamefont {H.}~\bibnamefont {Krauter}}, \bibinfo
  {author} {\bibfnamefont {J.~J.}\ \bibnamefont {Renema}}, \bibinfo {author}
  {\bibfnamefont {M.~V.}\ \bibnamefont {Balabas}}, \ and\ \bibinfo {author}
  {\bibfnamefont {E.~S.}\ \bibnamefont {Polzik}},\ }\bibfield  {title}
  {\enquote {\bibinfo {title} {{Quantum Noise Limited and Entanglement-Assisted
  Magnetometry}},}\ }\href {\doibase 10.1103/PhysRevLett.104.133601} {\bibfield
   {journal} {\bibinfo  {journal} {Physical Review Letters}\ }\textbf {\bibinfo
  {volume} {104}},\ \bibinfo {pages} {133601} (\bibinfo {year}
  {2010})}\BibitemShut {NoStop}%
\bibitem [{\citenamefont {Dhombridge}\ \emph {et~al.}(2022)\citenamefont
  {Dhombridge}, \citenamefont {Claussen}, \citenamefont {Iivanainen},\ and\
  \citenamefont {Schwindt}}]{Dho2022}%
  \BibitemOpen
  \bibfield  {author} {\bibinfo {author} {\bibfnamefont {J.~E.}\ \bibnamefont
  {Dhombridge}}, \bibinfo {author} {\bibfnamefont {N.~R.}\ \bibnamefont
  {Claussen}}, \bibinfo {author} {\bibfnamefont {J.}~\bibnamefont
  {Iivanainen}}, \ and\ \bibinfo {author} {\bibfnamefont {P.~D.~D.}\
  \bibnamefont {Schwindt}},\ }\bibfield  {title} {\enquote {\bibinfo {title}
  {{High-Sensitivity rf Detection Using an Optically Pumped Comagnetometer
  Based on Natural-Abundance Rubidium with Active Ambient-Field
  Cancellation}},}\ }\href {\doibase 10.1103/PhysRevApplied.18.044052}
  {\bibfield  {journal} {\bibinfo  {journal} {Physical Review Applied}\
  }\textbf {\bibinfo {volume} {10}},\ \bibinfo {pages} {1} (\bibinfo {year}
  {2022})}\BibitemShut {NoStop}%
\bibitem [{\citenamefont {Chalupczak}\ \emph {et~al.}(2012)\citenamefont
  {Chalupczak}, \citenamefont {Godun}, \citenamefont {Pustelny},\ and\
  \citenamefont {Gawlik}}]{Cha2012}%
  \BibitemOpen
  \bibfield  {author} {\bibinfo {author} {\bibfnamefont {W.}~\bibnamefont
  {Chalupczak}}, \bibinfo {author} {\bibfnamefont {R.~M.}\ \bibnamefont
  {Godun}}, \bibinfo {author} {\bibfnamefont {S.}~\bibnamefont {Pustelny}}, \
  and\ \bibinfo {author} {\bibfnamefont {W.}~\bibnamefont {Gawlik}},\
  }\bibfield  {title} {\enquote {\bibinfo {title} {{Room temperature femtotesla
  radio-frequency atomic magnetometer}},}\ }\href {\doibase 10.1063/1.4729016}
  {\bibfield  {journal} {\bibinfo  {journal} {Applied Physics Letters}\
  }\textbf {\bibinfo {volume} {100}},\ \bibinfo {pages} {242401} (\bibinfo
  {year} {2012})}\BibitemShut {NoStop}%
\bibitem [{\citenamefont {Keder}\ \emph {et~al.}(2014)\citenamefont {Keder},
  \citenamefont {Prescott}, \citenamefont {Conovaloff},\ and\ \citenamefont
  {Sauer}}]{Ked2014}%
  \BibitemOpen
  \bibfield  {author} {\bibinfo {author} {\bibfnamefont {D.~A.}\ \bibnamefont
  {Keder}}, \bibinfo {author} {\bibfnamefont {D.~W.}\ \bibnamefont {Prescott}},
  \bibinfo {author} {\bibfnamefont {A.~W.}\ \bibnamefont {Conovaloff}}, \ and\
  \bibinfo {author} {\bibfnamefont {K.~L.}\ \bibnamefont {Sauer}},\ }\bibfield
  {title} {\enquote {\bibinfo {title} {{An unshielded radio-frequency atomic
  magnetometer with sub-femtoTesla sensitivity}},}\ }\href {\doibase
  10.1063/1.4905449} {\bibfield  {journal} {\bibinfo  {journal} {AIP Advances}\
  }\textbf {\bibinfo {volume} {4}},\ \bibinfo {pages} {127159} (\bibinfo {year}
  {2014})}\BibitemShut {NoStop}%
\bibitem [{\citenamefont {Clarke}(1980)}]{Cla1980}%
  \BibitemOpen
  \bibfield  {author} {\bibinfo {author} {\bibfnamefont {J.}~\bibnamefont
  {Clarke}},\ }\bibfield  {title} {\enquote {\bibinfo {title} {{Advances in
  SQUID Magnetometers}},}\ }\href {\doibase 10.1109/T-ED.1980.20127} {\bibfield
   {journal} {\bibinfo  {journal} {IEEE Transactions on Electron Devices}\
  }\textbf {\bibinfo {volume} {27}},\ \bibinfo {pages} {1896} (\bibinfo {year}
  {1980})}\BibitemShut {NoStop}%
\bibitem [{\citenamefont {Schmelz}\ \emph {et~al.}(2016)\citenamefont
  {Schmelz}, \citenamefont {Zakosarenko}, \citenamefont {Chwala}, \citenamefont
  {Sch{\"{o}}nau}, \citenamefont {Stolz}, \citenamefont {Anders}, \citenamefont
  {Linzen},\ and\ \citenamefont {Meyer}}]{Sch2016}%
  \BibitemOpen
  \bibfield  {author} {\bibinfo {author} {\bibfnamefont {M.}~\bibnamefont
  {Schmelz}}, \bibinfo {author} {\bibfnamefont {V.}~\bibnamefont
  {Zakosarenko}}, \bibinfo {author} {\bibfnamefont {A.}~\bibnamefont {Chwala}},
  \bibinfo {author} {\bibfnamefont {T.}~\bibnamefont {Sch{\"{o}}nau}}, \bibinfo
  {author} {\bibfnamefont {R.}~\bibnamefont {Stolz}}, \bibinfo {author}
  {\bibfnamefont {S.}~\bibnamefont {Anders}}, \bibinfo {author} {\bibfnamefont
  {S.}~\bibnamefont {Linzen}}, \ and\ \bibinfo {author} {\bibfnamefont {H.~G.}\
  \bibnamefont {Meyer}},\ }\bibfield  {title} {\enquote {\bibinfo {title}
  {{Thin-Film-Based Ultralow Noise SQUID Magnetometer}},}\ }\href {\doibase
  10.1109/TASC.2016.2530699} {\bibfield  {journal} {\bibinfo  {journal} {IEEE
  Transactions on Applied Superconductivity}\ }\textbf {\bibinfo {volume}
  {26}},\ \bibinfo {pages} {1} (\bibinfo {year} {2016})}\BibitemShut {NoStop}%
\bibitem [{\citenamefont {Storm}\ \emph {et~al.}(2017)\citenamefont {Storm},
  \citenamefont {H{\"{o}}mmen}, \citenamefont {Drung},\ and\ \citenamefont
  {K{\"{o}}rber}}]{Sto2017}%
  \BibitemOpen
  \bibfield  {author} {\bibinfo {author} {\bibfnamefont {J.-H.}\ \bibnamefont
  {Storm}}, \bibinfo {author} {\bibfnamefont {P.}~\bibnamefont {H{\"{o}}mmen}},
  \bibinfo {author} {\bibfnamefont {D.}~\bibnamefont {Drung}}, \ and\ \bibinfo
  {author} {\bibfnamefont {R.}~\bibnamefont {K{\"{o}}rber}},\ }\bibfield
  {title} {\enquote {\bibinfo {title} {{An ultra-sensitive and wideband
  magnetometer based on a superconducting quantum interference device}},}\
  }\href {\doibase 10.1063/1.4976823} {\bibfield  {journal} {\bibinfo
  {journal} {Applied Physics Letters}\ }\textbf {\bibinfo {volume} {110}},\
  \bibinfo {pages} {072603} (\bibinfo {year} {2017})}\BibitemShut {NoStop}%
\bibitem [{\citenamefont {Li}\ \emph {et~al.}(2021)\citenamefont {Li},
  \citenamefont {Wang}, \citenamefont {Guo}, \citenamefont {Guo}, \citenamefont
  {Wen}, \citenamefont {Tang},\ and\ \citenamefont {Liu}}]{Li2021}%
  \BibitemOpen
  \bibfield  {author} {\bibinfo {author} {\bibfnamefont {Z.-H.}\ \bibnamefont
  {Li}}, \bibinfo {author} {\bibfnamefont {T.-Y.}\ \bibnamefont {Wang}},
  \bibinfo {author} {\bibfnamefont {Q.}~\bibnamefont {Guo}}, \bibinfo {author}
  {\bibfnamefont {H.}~\bibnamefont {Guo}}, \bibinfo {author} {\bibfnamefont
  {H.-F.}\ \bibnamefont {Wen}}, \bibinfo {author} {\bibfnamefont
  {J.}~\bibnamefont {Tang}}, \ and\ \bibinfo {author} {\bibfnamefont
  {J.}~\bibnamefont {Liu}},\ }\bibfield  {title} {\enquote {\bibinfo {title}
  {{Enhancement of magnetic detection by ensemble NV color center based on
  magnetic flux concentration effect}},}\ }\href {\doibase
  10.7498/aps.70.20210129} {\bibfield  {journal} {\bibinfo  {journal} {Acta
  Physica Sinica}\ }\textbf {\bibinfo {volume} {70}},\ \bibinfo {pages}
  {147601} (\bibinfo {year} {2021})}\BibitemShut {NoStop}%
\bibitem [{\citenamefont {Xie}\ \emph {et~al.}(2022)\citenamefont {Xie},
  \citenamefont {Xie}, \citenamefont {Zhu}, \citenamefont {Jing}, \citenamefont
  {Tong}, \citenamefont {Qin}, \citenamefont {Guan}, \citenamefont {Duan},
  \citenamefont {Wang}, \citenamefont {Rong},\ and\ \citenamefont
  {Du}}]{Xie2022}%
  \BibitemOpen
  \bibfield  {author} {\bibinfo {author} {\bibfnamefont {Y.}~\bibnamefont
  {Xie}}, \bibinfo {author} {\bibfnamefont {C.}~\bibnamefont {Xie}}, \bibinfo
  {author} {\bibfnamefont {Y.}~\bibnamefont {Zhu}}, \bibinfo {author}
  {\bibfnamefont {K.}~\bibnamefont {Jing}}, \bibinfo {author} {\bibfnamefont
  {Y.}~\bibnamefont {Tong}}, \bibinfo {author} {\bibfnamefont {X.}~\bibnamefont
  {Qin}}, \bibinfo {author} {\bibfnamefont {H.}~\bibnamefont {Guan}}, \bibinfo
  {author} {\bibfnamefont {C.-K.}\ \bibnamefont {Duan}}, \bibinfo {author}
  {\bibfnamefont {Y.}~\bibnamefont {Wang}}, \bibinfo {author} {\bibfnamefont
  {X.}~\bibnamefont {Rong}}, \ and\ \bibinfo {author} {\bibfnamefont
  {J.}~\bibnamefont {Du}},\ }\bibfield  {title} {\enquote {\bibinfo {title}
  {{$T_2$-limited dc Quantum Magnetometry via Flux Modulation}},}\ }\href
  {http://arxiv.org/abs/2204.07343} {\bibfield  {journal} {\bibinfo  {journal}
  {arXiv:2204.07343}\ } (\bibinfo {year} {2022})}\BibitemShut {NoStop}%
\bibitem [{\citenamefont {Taylor}\ \emph {et~al.}(2008)\citenamefont {Taylor},
  \citenamefont {Cappellaro}, \citenamefont {Childress}, \citenamefont {Jiang},
  \citenamefont {Budker}, \citenamefont {Hemmer}, \citenamefont {Yacoby},
  \citenamefont {Walsworth},\ and\ \citenamefont {Lukin}}]{Tay2008}%
  \BibitemOpen
  \bibfield  {author} {\bibinfo {author} {\bibfnamefont {J.~M.}\ \bibnamefont
  {Taylor}}, \bibinfo {author} {\bibfnamefont {P.}~\bibnamefont {Cappellaro}},
  \bibinfo {author} {\bibfnamefont {L.}~\bibnamefont {Childress}}, \bibinfo
  {author} {\bibfnamefont {L.}~\bibnamefont {Jiang}}, \bibinfo {author}
  {\bibfnamefont {D.}~\bibnamefont {Budker}}, \bibinfo {author} {\bibfnamefont
  {P.~R.}\ \bibnamefont {Hemmer}}, \bibinfo {author} {\bibfnamefont
  {A.}~\bibnamefont {Yacoby}}, \bibinfo {author} {\bibfnamefont
  {R.}~\bibnamefont {Walsworth}}, \ and\ \bibinfo {author} {\bibfnamefont
  {M.~D.}\ \bibnamefont {Lukin}},\ }\bibfield  {title} {\enquote {\bibinfo
  {title} {{High-sensitivity diamond magnetometer with nanoscale
  resolution}},}\ }\href {\doibase 10.1038/nphys1075} {\bibfield  {journal}
  {\bibinfo  {journal} {Nature Physics}\ }\textbf {\bibinfo {volume} {4}},\
  \bibinfo {pages} {810} (\bibinfo {year} {2008})}\BibitemShut {NoStop}%
\bibitem [{\citenamefont {Connor}\ \emph {et~al.}(1990)\citenamefont {Connor},
  \citenamefont {Chang},\ and\ \citenamefont {Pines}}]{Con1990}%
  \BibitemOpen
  \bibfield  {author} {\bibinfo {author} {\bibfnamefont {C.}~\bibnamefont
  {Connor}}, \bibinfo {author} {\bibfnamefont {J.}~\bibnamefont {Chang}}, \
  and\ \bibinfo {author} {\bibfnamefont {A.}~\bibnamefont {Pines}},\ }\bibfield
   {title} {\enquote {\bibinfo {title} {{Magnetic resonance spectrometer with a
  dc SQUID detector}},}\ }\href {\doibase 10.1063/1.1141476} {\bibfield
  {journal} {\bibinfo  {journal} {Review of Scientific Instruments}\ }\textbf
  {\bibinfo {volume} {61}},\ \bibinfo {pages} {1059} (\bibinfo {year}
  {1990})}\BibitemShut {NoStop}%
\bibitem [{\citenamefont {Augustine}\ \emph {et~al.}(1998)\citenamefont
  {Augustine}, \citenamefont {TonThat},\ and\ \citenamefont
  {Clarke}}]{Aug1998}%
  \BibitemOpen
  \bibfield  {author} {\bibinfo {author} {\bibfnamefont {M.~P.}\ \bibnamefont
  {Augustine}}, \bibinfo {author} {\bibfnamefont {D.~M.}\ \bibnamefont
  {TonThat}}, \ and\ \bibinfo {author} {\bibfnamefont {J.}~\bibnamefont
  {Clarke}},\ }\bibfield  {title} {\enquote {\bibinfo {title} {{SQUID detected
  NMR and NQR}},}\ }\href {\doibase 10.1016/S0926-2040(97)00103-3} {\bibfield
  {journal} {\bibinfo  {journal} {Solid State Nuclear Magnetic Resonance}\
  }\textbf {\bibinfo {volume} {11}},\ \bibinfo {pages} {139} (\bibinfo {year}
  {1998})}\BibitemShut {NoStop}%
\bibitem [{\citenamefont {Lee}\ \emph {et~al.}(2006)\citenamefont {Lee},
  \citenamefont {Sauer}, \citenamefont {Seltzer}, \citenamefont {Alem},\ and\
  \citenamefont {Romalis}}]{Lee2006}%
  \BibitemOpen
  \bibfield  {author} {\bibinfo {author} {\bibfnamefont {S.~K.}\ \bibnamefont
  {Lee}}, \bibinfo {author} {\bibfnamefont {K.~L.}\ \bibnamefont {Sauer}},
  \bibinfo {author} {\bibfnamefont {S.~J.}\ \bibnamefont {Seltzer}}, \bibinfo
  {author} {\bibfnamefont {O.}~\bibnamefont {Alem}}, \ and\ \bibinfo {author}
  {\bibfnamefont {M.~V.}\ \bibnamefont {Romalis}},\ }\bibfield  {title}
  {\enquote {\bibinfo {title} {{Subfemtotesla radio-frequency atomic
  magnetometer for detection of nuclear quadrupole resonance}},}\ }\href
  {\doibase 10.1063/1.2390643} {\bibfield  {journal} {\bibinfo  {journal}
  {Applied Physics Letters}\ }\textbf {\bibinfo {volume} {89}},\ \bibinfo
  {pages} {23} (\bibinfo {year} {2006})}\BibitemShut {NoStop}%
\bibitem [{\citenamefont {Das}\ and\ \citenamefont {Hahn}(1958)}]{Das1958}%
  \BibitemOpen
  \bibfield  {author} {\bibinfo {author} {\bibfnamefont {T.}~\bibnamefont
  {Das}}\ and\ \bibinfo {author} {\bibfnamefont {E.}~\bibnamefont {Hahn}},\
  }\href
  {https://www.worldcat.org/title/nuclear-quadrupole-resonance-spectroscopy/oclc/545815}
  {\emph {\bibinfo {title} {{Nuclear quadrupole resonance spectroscopy}}}}\
  (\bibinfo  {publisher} {Academic Press},\ \bibinfo {year} {1958})\BibitemShut
  {NoStop}%
\bibitem [{\citenamefont {Zax}\ \emph {et~al.}(1985)\citenamefont {Zax},
  \citenamefont {Bielecki}, \citenamefont {Zilm}, \citenamefont {Pines},\ and\
  \citenamefont {Weitekamp}}]{Zax1985}%
  \BibitemOpen
  \bibfield  {author} {\bibinfo {author} {\bibfnamefont {D.~B.}\ \bibnamefont
  {Zax}}, \bibinfo {author} {\bibfnamefont {A.}~\bibnamefont {Bielecki}},
  \bibinfo {author} {\bibfnamefont {K.~W.}\ \bibnamefont {Zilm}}, \bibinfo
  {author} {\bibfnamefont {A.}~\bibnamefont {Pines}}, \ and\ \bibinfo {author}
  {\bibfnamefont {D.~P.}\ \bibnamefont {Weitekamp}},\ }\bibfield  {title}
  {\enquote {\bibinfo {title} {{Zero field NMR and NQR}},}\ }\href {\doibase
  10.1063/1.449748} {\bibfield  {journal} {\bibinfo  {journal} {The Journal of
  Chemical Physics}\ }\textbf {\bibinfo {volume} {83}},\ \bibinfo {pages}
  {4877} (\bibinfo {year} {1985})}\BibitemShut {NoStop}%
\bibitem [{\citenamefont {Garroway}\ \emph {et~al.}(2001)\citenamefont
  {Garroway}, \citenamefont {Buess}, \citenamefont {Miller}, \citenamefont
  {Suits}, \citenamefont {Hibbs}, \citenamefont {Barrall}, \citenamefont
  {Matthews},\ and\ \citenamefont {Burnett}}]{Gar2001}%
  \BibitemOpen
  \bibfield  {author} {\bibinfo {author} {\bibfnamefont {A.~N.}\ \bibnamefont
  {Garroway}}, \bibinfo {author} {\bibfnamefont {M.~L.}\ \bibnamefont {Buess}},
  \bibinfo {author} {\bibfnamefont {J.~B.}\ \bibnamefont {Miller}}, \bibinfo
  {author} {\bibfnamefont {B.~H.}\ \bibnamefont {Suits}}, \bibinfo {author}
  {\bibfnamefont {A.~D.}\ \bibnamefont {Hibbs}}, \bibinfo {author}
  {\bibfnamefont {G.~A.}\ \bibnamefont {Barrall}}, \bibinfo {author}
  {\bibfnamefont {R.}~\bibnamefont {Matthews}}, \ and\ \bibinfo {author}
  {\bibfnamefont {L.~J.}\ \bibnamefont {Burnett}},\ }\bibfield  {title}
  {\enquote {\bibinfo {title} {{Remote sensing by nuclear quadrupole
  resonance}},}\ }\href {\doibase 10.1109/36.927420} {\bibfield  {journal}
  {\bibinfo  {journal} {IEEE Transactions on Geoscience and Remote Sensing}\
  }\textbf {\bibinfo {volume} {39}},\ \bibinfo {pages} {1108} (\bibinfo {year}
  {2001})}\BibitemShut {NoStop}%
\bibitem [{\citenamefont {Kim}\ \emph {et~al.}(2014)\citenamefont {Kim},
  \citenamefont {Karaulanov}, \citenamefont {Matlashov}, \citenamefont
  {Newman}, \citenamefont {Urbaitis}, \citenamefont {Volegov}, \citenamefont
  {Yoder},\ and\ \citenamefont {Espy}}]{Kim2014}%
  \BibitemOpen
  \bibfield  {author} {\bibinfo {author} {\bibfnamefont {Y.~J.}\ \bibnamefont
  {Kim}}, \bibinfo {author} {\bibfnamefont {T.}~\bibnamefont {Karaulanov}},
  \bibinfo {author} {\bibfnamefont {A.~N.}\ \bibnamefont {Matlashov}}, \bibinfo
  {author} {\bibfnamefont {S.}~\bibnamefont {Newman}}, \bibinfo {author}
  {\bibfnamefont {A.}~\bibnamefont {Urbaitis}}, \bibinfo {author}
  {\bibfnamefont {P.}~\bibnamefont {Volegov}}, \bibinfo {author} {\bibfnamefont
  {J.}~\bibnamefont {Yoder}}, \ and\ \bibinfo {author} {\bibfnamefont {M.~A.}\
  \bibnamefont {Espy}},\ }\bibfield  {title} {\enquote {\bibinfo {title}
  {{Polarization enhancement technique for nuclear quadrupole resonance
  detection}},}\ }\href {\doibase 10.1016/j.ssnmr.2014.05.002} {\bibfield
  {journal} {\bibinfo  {journal} {Solid State Nuclear Magnetic Resonance}\
  }\textbf {\bibinfo {volume} {61-62}},\ \bibinfo {pages} {35} (\bibinfo {year}
  {2014})}\BibitemShut {NoStop}%
\bibitem [{\citenamefont {Malone}\ \emph {et~al.}(2020)\citenamefont {Malone},
  \citenamefont {Espy}, \citenamefont {He}, \citenamefont {Janicke},\ and\
  \citenamefont {Williams}}]{Mal2020}%
  \BibitemOpen
  \bibfield  {author} {\bibinfo {author} {\bibfnamefont {M.~W.}\ \bibnamefont
  {Malone}}, \bibinfo {author} {\bibfnamefont {M.~A.}\ \bibnamefont {Espy}},
  \bibinfo {author} {\bibfnamefont {S.}~\bibnamefont {He}}, \bibinfo {author}
  {\bibfnamefont {M.~T.}\ \bibnamefont {Janicke}}, \ and\ \bibinfo {author}
  {\bibfnamefont {R.~F.}\ \bibnamefont {Williams}},\ }\bibfield  {title}
  {\enquote {\bibinfo {title} {{The 1H T1 dispersion curve of fentanyl citrate
  to identify NQR parameters}},}\ }\href {\doibase 10.1016/j.ssnmr.2020.101697}
  {\bibfield  {journal} {\bibinfo  {journal} {Solid State Nuclear Magnetic
  Resonance}\ }\textbf {\bibinfo {volume} {110}},\ \bibinfo {pages} {101697}
  (\bibinfo {year} {2020})}\BibitemShut {NoStop}%
\bibitem [{\citenamefont {Balchin}\ \emph {et~al.}(2005)\citenamefont
  {Balchin}, \citenamefont {Malcolme-Lawes}, \citenamefont {Poplett},
  \citenamefont {Rowe}, \citenamefont {Smith}, \citenamefont {Pearce},\ and\
  \citenamefont {Wren}}]{Bal2005}%
  \BibitemOpen
  \bibfield  {author} {\bibinfo {author} {\bibfnamefont {E.}~\bibnamefont
  {Balchin}}, \bibinfo {author} {\bibfnamefont {D.~J.}\ \bibnamefont
  {Malcolme-Lawes}}, \bibinfo {author} {\bibfnamefont {I.~J.}\ \bibnamefont
  {Poplett}}, \bibinfo {author} {\bibfnamefont {M.~D.}\ \bibnamefont {Rowe}},
  \bibinfo {author} {\bibfnamefont {J.~A.}\ \bibnamefont {Smith}}, \bibinfo
  {author} {\bibfnamefont {G.~E.}\ \bibnamefont {Pearce}}, \ and\ \bibinfo
  {author} {\bibfnamefont {S.~A.}\ \bibnamefont {Wren}},\ }\bibfield  {title}
  {\enquote {\bibinfo {title} {{Potential of nuclear quadrupole resonance in
  pharmaceutical analysis}},}\ }\href {\doibase 10.1021/ac0503658} {\bibfield
  {journal} {\bibinfo  {journal} {Analytical Chemistry}\ }\textbf {\bibinfo
  {volume} {77}},\ \bibinfo {pages} {3925} (\bibinfo {year}
  {2005})}\BibitemShut {NoStop}%
\bibitem [{\citenamefont {Lu{\v{z}}nik}\ \emph {et~al.}(2013)\citenamefont
  {Lu{\v{z}}nik}, \citenamefont {Pirnat}, \citenamefont {Jazbin{\v{s}}ek},
  \citenamefont {Lavri{\v{c}}}, \citenamefont {Sr{\v{c}}i{\v{c}}},\ and\
  \citenamefont {Trontelj}}]{Luz2013}%
  \BibitemOpen
  \bibfield  {author} {\bibinfo {author} {\bibfnamefont {J.}~\bibnamefont
  {Lu{\v{z}}nik}}, \bibinfo {author} {\bibfnamefont {J.}~\bibnamefont
  {Pirnat}}, \bibinfo {author} {\bibfnamefont {V.}~\bibnamefont
  {Jazbin{\v{s}}ek}}, \bibinfo {author} {\bibfnamefont {Z.}~\bibnamefont
  {Lavri{\v{c}}}}, \bibinfo {author} {\bibfnamefont {S.}~\bibnamefont
  {Sr{\v{c}}i{\v{c}}}}, \ and\ \bibinfo {author} {\bibfnamefont
  {Z.}~\bibnamefont {Trontelj}},\ }\bibfield  {title} {\enquote {\bibinfo
  {title} {{The Influence of Pressure in Paracetamol Tablet Compaction on 14N
  Nuclear Quadrupole Resonance Signal}},}\ }\href {\doibase
  10.1007/s00723-013-0440-3} {\bibfield  {journal} {\bibinfo  {journal}
  {Applied Magnetic Resonance}\ }\textbf {\bibinfo {volume} {44}},\ \bibinfo
  {pages} {735} (\bibinfo {year} {2013})}\BibitemShut {NoStop}%
\bibitem [{\citenamefont {Kyriakidou}\ \emph {et~al.}(2015)\citenamefont
  {Kyriakidou}, \citenamefont {Jakobsson}, \citenamefont {Althoefer},\ and\
  \citenamefont {Barras}}]{Kyr2015}%
  \BibitemOpen
  \bibfield  {author} {\bibinfo {author} {\bibfnamefont {G.}~\bibnamefont
  {Kyriakidou}}, \bibinfo {author} {\bibfnamefont {A.}~\bibnamefont
  {Jakobsson}}, \bibinfo {author} {\bibfnamefont {K.}~\bibnamefont
  {Althoefer}}, \ and\ \bibinfo {author} {\bibfnamefont {J.}~\bibnamefont
  {Barras}},\ }\bibfield  {title} {\enquote {\bibinfo {title} {{Batch-Specific
  Discrimination Using Nuclear Quadrupole Resonance Spectroscopy}},}\ }\href
  {\doibase 10.1021/ac5044658} {\bibfield  {journal} {\bibinfo  {journal}
  {Analytical Chemistry}\ }\textbf {\bibinfo {volume} {87}},\ \bibinfo {pages}
  {3806} (\bibinfo {year} {2015})}\BibitemShut {NoStop}%
\bibitem [{\citenamefont {Lyfar}\ \emph {et~al.}(1976)\citenamefont {Lyfar},
  \citenamefont {Goncharuk},\ and\ \citenamefont {Ryabchenko}}]{Lyf1976}%
  \BibitemOpen
  \bibfield  {author} {\bibinfo {author} {\bibfnamefont {D.~L.}\ \bibnamefont
  {Lyfar}}, \bibinfo {author} {\bibfnamefont {V.~E.}\ \bibnamefont
  {Goncharuk}}, \ and\ \bibinfo {author} {\bibfnamefont {S.~M.}\ \bibnamefont
  {Ryabchenko}},\ }\bibfield  {title} {\enquote {\bibinfo {title} {{Temperature
  Dependence of Nuclear Quadrupole Resonance in Layer-Type Crystals}},}\ }\href
  {\doibase 10.1002/pssb.2220760119} {\bibfield  {journal} {\bibinfo  {journal}
  {physica status solidi (b)}\ }\textbf {\bibinfo {volume} {76}},\ \bibinfo
  {pages} {183} (\bibinfo {year} {1976})}\BibitemShut {NoStop}%
\bibitem [{\citenamefont {Sharma}\ \emph {et~al.}(1986)\citenamefont {Sharma},
  \citenamefont {Paulus}, \citenamefont {Weiden},\ and\ \citenamefont
  {Weiss}}]{Sha1986}%
  \BibitemOpen
  \bibfield  {author} {\bibinfo {author} {\bibfnamefont {S.}~\bibnamefont
  {Sharma}}, \bibinfo {author} {\bibfnamefont {H.}~\bibnamefont {Paulus}},
  \bibinfo {author} {\bibfnamefont {N.}~\bibnamefont {Weiden}}, \ and\ \bibinfo
  {author} {\bibfnamefont {A.}~\bibnamefont {Weiss}},\ }\bibfield  {title}
  {\enquote {\bibinfo {title} {{Crystal Structure and Single Crystal 35 Cl NQR
  of 1,2-Dichloro-3-Nitrobenzene, Cl (1) Cl (2) (NO 2 ) (3) C 6 H 3}},}\ }\href
  {\doibase 10.1515/zna-1986-1-220} {\bibfield  {journal} {\bibinfo  {journal}
  {Zeitschrift f{\"{u}}r Naturforschung A}\ }\textbf {\bibinfo {volume} {41}},\
  \bibinfo {pages} {134} (\bibinfo {year} {1986})}\BibitemShut {NoStop}%
\bibitem [{\citenamefont {Horiuchi}\ \emph {et~al.}(1990)\citenamefont
  {Horiuchi}, \citenamefont {Shimizu}, \citenamefont {Iwafune}, \citenamefont
  {Asaji},\ and\ \citenamefont {Nakamura}}]{Hor1990}%
  \BibitemOpen
  \bibfield  {author} {\bibinfo {author} {\bibfnamefont {K.}~\bibnamefont
  {Horiuchi}}, \bibinfo {author} {\bibfnamefont {T.}~\bibnamefont {Shimizu}},
  \bibinfo {author} {\bibfnamefont {H.}~\bibnamefont {Iwafune}}, \bibinfo
  {author} {\bibfnamefont {T.}~\bibnamefont {Asaji}}, \ and\ \bibinfo {author}
  {\bibfnamefont {D.}~\bibnamefont {Nakamura}},\ }\bibfield  {title} {\enquote
  {\bibinfo {title} {{Temperature Dependences of NQR Frequencies and Nuclear
  Quadrupole Relaxation Times of Chlorine in 2, 6-Lutidinium
  Hexachlorotellurate(IV) as Studied by Pulsed NQR Techniques}},}\ }\href
  {\doibase 10.1515/zna-1990-3-448} {\bibfield  {journal} {\bibinfo  {journal}
  {Zeitschrift fur Naturforschung - Section A Journal of Physical Sciences}\
  }\textbf {\bibinfo {volume} {45}},\ \bibinfo {pages} {485} (\bibinfo {year}
  {1990})}\BibitemShut {NoStop}%
\bibitem [{\citenamefont {Huebner}\ \emph {et~al.}(1999)\citenamefont
  {Huebner}, \citenamefont {Leib},\ and\ \citenamefont {Eska}}]{Hue1999}%
  \BibitemOpen
  \bibfield  {author} {\bibinfo {author} {\bibfnamefont {M.}~\bibnamefont
  {Huebner}}, \bibinfo {author} {\bibfnamefont {J.}~\bibnamefont {Leib}}, \
  and\ \bibinfo {author} {\bibfnamefont {G.}~\bibnamefont {Eska}},\ }\bibfield
  {title} {\enquote {\bibinfo {title} {{NQR on Gallium Single Crystals for
  Absolute Thermometry at Very Low Temperatures}},}\ }\href {\doibase
  10.1023/a:1021810105778} {\bibfield  {journal} {\bibinfo  {journal} {Journal
  of Low Temperature Physics}\ }\textbf {\bibinfo {volume} {114}},\ \bibinfo
  {pages} {203} (\bibinfo {year} {1999})}\BibitemShut {NoStop}%
\bibitem [{\citenamefont {Lang}\ \emph {et~al.}(2010)\citenamefont {Lang},
  \citenamefont {Grafe}, \citenamefont {Paar}, \citenamefont {Hammerath},
  \citenamefont {Manthey}, \citenamefont {Behr}, \citenamefont {Werner},\ and\
  \citenamefont {B{\"{u}}chner}}]{Lan2010}%
  \BibitemOpen
  \bibfield  {author} {\bibinfo {author} {\bibfnamefont {G.}~\bibnamefont
  {Lang}}, \bibinfo {author} {\bibfnamefont {H.~J.}\ \bibnamefont {Grafe}},
  \bibinfo {author} {\bibfnamefont {D.}~\bibnamefont {Paar}}, \bibinfo {author}
  {\bibfnamefont {F.}~\bibnamefont {Hammerath}}, \bibinfo {author}
  {\bibfnamefont {K.}~\bibnamefont {Manthey}}, \bibinfo {author} {\bibfnamefont
  {G.}~\bibnamefont {Behr}}, \bibinfo {author} {\bibfnamefont {J.}~\bibnamefont
  {Werner}}, \ and\ \bibinfo {author} {\bibfnamefont {B.}~\bibnamefont
  {B{\"{u}}chner}},\ }\bibfield  {title} {\enquote {\bibinfo {title}
  {{Nanoscale electronic order in iron pnictides}},}\ }\href {\doibase
  10.1103/PhysRevLett.104.097001} {\bibfield  {journal} {\bibinfo  {journal}
  {Physical Review Letters}\ }\textbf {\bibinfo {volume} {104}},\ \bibinfo
  {pages} {3} (\bibinfo {year} {2010})}\BibitemShut {NoStop}%
\bibitem [{\citenamefont {Malone}\ \emph {et~al.}(2016)\citenamefont {Malone},
  \citenamefont {Barrall}, \citenamefont {Espy}, \citenamefont {Monti},
  \citenamefont {Alexson},\ and\ \citenamefont {Okamitsu}}]{Mal2016}%
  \BibitemOpen
  \bibfield  {author} {\bibinfo {author} {\bibfnamefont {M.~W.}\ \bibnamefont
  {Malone}}, \bibinfo {author} {\bibfnamefont {G.~A.}\ \bibnamefont {Barrall}},
  \bibinfo {author} {\bibfnamefont {M.~A.}\ \bibnamefont {Espy}}, \bibinfo
  {author} {\bibfnamefont {M.~C.}\ \bibnamefont {Monti}}, \bibinfo {author}
  {\bibfnamefont {D.~A.}\ \bibnamefont {Alexson}}, \ and\ \bibinfo {author}
  {\bibfnamefont {J.~K.}\ \bibnamefont {Okamitsu}},\ }\bibfield  {title}
  {\enquote {\bibinfo {title} {{Polarization enhanced Nuclear Quadrupole
  Resonance with an atomic magnetometer}},}\ }in\ \href {\doibase
  10.1117/12.2224070} {\emph {\bibinfo {booktitle} {Detection and Sensing of
  Mines, Explosive Objects, and Obscured Targets XXI}}},\ Vol.\ \bibinfo
  {volume} {9823},\ \bibinfo {editor} {edited by\ \bibinfo {editor}
  {\bibfnamefont {S.~S.}\ \bibnamefont {Bishop}}\ and\ \bibinfo {editor}
  {\bibfnamefont {J.~C.}\ \bibnamefont {Isaacs}}}\ (\bibinfo {year} {2016})\
  p.\ \bibinfo {pages} {98230Z}\BibitemShut {NoStop}%
\bibitem [{\citenamefont {Lovchinsky}\ \emph {et~al.}(2017)\citenamefont
  {Lovchinsky}, \citenamefont {Sanchez-Yamagishi}, \citenamefont {Urbach},
  \citenamefont {Choi}, \citenamefont {Fang}, \citenamefont {Andersen},
  \citenamefont {Watanabe}, \citenamefont {Taniguchi}, \citenamefont
  {Bylinskii}, \citenamefont {Kaxiras}, \citenamefont {Kim}, \citenamefont
  {Park},\ and\ \citenamefont {Lukin}}]{Lov2017}%
  \BibitemOpen
  \bibfield  {author} {\bibinfo {author} {\bibfnamefont {I.}~\bibnamefont
  {Lovchinsky}}, \bibinfo {author} {\bibfnamefont {J.~D.}\ \bibnamefont
  {Sanchez-Yamagishi}}, \bibinfo {author} {\bibfnamefont {E.~K.}\ \bibnamefont
  {Urbach}}, \bibinfo {author} {\bibfnamefont {S.}~\bibnamefont {Choi}},
  \bibinfo {author} {\bibfnamefont {S.}~\bibnamefont {Fang}}, \bibinfo {author}
  {\bibfnamefont {T.~I.}\ \bibnamefont {Andersen}}, \bibinfo {author}
  {\bibfnamefont {K.}~\bibnamefont {Watanabe}}, \bibinfo {author}
  {\bibfnamefont {T.}~\bibnamefont {Taniguchi}}, \bibinfo {author}
  {\bibfnamefont {A.}~\bibnamefont {Bylinskii}}, \bibinfo {author}
  {\bibfnamefont {E.}~\bibnamefont {Kaxiras}}, \bibinfo {author} {\bibfnamefont
  {P.}~\bibnamefont {Kim}}, \bibinfo {author} {\bibfnamefont {H.}~\bibnamefont
  {Park}}, \ and\ \bibinfo {author} {\bibfnamefont {M.~D.}\ \bibnamefont
  {Lukin}},\ }\bibfield  {title} {\enquote {\bibinfo {title} {{Magnetic
  resonance spectroscopy of an atomically thin material using a single-spin
  qubit}},}\ }\href {\doibase 10.1126/science.aal2538} {\bibfield  {journal}
  {\bibinfo  {journal} {Science}\ }\textbf {\bibinfo {volume} {355}},\ \bibinfo
  {pages} {503} (\bibinfo {year} {2017})}\BibitemShut {NoStop}%
\bibitem [{\citenamefont {Henshaw}\ \emph {et~al.}(2022)\citenamefont
  {Henshaw}, \citenamefont {Kehayias}, \citenamefont {{Saleh Ziabari}},
  \citenamefont {Titze}, \citenamefont {Morissette}, \citenamefont {Watanabe},
  \citenamefont {Taniguchi}, \citenamefont {Li}, \citenamefont {Acosta},
  \citenamefont {Bielejec}, \citenamefont {Lilly},\ and\ \citenamefont
  {Mounce}}]{Hen2022}%
  \BibitemOpen
  \bibfield  {author} {\bibinfo {author} {\bibfnamefont {J.}~\bibnamefont
  {Henshaw}}, \bibinfo {author} {\bibfnamefont {P.}~\bibnamefont {Kehayias}},
  \bibinfo {author} {\bibfnamefont {M.}~\bibnamefont {{Saleh Ziabari}}},
  \bibinfo {author} {\bibfnamefont {M.}~\bibnamefont {Titze}}, \bibinfo
  {author} {\bibfnamefont {E.}~\bibnamefont {Morissette}}, \bibinfo {author}
  {\bibfnamefont {K.}~\bibnamefont {Watanabe}}, \bibinfo {author}
  {\bibfnamefont {T.}~\bibnamefont {Taniguchi}}, \bibinfo {author}
  {\bibfnamefont {J.~I.~A.}\ \bibnamefont {Li}}, \bibinfo {author}
  {\bibfnamefont {V.~M.}\ \bibnamefont {Acosta}}, \bibinfo {author}
  {\bibfnamefont {E.~S.}\ \bibnamefont {Bielejec}}, \bibinfo {author}
  {\bibfnamefont {M.~P.}\ \bibnamefont {Lilly}}, \ and\ \bibinfo {author}
  {\bibfnamefont {A.~M.}\ \bibnamefont {Mounce}},\ }\bibfield  {title}
  {\enquote {\bibinfo {title} {{Nanoscale solid-state nuclear quadrupole
  resonance spectroscopy using depth-optimized nitrogen-vacancy ensembles in
  diamond}},}\ }\href {\doibase 10.1063/5.0083774} {\bibfield  {journal}
  {\bibinfo  {journal} {Applied Physics Letters}\ }\textbf {\bibinfo {volume}
  {120}},\ \bibinfo {pages} {174002} (\bibinfo {year} {2022})}\BibitemShut
  {NoStop}%
\bibitem [{\citenamefont {Shin}\ \emph {et~al.}(2014)\citenamefont {Shin},
  \citenamefont {Butler}, \citenamefont {Wang}, \citenamefont {Avalos},
  \citenamefont {Seltzer}, \citenamefont {Liu}, \citenamefont {Pines},\ and\
  \citenamefont {Bajaj}}]{Shi2014}%
  \BibitemOpen
  \bibfield  {author} {\bibinfo {author} {\bibfnamefont {C.~S.}\ \bibnamefont
  {Shin}}, \bibinfo {author} {\bibfnamefont {M.~C.}\ \bibnamefont {Butler}},
  \bibinfo {author} {\bibfnamefont {H.~J.}\ \bibnamefont {Wang}}, \bibinfo
  {author} {\bibfnamefont {C.~E.}\ \bibnamefont {Avalos}}, \bibinfo {author}
  {\bibfnamefont {S.~J.}\ \bibnamefont {Seltzer}}, \bibinfo {author}
  {\bibfnamefont {R.~B.}\ \bibnamefont {Liu}}, \bibinfo {author} {\bibfnamefont
  {A.}~\bibnamefont {Pines}}, \ and\ \bibinfo {author} {\bibfnamefont {V.~S.}\
  \bibnamefont {Bajaj}},\ }\bibfield  {title} {\enquote {\bibinfo {title}
  {{Optically detected nuclear quadrupolar interaction of N 14 in
  nitrogen-vacancy centers in diamond}},}\ }\href {\doibase
  10.1103/PhysRevB.89.205202} {\bibfield  {journal} {\bibinfo  {journal}
  {Physical Review B - Condensed Matter and Materials Physics}\ }\textbf
  {\bibinfo {volume} {89}},\ \bibinfo {pages} {1} (\bibinfo {year}
  {2014})}\BibitemShut {NoStop}%
\bibitem [{\citenamefont {Oja}\ \emph {et~al.}(1967)\citenamefont {Oja},
  \citenamefont {Marino},\ and\ \citenamefont {Bray}}]{Oja1967}%
  \BibitemOpen
  \bibfield  {author} {\bibinfo {author} {\bibfnamefont {T.}~\bibnamefont
  {Oja}}, \bibinfo {author} {\bibfnamefont {R.~A.}\ \bibnamefont {Marino}}, \
  and\ \bibinfo {author} {\bibfnamefont {P.~J.}\ \bibnamefont {Bray}},\
  }\bibfield  {title} {\enquote {\bibinfo {title} {{14N nuclear quadrupole
  resonance in the ferroelectric phase of sodium nitrite}},}\ }\href {\doibase
  10.1016/0375-9601(67)90530-0} {\bibfield  {journal} {\bibinfo  {journal}
  {Physics Letters A}\ }\textbf {\bibinfo {volume} {26}},\ \bibinfo {pages}
  {11} (\bibinfo {year} {1967})}\BibitemShut {NoStop}%
\bibitem [{\citenamefont {Petersen}\ and\ \citenamefont
  {Bray}(1976)}]{Pet1976}%
  \BibitemOpen
  \bibfield  {author} {\bibinfo {author} {\bibfnamefont {G.}~\bibnamefont
  {Petersen}}\ and\ \bibinfo {author} {\bibfnamefont {P.~J.}\ \bibnamefont
  {Bray}},\ }\bibfield  {title} {\enquote {\bibinfo {title} {{14N nuclear
  quadrupole resonance and relaxation measurements of sodium nitrite}},}\
  }\href {\doibase 10.1063/1.432241} {\bibfield  {journal} {\bibinfo  {journal}
  {The Journal of Chemical Physics}\ }\textbf {\bibinfo {volume} {64}},\
  \bibinfo {pages} {522} (\bibinfo {year} {1976})}\BibitemShut {NoStop}%
\bibitem [{\citenamefont {{National Magnetics Group-Inc.}}(2019)}]{MN60}%
  \BibitemOpen
  \bibfield  {author} {\bibinfo {author} {\bibnamefont {{National Magnetics
  Group-Inc.}}},\ }\href
  {https://www.magneticsgroup.com/wp-content/uploads/2019/09/MN60-ISO-WEB-DATA.pdf}
  {\enquote {\bibinfo {title} {{High Permeability Mn-Zn Ferrite}},}\ }
  (\bibinfo {year} {2019})\BibitemShut {NoStop}%
\bibitem [{\citenamefont {Bolton}(2016)}]{Bol2016}%
  \BibitemOpen
  \bibfield  {author} {\bibinfo {author} {\bibfnamefont {T.}~\bibnamefont
  {Bolton}},\ }\emph {\bibinfo {title} {{Optimal design of electrically-small
  loop antenna including surrounding medium effects}}},\ \href
  {http://hdl.handle.net/1853/54836} {Ph.D. thesis} (\bibinfo {year}
  {2016})\BibitemShut {NoStop}%
\bibitem [{\citenamefont {Fisher}\ \emph {et~al.}(1999)\citenamefont {Fisher},
  \citenamefont {MacNamara}, \citenamefont {Santini},\ and\ \citenamefont
  {Raftery}}]{Fis1999}%
  \BibitemOpen
  \bibfield  {author} {\bibinfo {author} {\bibfnamefont {G.}~\bibnamefont
  {Fisher}}, \bibinfo {author} {\bibfnamefont {E.}~\bibnamefont {MacNamara}},
  \bibinfo {author} {\bibfnamefont {R.~E.}\ \bibnamefont {Santini}}, \ and\
  \bibinfo {author} {\bibfnamefont {D.}~\bibnamefont {Raftery}},\ }\bibfield
  {title} {\enquote {\bibinfo {title} {{A versatile computer-controlled pulsed
  nuclear quadrupole resonance spectrometer}},}\ }\href {\doibase
  10.1063/1.1150131} {\bibfield  {journal} {\bibinfo  {journal} {Review of
  Scientific Instruments}\ }\textbf {\bibinfo {volume} {70}},\ \bibinfo {pages}
  {4676} (\bibinfo {year} {1999})}\BibitemShut {NoStop}%
\bibitem [{\citenamefont {Hiblot}\ \emph {et~al.}(2008)\citenamefont {Hiblot},
  \citenamefont {Cordier}, \citenamefont {Ferrari}, \citenamefont {Retournard},
  \citenamefont {Grandclaude}, \citenamefont {Bedet}, \citenamefont {Leclerc},\
  and\ \citenamefont {Canet}}]{Hib2008}%
  \BibitemOpen
  \bibfield  {author} {\bibinfo {author} {\bibfnamefont {N.}~\bibnamefont
  {Hiblot}}, \bibinfo {author} {\bibfnamefont {B.}~\bibnamefont {Cordier}},
  \bibinfo {author} {\bibfnamefont {M.}~\bibnamefont {Ferrari}}, \bibinfo
  {author} {\bibfnamefont {A.}~\bibnamefont {Retournard}}, \bibinfo {author}
  {\bibfnamefont {D.}~\bibnamefont {Grandclaude}}, \bibinfo {author}
  {\bibfnamefont {J.}~\bibnamefont {Bedet}}, \bibinfo {author} {\bibfnamefont
  {S.}~\bibnamefont {Leclerc}}, \ and\ \bibinfo {author} {\bibfnamefont
  {D.}~\bibnamefont {Canet}},\ }\bibfield  {title} {\enquote {\bibinfo {title}
  {{A fully homemade 14N quadrupole resonance spectrometer}},}\ }\href
  {\doibase 10.1016/j.crci.2007.08.011} {\bibfield  {journal} {\bibinfo
  {journal} {Comptes Rendus Chimie}\ }\textbf {\bibinfo {volume} {11}},\
  \bibinfo {pages} {568} (\bibinfo {year} {2008})}\BibitemShut {NoStop}%
\bibitem [{\citenamefont {Vega}(1974)}]{Veg1974}%
  \BibitemOpen
  \bibfield  {author} {\bibinfo {author} {\bibfnamefont {S.}~\bibnamefont
  {Vega}},\ }\bibfield  {title} {\enquote {\bibinfo {title} {{Theory of T1
  relaxation measurements in pure nuclear quadrupole resonance for spins I =
  1}},}\ }\href {\doibase 10.1063/1.1681979} {\bibfield  {journal} {\bibinfo
  {journal} {The Journal of Chemical Physics}\ }\textbf {\bibinfo {volume}
  {61}},\ \bibinfo {pages} {1093} (\bibinfo {year} {1974})}\BibitemShut
  {NoStop}%
\bibitem [{\citenamefont {Suits}(2006)}]{Sui2006}%
  \BibitemOpen
  \bibfield  {author} {\bibinfo {author} {\bibfnamefont {B.~H.}\ \bibnamefont
  {Suits}},\ }\bibfield  {title} {\enquote {\bibinfo {title} {{NUCLEAR
  QUADRUPOLE RESONANCE SPECTROSCOPY}},}\ }in\ \href {\doibase
  10.1007/0-387-37590-2_2} {\emph {\bibinfo {booktitle} {Handbook of Applied
  Solid State Spectroscopy}}},\ \bibinfo {editor} {edited by\ \bibinfo {editor}
  {\bibfnamefont {D.~R.}\ \bibnamefont {Vij}}}\ (\bibinfo  {publisher}
  {Springer US},\ \bibinfo {address} {Boston, MA},\ \bibinfo {year} {2006})\
  pp.\ \bibinfo {pages} {65--96}\BibitemShut {NoStop}%
\bibitem [{\citenamefont {Bloom}\ \emph {et~al.}(1955)\citenamefont {Bloom},
  \citenamefont {Hahn},\ and\ \citenamefont {Herzog}}]{Blo1955}%
  \BibitemOpen
  \bibfield  {author} {\bibinfo {author} {\bibfnamefont {M.}~\bibnamefont
  {Bloom}}, \bibinfo {author} {\bibfnamefont {E.~L.}\ \bibnamefont {Hahn}}, \
  and\ \bibinfo {author} {\bibfnamefont {B.}~\bibnamefont {Herzog}},\
  }\bibfield  {title} {\enquote {\bibinfo {title} {{Free magnetic induction in
  nuclear quadrupole resonance}},}\ }\href {\doibase 10.1103/PhysRev.97.1699}
  {\bibfield  {journal} {\bibinfo  {journal} {Physical Review}\ }\textbf
  {\bibinfo {volume} {97}},\ \bibinfo {pages} {1699} (\bibinfo {year}
  {1955})}\BibitemShut {NoStop}%
\bibitem [{\citenamefont {Qiu}\ \emph {et~al.}(2007)\citenamefont {Qiu},
  \citenamefont {Zhang}, \citenamefont {Krause}, \citenamefont {Braginski},\
  and\ \citenamefont {Usoskin}}]{Qiu2007}%
  \BibitemOpen
  \bibfield  {author} {\bibinfo {author} {\bibfnamefont {L.}~\bibnamefont
  {Qiu}}, \bibinfo {author} {\bibfnamefont {Y.}~\bibnamefont {Zhang}}, \bibinfo
  {author} {\bibfnamefont {H.-J.}\ \bibnamefont {Krause}}, \bibinfo {author}
  {\bibfnamefont {A.~I.}\ \bibnamefont {Braginski}}, \ and\ \bibinfo {author}
  {\bibfnamefont {A.}~\bibnamefont {Usoskin}},\ }\bibfield  {title} {\enquote
  {\bibinfo {title} {{High-temperature superconducting quantum interference
  device with cooled LC resonant circuit for measuring alternating magnetic
  fields with improved signal-to-noise ratio}},}\ }\href {\doibase
  10.1063/1.2735561} {\bibfield  {journal} {\bibinfo  {journal} {Review of
  Scientific Instruments}\ }\textbf {\bibinfo {volume} {78}},\ \bibinfo {pages}
  {054701} (\bibinfo {year} {2007})}\BibitemShut {NoStop}%
\bibitem [{\citenamefont {Greer}\ \emph {et~al.}(2021)\citenamefont {Greer},
  \citenamefont {Ariando}, \citenamefont {Hurlimann}, \citenamefont {Song},\
  and\ \citenamefont {Mandal}}]{Gre2021}%
  \BibitemOpen
  \bibfield  {author} {\bibinfo {author} {\bibfnamefont {M.}~\bibnamefont
  {Greer}}, \bibinfo {author} {\bibfnamefont {D.}~\bibnamefont {Ariando}},
  \bibinfo {author} {\bibfnamefont {M.}~\bibnamefont {Hurlimann}}, \bibinfo
  {author} {\bibfnamefont {Y.~Q.}\ \bibnamefont {Song}}, \ and\ \bibinfo
  {author} {\bibfnamefont {S.}~\bibnamefont {Mandal}},\ }\bibfield  {title}
  {\enquote {\bibinfo {title} {{Analytical models of probe dynamics effects on
  NMR measurements}},}\ }\href {\doibase 10.1016/j.jmr.2021.106975} {\bibfield
  {journal} {\bibinfo  {journal} {Journal of Magnetic Resonance}\ }\textbf
  {\bibinfo {volume} {327}},\ \bibinfo {pages} {106975} (\bibinfo {year}
  {2021})}\BibitemShut {NoStop}%
\bibitem [{\citenamefont {Verber}\ \emph {et~al.}(1962)\citenamefont {Verber},
  \citenamefont {Mahon},\ and\ \citenamefont {Tanttila}}]{Ver1962}%
  \BibitemOpen
  \bibfield  {author} {\bibinfo {author} {\bibfnamefont {C.~M.}\ \bibnamefont
  {Verber}}, \bibinfo {author} {\bibfnamefont {H.~P.}\ \bibnamefont {Mahon}}, \
  and\ \bibinfo {author} {\bibfnamefont {W.~H.}\ \bibnamefont {Tanttila}},\
  }\bibfield  {title} {\enquote {\bibinfo {title} {{Nuclear Resonance of
  Aluminum in Synthetic Ruby}},}\ }\href {\doibase
  https://doi.org/10.1103/PhysRev.125.1149} {\bibfield  {journal} {\bibinfo
  {journal} {Physical Review}\ }\textbf {\bibinfo {volume} {125}},\ \bibinfo
  {pages} {1149} (\bibinfo {year} {1962})}\BibitemShut {NoStop}%
\bibitem [{\citenamefont {Marino}\ and\ \citenamefont
  {Klainer}(1977)}]{Mar1977}%
  \BibitemOpen
  \bibfield  {author} {\bibinfo {author} {\bibfnamefont {R.~A.}\ \bibnamefont
  {Marino}}\ and\ \bibinfo {author} {\bibfnamefont {S.~M.}\ \bibnamefont
  {Klainer}},\ }\bibfield  {title} {\enquote {\bibinfo {title} {{Multiple spin
  echoes in pure quadrupole resonance}},}\ }\href {\doibase 10.1063/1.435286}
  {\bibfield  {journal} {\bibinfo  {journal} {The Journal of Chemical Physics}\
  }\textbf {\bibinfo {volume} {67}},\ \bibinfo {pages} {3388} (\bibinfo {year}
  {1977})}\BibitemShut {NoStop}%
\bibitem [{\citenamefont {Cantor}\ and\ \citenamefont {Waugh}(1980)}]{Can1980}%
  \BibitemOpen
  \bibfield  {author} {\bibinfo {author} {\bibfnamefont {R.~S.}\ \bibnamefont
  {Cantor}}\ and\ \bibinfo {author} {\bibfnamefont {J.~S.}\ \bibnamefont
  {Waugh}},\ }\bibfield  {title} {\enquote {\bibinfo {title} {{Pulsed spin
  locking in pure nuclear quadrupole resonance}},}\ }\href {\doibase
  10.1063/1.440277} {\bibfield  {journal} {\bibinfo  {journal} {The Journal of
  Chemical Physics}\ }\textbf {\bibinfo {volume} {73}},\ \bibinfo {pages}
  {1054} (\bibinfo {year} {1980})}\BibitemShut {NoStop}%
\bibitem [{\citenamefont {Maricq}(1986)}]{Mar1986}%
  \BibitemOpen
  \bibfield  {author} {\bibinfo {author} {\bibfnamefont {M.~M.}\ \bibnamefont
  {Maricq}},\ }\bibfield  {title} {\enquote {\bibinfo {title} {{Quasistationary
  state and its decay to equilibrium in the pulsed spin locking of a nuclear
  quadrupole resonance}},}\ }\href {\doibase 10.1103/PhysRevB.33.4501}
  {\bibfield  {journal} {\bibinfo  {journal} {Physical Review B}\ }\textbf
  {\bibinfo {volume} {33}},\ \bibinfo {pages} {4501} (\bibinfo {year}
  {1986})}\BibitemShut {NoStop}%
\bibitem [{\citenamefont {Malone}\ \emph {et~al.}(2011)\citenamefont {Malone},
  \citenamefont {McGillvray},\ and\ \citenamefont {Sauer}}]{Mal2011}%
  \BibitemOpen
  \bibfield  {author} {\bibinfo {author} {\bibfnamefont {M.~W.}\ \bibnamefont
  {Malone}}, \bibinfo {author} {\bibfnamefont {M.}~\bibnamefont {McGillvray}},
  \ and\ \bibinfo {author} {\bibfnamefont {K.~L.}\ \bibnamefont {Sauer}},\
  }\bibfield  {title} {\enquote {\bibinfo {title} {{Revealing dipolar coupling
  with NQR off-resonant pulsed spin locking}},}\ }\href {\doibase
  10.1103/PhysRevB.84.214430} {\bibfield  {journal} {\bibinfo  {journal}
  {Physical Review B - Condensed Matter and Materials Physics}\ }\textbf
  {\bibinfo {volume} {84}},\ \bibinfo {pages} {1} (\bibinfo {year}
  {2011})}\BibitemShut {NoStop}%
\bibitem [{\citenamefont {Clevenson}\ \emph {et~al.}(2015)\citenamefont
  {Clevenson}, \citenamefont {Trusheim}, \citenamefont {Teale}, \citenamefont
  {Schr{\"{o}}der}, \citenamefont {Braje},\ and\ \citenamefont
  {Englund}}]{Cle2015}%
  \BibitemOpen
  \bibfield  {author} {\bibinfo {author} {\bibfnamefont {H.}~\bibnamefont
  {Clevenson}}, \bibinfo {author} {\bibfnamefont {M.~E.}\ \bibnamefont
  {Trusheim}}, \bibinfo {author} {\bibfnamefont {C.}~\bibnamefont {Teale}},
  \bibinfo {author} {\bibfnamefont {T.}~\bibnamefont {Schr{\"{o}}der}},
  \bibinfo {author} {\bibfnamefont {D.}~\bibnamefont {Braje}}, \ and\ \bibinfo
  {author} {\bibfnamefont {D.}~\bibnamefont {Englund}},\ }\bibfield  {title}
  {\enquote {\bibinfo {title} {{Broadband magnetometry and temperature sensing
  with a light-trapping diamond waveguide}},}\ }\href {\doibase
  10.1038/nphys3291} {\bibfield  {journal} {\bibinfo  {journal} {Nature
  Physics}\ }\textbf {\bibinfo {volume} {11}},\ \bibinfo {pages} {393}
  (\bibinfo {year} {2015})}\BibitemShut {NoStop}%
\bibitem [{\citenamefont {Xu}\ \emph {et~al.}(2019)\citenamefont {Xu},
  \citenamefont {Yuan}, \citenamefont {Zhang}, \citenamefont {Zhang},
  \citenamefont {Bian}, \citenamefont {Fan}, \citenamefont {Li}, \citenamefont
  {Zhang}, \citenamefont {Zhai},\ and\ \citenamefont {Fang}}]{Xu2019}%
  \BibitemOpen
  \bibfield  {author} {\bibinfo {author} {\bibfnamefont {L.}~\bibnamefont
  {Xu}}, \bibinfo {author} {\bibfnamefont {H.}~\bibnamefont {Yuan}}, \bibinfo
  {author} {\bibfnamefont {N.}~\bibnamefont {Zhang}}, \bibinfo {author}
  {\bibfnamefont {J.}~\bibnamefont {Zhang}}, \bibinfo {author} {\bibfnamefont
  {G.}~\bibnamefont {Bian}}, \bibinfo {author} {\bibfnamefont {P.}~\bibnamefont
  {Fan}}, \bibinfo {author} {\bibfnamefont {M.}~\bibnamefont {Li}}, \bibinfo
  {author} {\bibfnamefont {C.}~\bibnamefont {Zhang}}, \bibinfo {author}
  {\bibfnamefont {Y.}~\bibnamefont {Zhai}}, \ and\ \bibinfo {author}
  {\bibfnamefont {J.}~\bibnamefont {Fang}},\ }\bibfield  {title} {\enquote
  {\bibinfo {title} {{High-efficiency fluorescence collection for NV- center
  ensembles in diamond}},}\ }\href {\doibase 10.1364/OE.27.010787} {\bibfield
  {journal} {\bibinfo  {journal} {Opt. Express}\ }\textbf {\bibinfo {volume}
  {27}},\ \bibinfo {pages} {10787} (\bibinfo {year} {2019})}\BibitemShut
  {NoStop}%
\bibitem [{\citenamefont {{Le Sage}}\ \emph {et~al.}(2012)\citenamefont {{Le
  Sage}}, \citenamefont {Pham}, \citenamefont {Bar-Gill}, \citenamefont
  {Belthangady}, \citenamefont {Lukin}, \citenamefont {Yacoby},\ and\
  \citenamefont {Walsworth}}]{Les2012}%
  \BibitemOpen
  \bibfield  {author} {\bibinfo {author} {\bibfnamefont {D.}~\bibnamefont {{Le
  Sage}}}, \bibinfo {author} {\bibfnamefont {L.~M.}\ \bibnamefont {Pham}},
  \bibinfo {author} {\bibfnamefont {N.}~\bibnamefont {Bar-Gill}}, \bibinfo
  {author} {\bibfnamefont {C.}~\bibnamefont {Belthangady}}, \bibinfo {author}
  {\bibfnamefont {M.~D.}\ \bibnamefont {Lukin}}, \bibinfo {author}
  {\bibfnamefont {A.}~\bibnamefont {Yacoby}}, \ and\ \bibinfo {author}
  {\bibfnamefont {R.~L.}\ \bibnamefont {Walsworth}},\ }\bibfield  {title}
  {\enquote {\bibinfo {title} {{Efficient photon detection from color centers
  in a diamond optical waveguide}},}\ }\href {\doibase
  10.1103/PhysRevB.85.121202} {\bibfield  {journal} {\bibinfo  {journal} {Phys.
  Rev. B}\ }\textbf {\bibinfo {volume} {85}},\ \bibinfo {pages} {121202}
  (\bibinfo {year} {2012})}\BibitemShut {NoStop}%
\bibitem [{\citenamefont {Zhou}\ \emph {et~al.}(2020)\citenamefont {Zhou},
  \citenamefont {Choi}, \citenamefont {Choi}, \citenamefont {Landig},
  \citenamefont {Douglas}, \citenamefont {Isoya}, \citenamefont {Jelezko},
  \citenamefont {Onoda}, \citenamefont {Sumiya}, \citenamefont {Cappellaro},
  \citenamefont {Knowles}, \citenamefont {Park},\ and\ \citenamefont
  {Lukin}}]{Zho2020}%
  \BibitemOpen
  \bibfield  {author} {\bibinfo {author} {\bibfnamefont {H.}~\bibnamefont
  {Zhou}}, \bibinfo {author} {\bibfnamefont {J.}~\bibnamefont {Choi}}, \bibinfo
  {author} {\bibfnamefont {S.}~\bibnamefont {Choi}}, \bibinfo {author}
  {\bibfnamefont {R.}~\bibnamefont {Landig}}, \bibinfo {author} {\bibfnamefont
  {A.~M.}\ \bibnamefont {Douglas}}, \bibinfo {author} {\bibfnamefont
  {J.}~\bibnamefont {Isoya}}, \bibinfo {author} {\bibfnamefont
  {F.}~\bibnamefont {Jelezko}}, \bibinfo {author} {\bibfnamefont
  {S.}~\bibnamefont {Onoda}}, \bibinfo {author} {\bibfnamefont
  {H.}~\bibnamefont {Sumiya}}, \bibinfo {author} {\bibfnamefont
  {P.}~\bibnamefont {Cappellaro}}, \bibinfo {author} {\bibfnamefont {H.~S.}\
  \bibnamefont {Knowles}}, \bibinfo {author} {\bibfnamefont {H.}~\bibnamefont
  {Park}}, \ and\ \bibinfo {author} {\bibfnamefont {M.~D.}\ \bibnamefont
  {Lukin}},\ }\bibfield  {title} {\enquote {\bibinfo {title} {{Quantum
  Metrology with Strongly Interacting Spin Systems}},}\ }\href {\doibase
  10.1103/PhysRevX.10.031003} {\bibfield  {journal} {\bibinfo  {journal} {Phys.
  Rev. X}\ }\textbf {\bibinfo {volume} {10}},\ \bibinfo {pages} {31003}
  (\bibinfo {year} {2020})}\BibitemShut {NoStop}%
\bibitem [{\citenamefont {Arunkumar}\ \emph {et~al.}(2022)\citenamefont
  {Arunkumar}, \citenamefont {Olsson}, \citenamefont {Oon}, \citenamefont
  {Hart}, \citenamefont {Bucher}, \citenamefont {Glenn}, \citenamefont {Lukin},
  \citenamefont {Park}, \citenamefont {Ham},\ and\ \citenamefont
  {Walsworth}}]{Aru2022}%
  \BibitemOpen
  \bibfield  {author} {\bibinfo {author} {\bibfnamefont {N.}~\bibnamefont
  {Arunkumar}}, \bibinfo {author} {\bibfnamefont {K.~S.}\ \bibnamefont
  {Olsson}}, \bibinfo {author} {\bibfnamefont {J.~T.}\ \bibnamefont {Oon}},
  \bibinfo {author} {\bibfnamefont {C.}~\bibnamefont {Hart}}, \bibinfo {author}
  {\bibfnamefont {D.~B.}\ \bibnamefont {Bucher}}, \bibinfo {author}
  {\bibfnamefont {D.}~\bibnamefont {Glenn}}, \bibinfo {author} {\bibfnamefont
  {M.~D.}\ \bibnamefont {Lukin}}, \bibinfo {author} {\bibfnamefont
  {H.}~\bibnamefont {Park}}, \bibinfo {author} {\bibfnamefont {D.}~\bibnamefont
  {Ham}}, \ and\ \bibinfo {author} {\bibfnamefont {R.~L.}\ \bibnamefont
  {Walsworth}},\ }\bibfield  {title} {\enquote {\bibinfo {title} {{Quantum
  Logic Enhanced Sensing in Solid-State Spin Ensembles}},}\ }\href
  {https://arxiv.org/abs/2203.12501} {\bibfield  {journal} {\bibinfo  {journal}
  {arXiv:2203.12501}\ } (\bibinfo {year} {2022})}\BibitemShut {NoStop}%
\bibitem [{\citenamefont {Lee}\ and\ \citenamefont {Romalis}(2008)}]{Lee2008}%
  \BibitemOpen
  \bibfield  {author} {\bibinfo {author} {\bibfnamefont {S.-K.}\ \bibnamefont
  {Lee}}\ and\ \bibinfo {author} {\bibfnamefont {M.~V.}\ \bibnamefont
  {Romalis}},\ }\bibfield  {title} {\enquote {\bibinfo {title} {{Calculation of
  magnetic field noise from high-permeability magnetic shields and conducting
  objects with simple geometry}},}\ }\href {\doibase 10.1063/1.2885711}
  {\bibfield  {journal} {\bibinfo  {journal} {Journal of Applied Physics}\
  }\textbf {\bibinfo {volume} {103}},\ \bibinfo {pages} {84904} (\bibinfo
  {year} {2008})}\BibitemShut {NoStop}%
\bibitem [{\citenamefont {Kim}\ and\ \citenamefont {Savukov}(2016)}]{Kim2016}%
  \BibitemOpen
  \bibfield  {author} {\bibinfo {author} {\bibfnamefont {Y.~J.}\ \bibnamefont
  {Kim}}\ and\ \bibinfo {author} {\bibfnamefont {I.}~\bibnamefont {Savukov}},\
  }\bibfield  {title} {\enquote {\bibinfo {title} {{Ultra-sensitive Magnetic
  Microscopy with an Optically Pumped Magnetometer}},}\ }\href {\doibase
  10.1038/srep24773} {\bibfield  {journal} {\bibinfo  {journal} {Scientific
  Reports}\ }\textbf {\bibinfo {volume} {6}},\ \bibinfo {pages} {24773}
  (\bibinfo {year} {2016})}\BibitemShut {NoStop}%
\bibitem [{\citenamefont {Hoult}\ and\ \citenamefont
  {Richards}(1976)}]{Hou1976}%
  \BibitemOpen
  \bibfield  {author} {\bibinfo {author} {\bibfnamefont {D.}~\bibnamefont
  {Hoult}}\ and\ \bibinfo {author} {\bibfnamefont {R.}~\bibnamefont
  {Richards}},\ }\bibfield  {title} {\enquote {\bibinfo {title} {{The
  signal-to-noise ratio of the nuclear magnetic resonance experiment}},}\
  }\href {\doibase 10.1016/0022-2364(76)90233-X} {\bibfield  {journal}
  {\bibinfo  {journal} {Journal of Magnetic Resonance (1969)}\ }\textbf
  {\bibinfo {volume} {24}},\ \bibinfo {pages} {71} (\bibinfo {year}
  {1976})}\BibitemShut {NoStop}%
\bibitem [{\citenamefont {Chatzidrosos}\ \emph {et~al.}(2019)\citenamefont
  {Chatzidrosos}, \citenamefont {Wickenbrock}, \citenamefont {Bougas},
  \citenamefont {Zheng}, \citenamefont {Tretiak}, \citenamefont {Yang},\ and\
  \citenamefont {Budker}}]{Cha2019}%
  \BibitemOpen
  \bibfield  {author} {\bibinfo {author} {\bibfnamefont {G.}~\bibnamefont
  {Chatzidrosos}}, \bibinfo {author} {\bibfnamefont {A.}~\bibnamefont
  {Wickenbrock}}, \bibinfo {author} {\bibfnamefont {L.}~\bibnamefont {Bougas}},
  \bibinfo {author} {\bibfnamefont {H.}~\bibnamefont {Zheng}}, \bibinfo
  {author} {\bibfnamefont {O.}~\bibnamefont {Tretiak}}, \bibinfo {author}
  {\bibfnamefont {Y.}~\bibnamefont {Yang}}, \ and\ \bibinfo {author}
  {\bibfnamefont {D.}~\bibnamefont {Budker}},\ }\bibfield  {title} {\enquote
  {\bibinfo {title} {{Eddy-Current Imaging with Nitrogen-Vacancy Centers in
  Diamond}},}\ }\href {\doibase 10.1103/PhysRevApplied.11.014060} {\bibfield
  {journal} {\bibinfo  {journal} {Phys. Rev. Appl.}\ }\textbf {\bibinfo
  {volume} {11}},\ \bibinfo {pages} {14060} (\bibinfo {year}
  {2019})}\BibitemShut {NoStop}%
\bibitem [{\citenamefont {Rushton}\ \emph {et~al.}(2022)\citenamefont
  {Rushton}, \citenamefont {Pyragius}, \citenamefont {Meraki}, \citenamefont
  {Elson},\ and\ \citenamefont {Jensen}}]{Rus2023}%
  \BibitemOpen
  \bibfield  {author} {\bibinfo {author} {\bibfnamefont {L.~M.}\ \bibnamefont
  {Rushton}}, \bibinfo {author} {\bibfnamefont {T.}~\bibnamefont {Pyragius}},
  \bibinfo {author} {\bibfnamefont {A.}~\bibnamefont {Meraki}}, \bibinfo
  {author} {\bibfnamefont {L.}~\bibnamefont {Elson}}, \ and\ \bibinfo {author}
  {\bibfnamefont {K.}~\bibnamefont {Jensen}},\ }\bibfield  {title} {\enquote
  {\bibinfo {title} {{Unshielded portable optically pumped magnetometer for the
  remote detection of conductive objects using eddy current measurements}},}\
  }\href {\doibase 10.1063/5.0102402} {\bibfield  {journal} {\bibinfo
  {journal} {Review of Scientific Instruments}\ }\textbf {\bibinfo {volume}
  {93}},\ \bibinfo {pages} {125103} (\bibinfo {year} {2022})}\BibitemShut
  {NoStop}%
\bibitem [{\citenamefont {Gerginov}\ \emph {et~al.}(2017)\citenamefont
  {Gerginov}, \citenamefont {da~Silva},\ and\ \citenamefont {Howe}}]{Ger2017}%
  \BibitemOpen
  \bibfield  {author} {\bibinfo {author} {\bibfnamefont {V.}~\bibnamefont
  {Gerginov}}, \bibinfo {author} {\bibfnamefont {F.~C.~S.}\ \bibnamefont
  {da~Silva}}, \ and\ \bibinfo {author} {\bibfnamefont {D.}~\bibnamefont
  {Howe}},\ }\bibfield  {title} {\enquote {\bibinfo {title} {{Prospects for
  magnetic field communications and location using quantum sensors}},}\ }\href
  {\doibase 10.1063/1.5003821} {\bibfield  {journal} {\bibinfo  {journal}
  {Review of Scientific Instruments}\ }\textbf {\bibinfo {volume} {88}},\
  \bibinfo {pages} {125005} (\bibinfo {year} {2017})}\BibitemShut {NoStop}%
\bibitem [{\citenamefont {Deans}\ \emph {et~al.}(2018)\citenamefont {Deans},
  \citenamefont {Marmugi},\ and\ \citenamefont {Renzoni}}]{Dea2018}%
  \BibitemOpen
  \bibfield  {author} {\bibinfo {author} {\bibfnamefont {C.}~\bibnamefont
  {Deans}}, \bibinfo {author} {\bibfnamefont {L.}~\bibnamefont {Marmugi}}, \
  and\ \bibinfo {author} {\bibfnamefont {F.}~\bibnamefont {Renzoni}},\
  }\bibfield  {title} {\enquote {\bibinfo {title} {{Active underwater detection
  with an array of atomic magnetometers}},}\ }\href {\doibase
  10.1364/AO.57.002346} {\bibfield  {journal} {\bibinfo  {journal} {Appl.
  Opt.}\ }\textbf {\bibinfo {volume} {57}},\ \bibinfo {pages} {2346} (\bibinfo
  {year} {2018})}\BibitemShut {NoStop}%
\bibitem [{\citenamefont {{Jackson Kimball}}\ \emph {et~al.}(2016)\citenamefont
  {{Jackson Kimball}}, \citenamefont {Dudley}, \citenamefont {Li},
  \citenamefont {Thulasi}, \citenamefont {Pustelny}, \citenamefont {Budker},\
  and\ \citenamefont {Zolotorev}}]{Jac2016}%
  \BibitemOpen
  \bibfield  {author} {\bibinfo {author} {\bibfnamefont {D.~F.}\ \bibnamefont
  {{Jackson Kimball}}}, \bibinfo {author} {\bibfnamefont {J.}~\bibnamefont
  {Dudley}}, \bibinfo {author} {\bibfnamefont {Y.}~\bibnamefont {Li}}, \bibinfo
  {author} {\bibfnamefont {S.}~\bibnamefont {Thulasi}}, \bibinfo {author}
  {\bibfnamefont {S.}~\bibnamefont {Pustelny}}, \bibinfo {author}
  {\bibfnamefont {D.}~\bibnamefont {Budker}}, \ and\ \bibinfo {author}
  {\bibfnamefont {M.}~\bibnamefont {Zolotorev}},\ }\bibfield  {title} {\enquote
  {\bibinfo {title} {{Magnetic shielding and exotic spin-dependent
  interactions}},}\ }\href {\doibase 10.1103/PhysRevD.94.082005} {\bibfield
  {journal} {\bibinfo  {journal} {Phys. Rev. D}\ }\textbf {\bibinfo {volume}
  {94}},\ \bibinfo {pages} {82005} (\bibinfo {year} {2016})}\BibitemShut
  {NoStop}%
\bibitem [{\citenamefont {Chu}\ \emph {et~al.}(2022)\citenamefont {Chu},
  \citenamefont {Ristoff}, \citenamefont {Smits}, \citenamefont {Jackson},
  \citenamefont {Kim}, \citenamefont {Savukov},\ and\ \citenamefont
  {Acosta}}]{Chu2022}%
  \BibitemOpen
  \bibfield  {author} {\bibinfo {author} {\bibfnamefont {P.-H.}\ \bibnamefont
  {Chu}}, \bibinfo {author} {\bibfnamefont {N.}~\bibnamefont {Ristoff}},
  \bibinfo {author} {\bibfnamefont {J.}~\bibnamefont {Smits}}, \bibinfo
  {author} {\bibfnamefont {N.}~\bibnamefont {Jackson}}, \bibinfo {author}
  {\bibfnamefont {Y.~J.}\ \bibnamefont {Kim}}, \bibinfo {author} {\bibfnamefont
  {I.}~\bibnamefont {Savukov}}, \ and\ \bibinfo {author} {\bibfnamefont
  {V.~M.}\ \bibnamefont {Acosta}},\ }\bibfield  {title} {\enquote {\bibinfo
  {title} {{Proposal for the search for new spin interactions at the micrometer
  scale using diamond quantum sensors}},}\ }\href {\doibase
  10.1103/PhysRevResearch.4.023162} {\bibfield  {journal} {\bibinfo  {journal}
  {Phys. Rev. Res.}\ }\textbf {\bibinfo {volume} {4}},\ \bibinfo {pages}
  {23162} (\bibinfo {year} {2022})}\BibitemShut {NoStop}%
\bibitem [{\citenamefont {Liang}\ \emph {et~al.}(2022)\citenamefont {Liang},
  \citenamefont {Jiao}, \citenamefont {Huang}, \citenamefont {Yu},
  \citenamefont {Ye}, \citenamefont {Wang}, \citenamefont {Xie}, \citenamefont
  {Cai}, \citenamefont {Rong},\ and\ \citenamefont {Du}}]{Lia2022}%
  \BibitemOpen
  \bibfield  {author} {\bibinfo {author} {\bibfnamefont {H.}~\bibnamefont
  {Liang}}, \bibinfo {author} {\bibfnamefont {M.}~\bibnamefont {Jiao}},
  \bibinfo {author} {\bibfnamefont {Y.}~\bibnamefont {Huang}}, \bibinfo
  {author} {\bibfnamefont {P.}~\bibnamefont {Yu}}, \bibinfo {author}
  {\bibfnamefont {X.}~\bibnamefont {Ye}}, \bibinfo {author} {\bibfnamefont
  {Y.}~\bibnamefont {Wang}}, \bibinfo {author} {\bibfnamefont {Y.}~\bibnamefont
  {Xie}}, \bibinfo {author} {\bibfnamefont {Y.-F.}\ \bibnamefont {Cai}},
  \bibinfo {author} {\bibfnamefont {X.}~\bibnamefont {Rong}}, \ and\ \bibinfo
  {author} {\bibfnamefont {J.}~\bibnamefont {Du}},\ }\bibfield  {title}
  {\enquote {\bibinfo {title} {{New Constraints on Exotic Spin-Dependent
  Interactions with an Ensemble-NV-Diamond Magnetometer}},}\ }\href
  {https://academic.oup.com/nsr/advance-article/doi/10.1093/nsr/nwac262/6832283}
  {\bibfield  {journal} {\bibinfo  {journal} {National Science Review}\ }
  (\bibinfo {year} {2022})}\BibitemShut {NoStop}%
\bibitem [{\citenamefont {Gregorovi{\v{c}}}\ and\ \citenamefont
  {Apih}(2008)}]{Gre2008}%
  \BibitemOpen
  \bibfield  {author} {\bibinfo {author} {\bibfnamefont {A.}~\bibnamefont
  {Gregorovi{\v{c}}}}\ and\ \bibinfo {author} {\bibfnamefont {T.}~\bibnamefont
  {Apih}},\ }\bibfield  {title} {\enquote {\bibinfo {title} {{Relaxation during
  spin-lock spin-echo pulse sequence in N14 nuclear quadrupole resonance}},}\
  }\href {\doibase 10.1063/1.3023091} {\bibfield  {journal} {\bibinfo
  {journal} {The Journal of Chemical Physics}\ }\textbf {\bibinfo {volume}
  {129}},\ \bibinfo {pages} {214504} (\bibinfo {year} {2008})}\BibitemShut
  {NoStop}%
\end{thebibliography}
\bibliographystyle{apsrev4-1}
\end{document}